\documentclass[final,3p,authoryear]{elsarticle}
\usepackage{amssymb}
\usepackage{amsthm}
\usepackage{graphics} 
\usepackage{epsfig} 
\usepackage{amsmath} 
\usepackage{hyperref}
\usepackage{multirow}
\usepackage{graphicx}
\usepackage{epstopdf}
\usepackage{subcaption}
\usepackage{siunitx}
\usepackage{bm}
\usepackage{float}
\usepackage{wasysym}
\usepackage{xcolor}
\usepackage[flushleft]{threeparttable}
\definecolor{carmine}{rgb}{0.59, 0.0, 0.09}

\usepackage{color}
\graphicspath{{./fig/}}
\usepackage{lineno}
\setpagewiselinenumbers
\usepackage{natbib}
\usepackage{algorithm}
\usepackage{algpseudocode}
\usepackage{algpascal}
\usepackage{soul}
\usepackage{titlesec}
\usepackage{longtable}
\usepackage{array}
\usepackage{sectsty}
\newcolumntype{M}[1]{>{\centering\arraybackslash}m{#1}}
  
\journal{}
\makeatletter
\def\ps@pprintTitle{%
  \let\@oddhead\@empty
  \let\@evenhead\@empty
  \def\@oddfoot{}%
  \let\@evenfoot\@oddfoot}
\makeatother
\begin{document}

\begin{frontmatter}



\title{\textbf{A new computational perceived risk model for automated vehicles based on potential collision avoidance difficulty (PCAD)}}


\author[1]{Xiaolin He\corref{cor1}} \ead{x.he-2@tudelft.nl}
\author[1]{Riender Happee} 
\author[2]{Meng Wang}

\address[1]{Department of Cognitive Robotics, Delft University of Technology, Mekelweg 2, 2628 CD, Delft, the Netherlands}
\address[2]{Chair of Traffic Process Automation,``Friedrich List" Faculty of Transport and Traffic Sciences, Technische Universität Dresden, Hettnerstraße 3,01069 Dresden, Germany}
\begin{abstract}
{Perceived risk is crucial in designing trustworthy and acceptable vehicle automation systems.
However, our understanding of its dynamics is limited, and models for perceived risk dynamics are scarce in the literature. 
This study formulates a new computational perceived risk model based on potential collision avoidance difficulty (PCAD) for drivers of SAE level 2 driving automation. PCAD uses the 2D safe velocity gap as the potential collision avoidance difficulty, and takes into account collision severity. The safe velocity gap is defined as the 2D gap between the current velocity and the safe velocity region, and represents the amount of braking and steering needed, considering behavioural uncertainty of neighbouring vehicles and imprecise control of the subject vehicle. The PCAD predicts perceived risk both in continuous time and per event. We compare the PCAD model with three state-of-the-art models and analyse the models both theoretically and empirically with two unique datasets: Dataset Merging and Dataset Obstacle Avoidance. The PCAD model generally outperforms the other models in terms of model error, detection rate, and the ability to accurately capture the tendencies of human drivers' perceived risk, albeit at a longer computation time. Additionally, the study shows that the perceived risk is not static and varies with the surrounding traffic conditions. This research advances our understanding of perceived risk in automated driving and paves the way for improved safety and acceptance of driving automation systems.}
\end{abstract}

\cortext[cor1]{Corresponding author. Tel: +31 15 27 84578}
\end{frontmatter}

\section{Introduction}\label{intro}
Road crashes are a leading cause of injury and death worldwide, resulting in approximately 1.35 million deaths and 20-50 million non-fatal injuries each year \citep{WHO2020RoadInjuries}. Most traffic accidents arise from human misjudgements \citep{Nadimi2016CalibrationSystem}. Specifically, distorted perception of driving risk by human drivers is one of the important causes of road accidents \citep{Eboli2017HowStyle}. 
 
Perceived risk captures the level of risk experienced by drivers, which can differ from operational (or actual) risk \citep{Griffin2020PatternsExperience,Kolekar2020Human-likeModel}. A low perceived risk leads to feeling safe, relaxed, and comfortable, while a high-risk perception results in cautious behaviour \citep{Griffin2020PatternsExperience}. The advent of active safety and driving automation systems has reduced actual risk, but changes in drivers' risk perception have been observed. Human drivers will inversely perceive a high level of risk if the driving automation shows inappropriate driving behaviours, causing decreased trust, low acceptance, and even refusal of vehicle automation \citep{Xu2018WhatExperiment}. In manual driving, maintaining perceived risk below a specific threshold motivates drivers' actions, such as steering and braking \citep{Summala1988RiskImplications}. Consequently, misperception of risk during automated driving may cause drivers’ to distrust and intervene unnecessarily while in other cases drivers may fail to recognise dangerous situations that require drivers’ intervention. Therefore, it is essential to comprehend and quantify drivers’ perceived risk in driving automation and in turn, use it to design driving automation which is not only technically safe, but is also perceived as safe.

Computational models for estimating perceived risk have been developed, falling into two categories: empirical models reliant on data, and mechanistic models grounded in first principles. In the first category, \cite{Kolekar2020Human-likeModel} established a driving risk field (DRF) model considering the probability of an event occurring and the event consequence based on drivers' subjective risk ratings and steering responses. \cite{Ping2018ModelingLearning} used deep learning methods to model perceived risk in urban scenarios with factors related to the subject vehicle and the driving environment. Our previous study \citep{He2022} built a regression-based perceived risk model to explain and compute event-based perceived risk in highway merging and braking scenarios. Among other factors, the model captures the influence of relative motion with respect to other road users on drivers’ subjective perceived risk ratings. 

Mechanistic perceived risk models typically rely on surrogate measures of safety (SMoS). The minimum time to collision (TTC) can show the drivers' threshold of perceived risk when they take last-moment braking actions \citep{Kiefer2005DevelopingJudgments}. The inverse TTC represents drivers' relative visual expansion of an obstacle, which can indicate drivers' risk perception \citep{Lee1976ACollision}. Additionally, \cite{kondoh2008identification, Kondoh2014DirectSituations} further analysed the relationship between drivers' risk perception regarding the leading vehicle and inverse TTC and time headway (THW) in car-following situations. Models using TTC and THW only capture one-dimensional (1D) interaction and are mainly validated for car following. Attempts have been made to model risk for two-dimensional (2D) motion based on the driving risk field theory \citep{Wang2016DrivingSystem} and develop collision warning algorithms  \citep{Li2020RiskTheory}. This research line is advanced by the probabilistic driving risk field model (PDRF) \citep{Mullakkal-Babu2020ProbabilisticAssessment} by considering motion probability distributions of other road users and the collision severity to estimate the collision risk. Although the above-mentioned models estimate the actual collision risk rather than the perceived risk, they are promising to predict human drivers' risk perception thanks to the strong connection between the actual risk and the perceived risk. 

The empirical models reviewed above accurately quantify perceived risk in certain scenarios, but lack validation across diverse situations and are not fully explainable. Mechanistic models, while explainable, can compute the actual risk. However, the mapping between the actual collision risk and perceived risk remains ambiguous and the thresholds of the SMoSs lack empirical support. Hence, an explainable and validated computational perceived risk model is still lacking. 

This study has two primary objectives: \textbf{Objective 1} is to formulate an explainable computational perceived risk model grounded in the human drivers' risk perception mechanism applicable to general 2D movements.  \textbf{Objective 2} is to analyse and compare our new model against existing models both theoretically and empirically. The model uses the velocity gap to the safe velocity region as the potential collision avoidance difficulty to quantify perceived risk. The safe velocity region accounts for vehicles' kinematics with uncertainty, as well as collision severity.  The model describes perceived risk in continuous time and per event, and is validated using event-based reported perceived risk. The model is developed towards the general driver population instead of personalised modelling but can capture individual differences by tuning model parameters. 

The remainder of this paper is structured as follows. We first revisit three computational perceived risk models from literature in Section \ref{chap:existing models}, and then present the formulation of the new model in Section \ref{chap:PCAD modelling}. Perceived risk data, model calibration approach and model performance indices are introduced in Section \ref{chap:Methodology}. The model evaluation results are represented in Section \ref{chap:Results} followed by a discussion in Section \ref{chap:discussions}, and conclusions in Section \ref{chap:Conclusions and future work}. 

\section{Related perceived risk models} \label{chap:existing models}
This section introduces the preliminaries for perceived risk modelling and three models that provide a baseline for model performance evaluation.  

\subsection{Coordinate system, reference points definition and vehicle model} \label{chap:coordinate system}
\begin{figure}[H]
    \centering
    \includegraphics[width=0.5\textwidth]{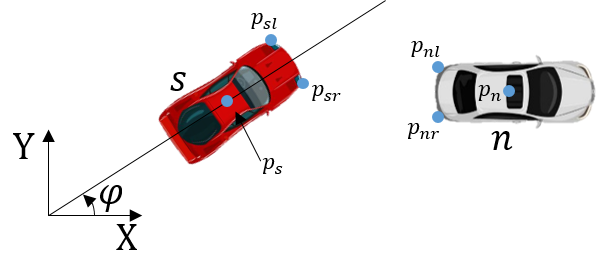}
    \caption{The coordinate system in the definition}
    \label{fig:coordinate system}
\end{figure}

All models in this study employ the same coordinate system. The road space is modelled as a flat Euclidian plane. The $X$-axis aligns with the direction of the road, while the $Y$-axis is perpendicular to it, oriented counter-clockwise, as illustrated in Figure \ref{fig:coordinate system}. Given our focus on perceived risk based on relative motion, rather than vehicle dynamics, we employ a simple point mass model incorporating vehicle dimension. According to the point mass model, the positions, velocities and accelerations of the geometric centre for both the subject vehicle $s$ and a neighbouring vehicle or an obstacle $n$ are $\boldsymbol{p}_s = [x_s, y_s]^T$, $\boldsymbol{p}_n = [x_n, y_n]^T$, $\boldsymbol{v}_s = [v_{s,X}, v_{s,Y}]^T$, $\boldsymbol{v}_n = [v_{n,X}, v_{n,Y}]^T$, $\boldsymbol{a}_s = [a_{s,X}, a_{s,Y}]^T$, $\boldsymbol{a}_n = [a_{n,X}, a_{n,Y}]^T$ respectively. The heading angle $\psi$ follows from the vehicle velocity direction for the point mass model. 

Vehicle dimensions are incorporated into the perceived risk models. Figure \ref{fig:coordinate system} illustrates that the leftmost and the rightmost points in the front side of the subject vehicle and the rear side of the neighbouring vehicle that is closest to the subject vehicle are the reference points. Given the vehicle's length $L$ and width $W$ in straight drive, the positions of the reference points are  $\boldsymbol{p}_{sl} = \boldsymbol{p}_s + [L/2 \; \; W/2]^T$ and $\boldsymbol{p}_{sr} = \boldsymbol{p}_s + [L/2 \; \; -W/2]^T$ for the subject vehicle, $\boldsymbol{p}_{nl} = \boldsymbol{p}_s + [-L/2 \; \; W/2]^T$ and $\boldsymbol{p}_{nr} = \boldsymbol{p}_s + [-L/2 \; \; -W/2]^T$ for the neighbouring vehicle.

\subsection{Regression Perceived Risk Model (RPR)} \label{chap: RPR}
The Regression Perceived Risk model (RPR) is an event-based perceived risk model derived from our previous simulator experiment, where 18 merging events with various merging distances and braking intensities on a 2-lane highway were simulated. RPR predicts human drivers' event-based perceived risk ratings ranging from 0-10 regarding merging events based on the corresponding kinematic data from the simulator drive. The details of the experiment data will be introduced later. 

The RPR model builds on several assumptions:
\begin{itemize}
    \item Perceived risk stems from the vehicles directly in front of the subject vehicle, which means the merging vehicles cause perceived risk only after entering the current lane.
    \item Drivers can accurately estimate the motion information (e.g., relative position, velocity, acceleration, etc.) with the human sensory system. 
\end{itemize}

The initial model can predict event-based perceived risk \citep{He2022}, as shown in Equation \eqref{rpre}
\begin{linenomath}
\begin{equation}
    R_{RPR\_event} = 9.384 - 2.473\cdot \ln (gap_{min}) - 0.038 \cdot YDL - 0.201 \cdot BI_{max} + 0.470 \cdot GEN
    \label{rpre}
\end{equation}
\end{linenomath}
where $R_{RPR\_event}$ is the event-based perceived risk ranging from 0-10; $gap_{min}$ is the minimum clearance in metres to the leading vehicle during an event; $YDL$ represents the years with a valid driving license; $BI_{max}$ denotes the maximum braking intensity ($\SI{}{m/s^2}$) of the merging vehicle; $GEN$ represents the gender of the participants with $Female = 1$ and $Male = 0$. The model coefficients were originally calibrated based on a perceived risk dataset \citep{He2022}, which will be detailed in Section \ref{Chap: Dataset introduction}.

We extend the model to compute real-time perceived risk by replacing $min \_ gap$ and $max \_ BI$ with the real-time values. $YDL$ and $GEN$ are omitted since they remain constant for a certain group of participants, and can be accounted for by the intercept. In this way, RPR is formulated in the continuous time domain as 
\begin{linenomath}
\begin{equation}
   R_{RPR}(t) = C_{0}+C_{1} \cdot \ln (x_n(t)-x_s(t))+C_{2} \cdot a_{n,b}(t)
    \label{rprc}
\end{equation}
\end{linenomath}
where $x_n(t)$ and $x_s(t)$ are the real-time longitudinal position (m) of the neighbouring vehicle and the subject vehicle; $a_{n,b}(t)$ is the current acceleration (\SI{}{m/s^2}) of the neighbouring vehicle, which is the braking intensity in this model; According to the simulator experiment settings \citep{He2022}, the validity range of the model is that $x_n(t)-x_s(t)<\SI{33} {m}$ and $\SI{-8} {m/s^2} \leqslant a_{nb}(t) \leqslant \SI{-2} {m/s^2}$. Verification is required for the model outside this range. For enhanced performance, we will later calibrate parameters $C_0$, $C_1$ and $C_2$ using different datasets. 

\subsection{Perceived Probabilistic Driving Risk Field Model (PPDRF)} \label{chap: PPDRF}
Perceived Probabilistic Driving Risk Field Model (PPDRF) enhances Probabilistic Driving Risk Field Model (PDRF) \citep{Mullakkal-Babu2020ProbabilisticAssessment} by accounting for diverse traffic scenarios and driver individuality. The model is inspired by artificial potential field used in driving automation \citep{Wang2016DrivingSystem, Li2020RiskTheory, Ni2013ATheory}. PDRF estimates collision risk by considering potential risk from non-moving vehicles/objects and kinetic risk from other road users. The former accounts for collision energy and probability with stationary obstacles, while the latter involves spatial overlap with neighbouring vehicles using predicted positions and stochastic accelerations. In stable highway driving, the longitudinal and lateral accelerations of neighbour follow a Gaussian distribution \citep{Wagner2016AnalyzingMethods, Ko2010AnalysisVehicles}. However, due to uncertainties and behavioural deviations, human drivers perceive risk differently, leading to a bias between objective and perceived risk. To address this, we introduce assumptions to extend PDRF into PPDRF for predicting perceived risk.

\begin{itemize}
    \item The future acceleration in longitudinal and lateral directions of neighbouring vehicles follows independent Gaussian distributions with the current acceleration as the mean value, remaining constant over the prediction horizon;
    \item The subject vehicle maintains the current acceleration over the prediction horizon;
\end{itemize}

The two assumptions simplify road users' motion. 

In PPDRF model, human drivers, at time $t$, perceive a total risk as a combination of kinetic risk and potential risk as follows
\begin{linenomath}
\begin{equation}
    R_{PPDRF}(t)=R_{n,s}(t)+R_{o,s}(t)
\label{eq: Total risk}
\end{equation}
\end{linenomath}

The kinetic risk in PPDRF concerning moving neighbouring vehicles is given by
\begin{linenomath}
\begin{equation}
    R_{n, s}(t)=0.5 M_{s} \beta^{2}\left|\Delta {v}_{s, n}(t+\tau)\right|^{2} \cdot \tilde{p}(n, s \mid t)
    \label{perceived probabilistic driving risk field}
\end{equation}
\end{linenomath}
where $R_{n, s}(t)$ is the kinetic collision risk between the subject vehicle $s$ and a neighbouring vehicle $n$ in Joules at time $t$. $\beta  = \frac{{{M_n}}}{{{M_s} + {M_n}}}$ denotes the mass ratio. $M_s$ and $M_n$ are the mass of the subject vehicle and the neighbouring vehicle. $\Delta {v}_{s, n}(t+\tau)$ is the relative velocity between the subject vehicle and the neighbouring vehicle at time $t+\tau$. $\tilde{p}(n, s \mid t)$ is the collision probability to the neighbouring vehicle estimated by drivers ranging on [0,1]. 

\begin{figure}[H]
    \centering
    \includegraphics[width=0.4\textwidth]{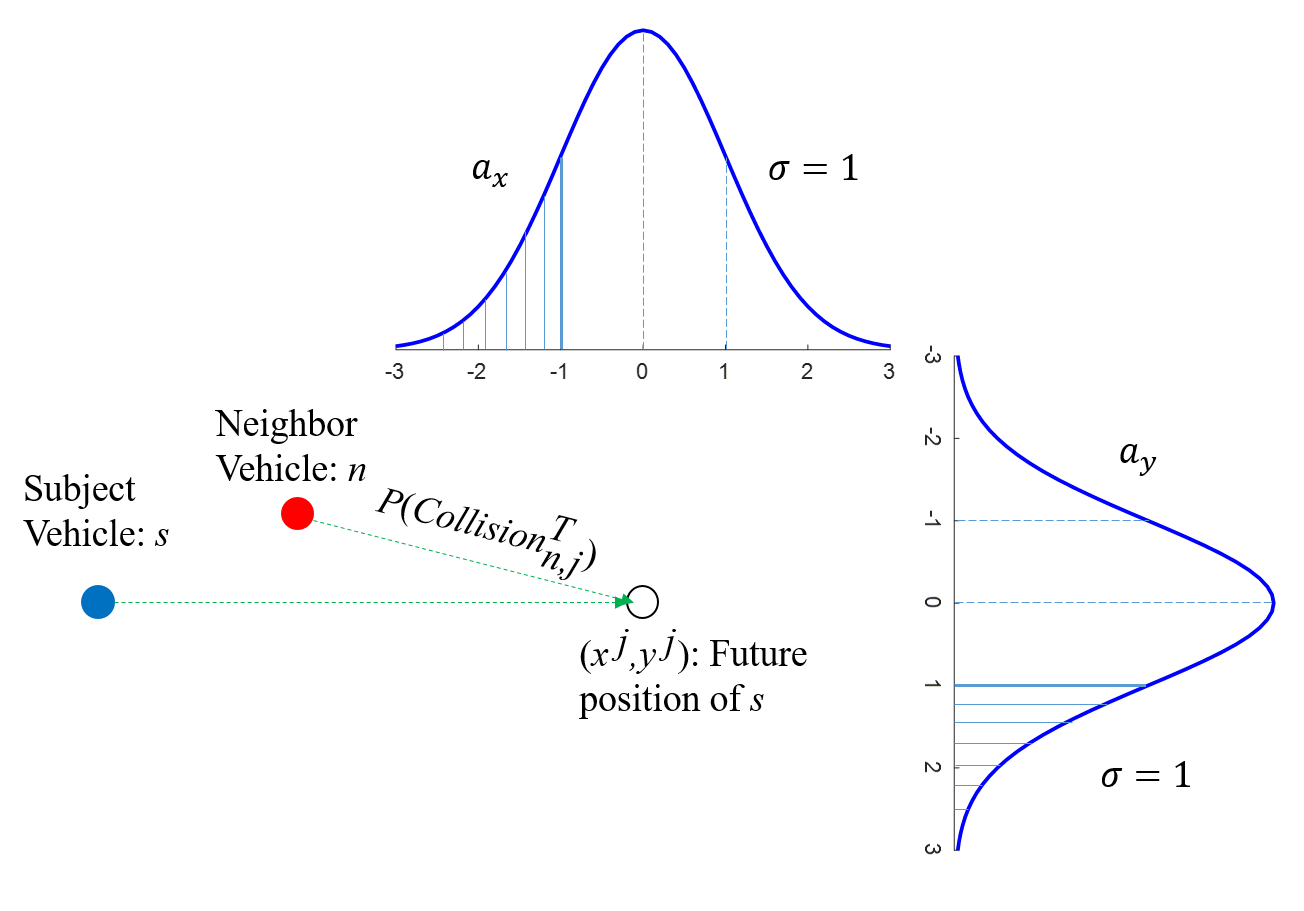}
    \caption{The acceleration distribution of the neighbouring vehicle and the relative spacing between the subject vehicle and the neighbouring vehicle}
    \label{fig:acceleration probability distribution}
\end{figure}

The collision probability $\tilde{p}(n, s \mid t)$ to the neighbouring vehicle estimated by human drivers is constructed as Equation \eqref{estimated collision probability}.
\begin{linenomath}
\begin{equation}
   \tilde{p}(n, s \mid t)=\mathcal{N}\left(\frac{\Delta x(t)-\Delta v_{X}(t) \tau}{0.5 \tau^{2}} \mid \mu_{X}(t), \tilde{\sigma}_{X}\right) \cdot \mathcal{N}\left(\frac{\Delta y(t)-\Delta v_{Y}(t) \tau}{0.5 \tau^{2}} \mid \mu_{Y}(t), \tilde{\sigma}_{Y}\right)
    \label{estimated collision probability}
\end{equation}
\end{linenomath}
where $\mathcal{N}$ is the assumed Gaussian collision probability density function (Figure \ref{fig:acceleration probability distribution}). $\mu_X(t)$ and $\mu_Y(t)$ represent the mean values for longitudinal and lateral acceleration distribution, while $\tilde{\sigma}{X}$ and $\tilde{\sigma}{Y}$ are the respective standard deviations. The relative spacing and velocities between the subject and neighbouring vehicles are denoted as $\Delta x(t)$, $\Delta y(t)$, $\Delta v_X (t)$, and $\Delta v_Y (t)$. PPDRF evaluates collision probability using multiple values of $\tau = \SI{0.5}{s}, \SI{1}{s}, \SI{2}{s}, \SI{3}{s}$, with the model employing all these values to maximise the computed collision probability. Using the constant acceleration assumption, the predicted position of the subject vehicle and stochastic positions of neighbouring vehicles are calculated over a prediction horizon. Spatial overlap and collision predictions are determined accordingly. The actual $\tilde{p}(n, s \mid t)$ is obtained through integration over the expected accelerations.

The potential risk posed to vehicle $s$ by a static object $o$ can be modelled as
\begin{linenomath}
\begin{equation}
    R_{o, s}(t)=
0.5 k M\left(\Delta v_{s, o}\right (t))^{2} \cdot \max \left(e^{-\left|r_{s, o}\right| / D}, 0.001\right) 
\label{eq:potential field}
\end{equation}
\end{linenomath}
where $R_{o,s}(t)$ denotes the potential risk caused by the static object $o$; $M$ denotes the mass of $s$; $\left|r_{s,o} \right| = ||\boldsymbol{p}_s-\boldsymbol{p}_n||$ is the distance between the subject vehicle $s$ and the non-moving object $o$; $V_{s,o}$ denotes the relative velocity; $0.5kM(V_{s,o})^2$ represents the expected crash energy scaled by the parameter $k$, with range $[0-1]$, which is set to 1 in this study representing the neighbour is immovable; the term $e^{-\left|r_{s, o}\right| / D}$ is the collision probability ranging between [0-1], where $D$ determines the steepness of descent of the potential field, varying among different drivers. 

It is noteworthy that $R_{PPDRF}(t)$ represents a probabilistic energy value, and can attain up to \SI{3e4}{J} under stable motorway driving conditions \citep{Mullakkal-Babu2020ProbabilisticAssessment}. 

\subsection{Driving risk field model (DRF)}\label{chap: DRF}
The DRF represents human drivers' risk perception as a 2D field, combining the probability (probability field) and consequence (severity field) of an event \citep{Kolekar2020Human-likeModel}, the product of which provides an estimation of driver’s perceived risk. The DRF model was derived from a simulator experiment involving obstacle avoidance with 77 obstacles distributed on a 2D plane in front of the subject vehicle. During each drive, one obstacle was randomly chosen and suddenly appeared, after which participants needed to steer to avoid the obstacle and gave a non-negative number indicating required steering effort. Based on the position information of the obstacles, the steering angle, and the subjective ratings, the DRF model fitted to the data, and thereby it is essentially an empirical model. 
The DRF is based on the following assumptions:
\begin{itemize}
    \item Perceived risk is the product of the probability of a hazardous event occurring estimated by drivers and the event severity;
    \item The perceived risk field widens as the longitudinal distance from the subject vehicle increases;
    \item The height of the perceived risk field decays as the lateral and longitudinal distance from the vehicle increases; 
\end{itemize}

The DRF model quantifies overall perceived risk as 
\begin{linenomath}
\begin{equation}
    R_{DRF}(t)=\sum p(x(t), y(t)) \cdot {sev}(t)
\label{DRF_RDRF}
\end{equation}
\end{linenomath}
where $p(x(t), y(t))$ is the probability of an event happening at position $(x(t),y(t))$;  ${sev}(t)$ is the severity field of events. Specifically, in straight drive, the probability field can be simplified as
\begin{linenomath}
\begin{equation}
    p(x(t), y(t))=h \cdot \exp \left(\frac{-y(t)^2}{2 \sigma^{2}}\right)
\label{DRF_zx}
\end{equation}
\end{linenomath}
\begin{linenomath}
\begin{equation}
    h=s \cdot \left(x(t)-v_{s,X}(t) \cdot t_{l a}\right)^{2}
\label{DRF_ax}
\end{equation}
\end{linenomath}
\begin{linenomath}
\begin{equation}
    \sigma = m \cdot x(t)+c
\label{DRF_sigma}
\end{equation}
\end{linenomath}
where the subject vehicle is at the origin $(0,0)$ with $h$ and $\sigma$ representing the height and the width of the Gaussian at longitudinal position $x(t)$; $s$ defines the steepness of the height parabola; $t_{la}$ is the human driver's preview time (s); $m$ defines the widening rate of the 2D probability field; $c$ is the quarter width of the subject vehicle (m). $v_{s,X}(t)$ is the subject vehicle's velocity (m/s). The lateral cross-section of the 2D probability field is a Gaussian. Note that the height of the Gaussian $h$ and the width $\sigma$ are separately modelled as a parabola and linear function of longitudinal distance $x$ in front of the subject vehicle. 

The severity field of the events in this study can be defined as 
\begin{linenomath}
\begin{equation}
    {sev}(t) = \begin{cases}C_{sev}, & (x(t), y(t)) \in A^O, \\ 0, & (x(t), y(t)) \notin A^O.\end{cases}
\label{DRF_costmap}
\end{equation}
\end{linenomath}
where $C_{sev}$ is the severity value that is set empirically and $A^O$ represents a neighbouring vehicle's spatial area. 

\section{Potential collision avoidance difficulty model (PCAD)} \label{chap:PCAD modelling}
Our proposed model is grounded in Fuller's Risk Allostasis Theory which proposes that a feeling of risk can be indicated by the driving task difficulty \citep{Fuller2011DriverAllostasis} and drivers' primary driving task is to perform avoidance actions to moderate the perceived risk to a preferred range \citep{Fuller1984}. Consequently, we develop a dynamic perceived risk model by quantifying the driving task difficulty, which computes real-time perceived risk and explain its underlying mechanism. We quantify the task difficulty considering the 2D velocity change to avoid a potential collision. Additionally, the model accounts for the behaviour uncertainties of other road users reflected in their velocities and imprecision in longitudinal and lateral control as motion uncertainties of the subject vehicle. In this section, we introduce the primary assumptions and the general structure of the model followed by a detailed explanation of each component, including the potential collision judgement method and the perceived velocity of a neighbouring vehicle and the subject vehicle, and a weighting function that considers the collision severity. 

\subsection{Assumptions} \label{chap:assumptions and parameters}
To operationalise the model, we adopt several simplifying assumptions: 
\begin{itemize}
    \item \textbf{Assumption 1}: Human drivers perceive risk based on an estimation of the difficulty in avoiding a potential collision according to their visual perception of the relative motion of the subject vehicle and neighbouring vehicles. They judge whether a vehicle is on a collision course based on looming \citep{Ward2015ExtendingScenarios}. 
    \item \textbf{Assumption 2}: Motion uncertainties of neighbouring and subject vehicles cause extra perceived risk and are presented as uncertain accelerations following a specific distribution (e.g., Gaussian with zero means in this study) and the acceleration will stay constant in a short time period. 
    \begin{itemize}
        \item \textbf{2a}: The uncertain acceleration of neighbouring vehicles comes from manoeuvre uncertainties (e.g., a sudden brake). 
        \item \textbf{2b}: The uncertain acceleration of the subject vehicle is caused by the imprecise control in steering and throttle/braking, which is relevant to human drivers' control ability or driving automation's performance. 
    \end{itemize}
    \item \textbf{Assumption 3}: Human drivers feel higher perceived risk with higher subject velocity due to more severe collision consequences.
\end{itemize}

\subsection{General structure of PCAD}
Let $\boldsymbol{x}_s = (\boldsymbol{p}_s, \boldsymbol{v}_s, \boldsymbol{a}_s)^T$ and $\boldsymbol{x}_n = (\boldsymbol{p}_n, \boldsymbol{v}_n, \boldsymbol{a}_n)^T$ denote the state of the subject vehicle $s$ and the neighbouring vehicle $n$ respectively, with $\boldsymbol{p}_s$ and $\boldsymbol{p}_n$, $\boldsymbol{v}_s$ and $\boldsymbol{v}_n$,  $\boldsymbol{a}_s$ and $\boldsymbol{a}_n$ being the position, velocity and acceleration vectors, and $T$ the transpose of a vector. The PCAD is formulated as Equation \ref{eq:general structure}
\begin{equation}
    R_{PCAD}(t) = \mathcal{A}(\boldsymbol{p}_s, \boldsymbol{p}_n, \mathcal{V}_s(\boldsymbol{x}_s, \boldsymbol{x}_n), \mathcal{V}_n(\boldsymbol{x}_s, \boldsymbol{x}_n))\cdot \mathcal{W}(\boldsymbol{v}_s)
\label{eq:general structure}
\end{equation}
Here, $\mathcal{A}$ represents the avoidance difficulty function. This function quantifies the required 2D velocity change to bring the subject vehicle to the safe velocity region in the velocity domain to avoid a potential collision with the neighbouring vehicle, considering factors such as their relative positions, velocities and accelerations. $\mathcal{V}_i$ denotes the 2D perceived velocity for vehicle $i \in \{s,n\}$, thereby capturing absolute and relative motion of the interacting vehicles. Finally, $\mathcal{W}$ is the weighting function, being a Power function with $v_s$, which accounts for the influence of the subject vehicle's speed on perceived risk. Higher speeds generally increase the perceived risk, as the consequence of a potential collision is more severe.

The perceived velocity function $\mathcal{V}$ can be represented as
\begin{equation}
    \mathcal{V}_i(\boldsymbol{x}_s,\boldsymbol{x}_n)=\boldsymbol{v}_i'=\boldsymbol{v}_i+\Delta\boldsymbol{v}_{i,a}+\boldsymbol{v}_{i,I}
    \label{eq:perceived velocity}
\end{equation}
where $\mathcal{V}_i$ is the functional operator to compute the perceived velocity $\boldsymbol{v}_i'$ of the vehicle $i \in \{s,n\}$ by human drivers for perceived risk computation.  The perceived velocity $\boldsymbol{v}_i'$ combines three components: the actual velocity $\boldsymbol{v}_i$ ($i \in \{s,n\}$), an adjustment $\Delta \boldsymbol{v}_{i,a} (i \in \{s,n\})$ based on the vehicle's acceleration, which is relevant to an accumulation time $t_{i,a}$, and an ``imaginary'' velocity $\boldsymbol{v}_{i,I}$ that accounts for uncertainties in vehicle motion  (\textbf{Assumption 2a} and \textbf{Assumption 2b}). For example, consider a driver who notices a car ahead is accelerating rapidly. He might perceive the car's velocity to be higher than it actually is because he anticipates its future motion based on the acceleration. The imaginary velocity component captures uncertainties in vehicle motion, such as a neighbouring vehicle suddenly swerving or the subject vehicle's imprecise control. According to \textbf{Assumption 2}, the uncertainties or the imprecise control are uncertain accelerations, which are assumed to follow a Gaussian distribution. Given an accumulation time period, the imaginary velocity $\boldsymbol{v}_{i,I}$ also follows a Gaussian distribution, which is $\mathcal{D}_i$ in this study involving parameters $\sigma_{i,X}$ and $\sigma_{i,Y}$ as its standard deviations both in longitudinal and lateral directions with zero mean. 

\subsection{Collision avoidance difficulty function $\mathcal{A}$ in deterministic conditions}\label{chap: avoidance difficulty}
In this section, the collision avoidance difficulty is formulated to capture part of human drivers' perceived risk under constant speed and deterministic motion conditions. The perceived velocity (\ref{eq:perceived velocity}) relaxes to the actual velocity $\boldsymbol{v}$ under such conditions. Uncertainties and acceleration are incorporated in the next section. 

\subsubsection{Potential collision judgement of human drivers \textemdash Looming detection}
A precedent step for collision avoidance is to detect a potential collision based on the current environment information. One observation from aircraft pilots is that two aircraft are on a crossing course if they remain in the same position in their field of view. Similarly, in road traffic, one vehicle lies on a crossing course at a specific moment, if its relative bearing to you does not change \citep{Ward2015ExtendingScenarios}. Additionally, if the vehicle is simultaneously approaching, a phenomenon known as \textit{looming} is occurring. This situation indicates a risk of collision (see two vehicles in interaction in Figure \ref{fig:collison_course}). To identify this phenomenon and anticipate a potential collision is referred to as \textit{looming detection}.

To mathematically define \textit{looming}, the same coordinate system (Figure \ref{fig:coordinate system}) as in Section \ref{chap:coordinate system} is used. We take into account vehicle size (Figure \ref{fig:collison_course}) to calculate \textit{looming} with the reference points located at the leftmost and rightmost points of the subject vehicle's front and the neighbouring vehicle's closest side. If leftmost points move to the left, and rightmost points move right, looming is detected, and a collision is expected. If all relative movements of different pairs of reference points are consistently in the same direction, either all to the left or all to the right, then no \textit{looming} is detected.

The first criterion for \textit{looming} is that a point of the vehicle stays at the same position in the subject vehicle's field of view. This can be assessed by calculating the relative bearing rate between the subject vehicle and the neighbouring vehicle using Equation \eqref{eq:angular velocity}. 
\begin{equation}
    \dot{\boldsymbol{\theta}}_{sj_1,nj_2}=\frac{\left(\boldsymbol{p}_{sj_1}-\boldsymbol{p}_{nj_2}\right) \times (\boldsymbol{v}_{sj_1}-\boldsymbol{v}_{nj_2})}{\left\|\boldsymbol{p}_{sj_1}-\boldsymbol{p}_{nj_2}\right\|^2}, \  \ j_1,j_2 \in \{l,r\}
\label{eq:angular velocity}
\end{equation}
where all combinations between four reference points should be considered\footnote{In straight drive, the velocity of reference points $\boldsymbol{v}_{i,j}$ ($i \in \{s,n\}, j \in \{l,r\}$) can be simplified as the vehicle's linear velocity  $\boldsymbol{v}_i$  ($i \in \{s,n\}$) without considering vehicle's yaw rate in straight drive. }. If we have 
\begin{equation}
    \min \dot{\boldsymbol{\theta}}_{sj_1,nj_2} \cdot \max \dot{\boldsymbol{\theta}}_{sj_1,nj_2} < 0, \;\; j_1,j_2 \in \{l,r\}, 
\label{eq:looming}
\end{equation}
the subject vehicle and the neighbouring vehicle are on a crossing course. Namely, the neighbouring vehicle stays at the same position in the subject vehicle's field of view.

The second criterion for \textit{looming} is that the neighbouring vehicle is approaching the subject vehicle. The distance between the subject vehicle and the neighbouring vehicle on a 2-D plane is give by Equation \eqref{eq:distance between reference points}
\begin{equation}\label{eq:distance between reference points}
    d_{s,n}= \sqrt{(\boldsymbol{p}_{s}-\boldsymbol{p}_{n})^T(\boldsymbol{p}_{s}-\boldsymbol{p}_{n})}
\end{equation}
Differentiating both sides of Equation \eqref{eq:distance between reference points}, we have the changing rate of the distance and if
\begin{equation}
    \dot{d}_{s,n}=\frac{1}{ d_{s,n}}(\boldsymbol{p}_{s}-\boldsymbol{p}_{n})^T(\boldsymbol{v}_{s}-\boldsymbol{v}_{n}) < 0, 
\label{eq:distance changing rate}
\end{equation}
the neighbouring vehicle is approaching us, which is the second criterion for \textit{looming}.

\begin{figure}[H]
\centering
\begin{subfigure}[t]{0.45\textwidth}
\centering
\includegraphics[width=\textwidth]{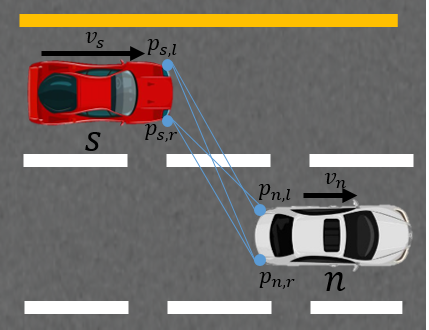}
\caption{The subject vehicle (Red) is overtaking a neighbouring vehicle (White). The neighbouring vehicle (White) is approaching the subject vehicle (Red) but is not remaining the same position (rotating around the subject vehicle). The situation meets Equation \eqref{eq:distance changing rate} but does not meet Equation \eqref{eq:looming}. Hence it is a non-looming case.}
\end{subfigure}
\hfill
\begin{subfigure}[b]{0.45\textwidth}
\centering
\includegraphics[width=\textwidth]{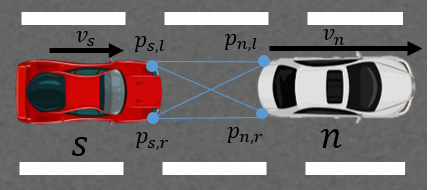}
\caption{The subject vehicle (red) is following a leading vehicle (white) with a slower speed. The neighbouring vehicle stays at the same position but is not approaching the subject vehicle. The situation meets Equation \eqref{eq:looming} but does not meet Equation \eqref{eq:distance changing rate}. It is a non-looming case.}
\end{subfigure}
\caption{Two examples of non-looming.}
\label{fig:Non_collison_course}
\end{figure}

If Equations  \eqref{eq:looming} and \eqref{eq:distance changing rate} are met at the same time, the neighbouring vehicle is looming (Figure \ref{fig:collison_course}). Conversely, if Equations \eqref{eq:looming} and \eqref{eq:distance changing rate} are not met simultaneously, the neighbouring vehicle is not looming (Figure \ref{fig:Non_collison_course}), namely 
\begin{equation}
    \min \dot{\boldsymbol{\theta}}_{sj_1,nj_2} \cdot \max \dot{\boldsymbol{\theta}}_{sj_1,nj_2} \geqslant 0, \;\; j_1,j_2 \in \{l,r\}, 
\label{eq:looming_NonCollision}
\end{equation}
or
\begin{equation}
    \dot{d}_{s,n}=\frac{1}{ d_{s,n}}(\boldsymbol{p}_{s}-\boldsymbol{p}_{n})^T(\boldsymbol{v}_{s}-\boldsymbol{v}_{n})\geqslant 0, 
\label{eq:distance changing rate_NonCollision}
\end{equation}

\begin{figure}[H]
\centering
\begin{subfigure}[t]{0.9\textwidth}
\centering
\includegraphics[width=\textwidth]{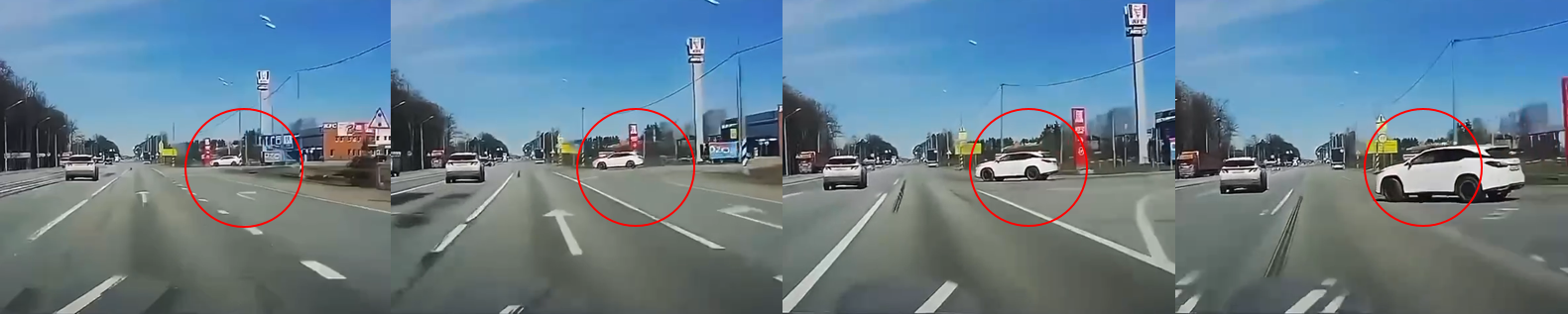}
\subcaption[]{The video stream for a potential collision. The neighbouring vehicle (white) stays at the same position (the red circle) and becomes larger when on a crossing course with the subject vehicle.}
\end{subfigure} 
\end{figure}

\begin{figure}[H] \ContinuedFloat
\centering
\begin{subfigure}[t]{0.9\textwidth}
\centering
\includegraphics[width=\textwidth]{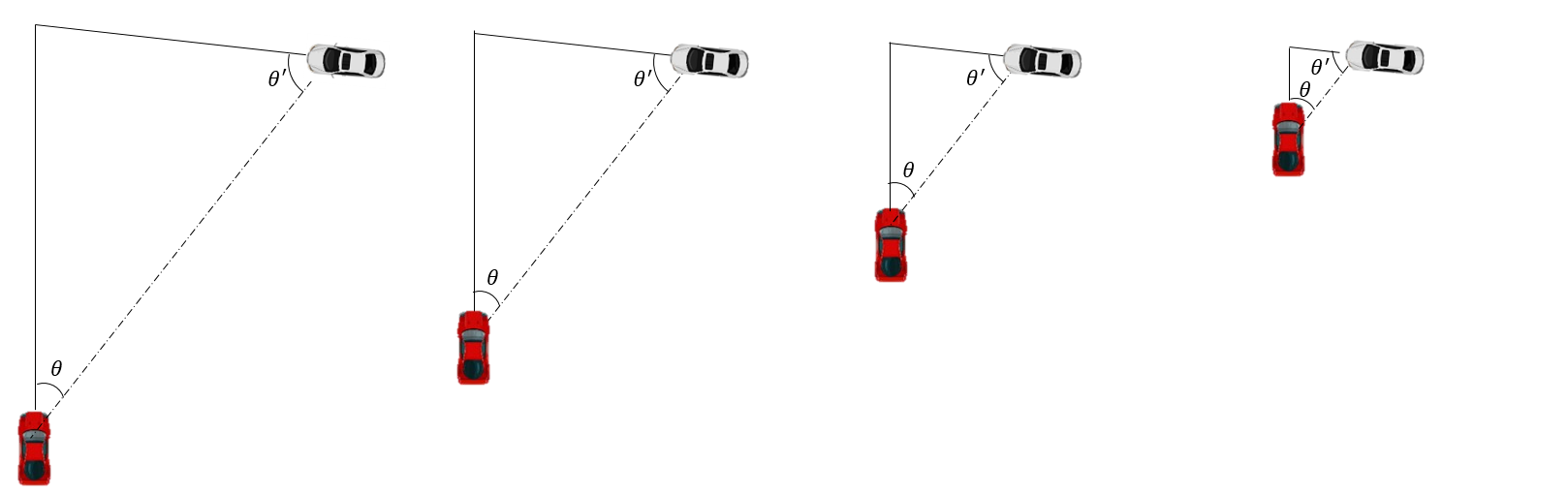}
\subcaption[]{The bird view for the potential collision above. The relative bearing between the subject vehicle (red) and the neighbouring vehicle (white) remains the same when they are on a crossing course.}
\end{subfigure}%
\caption{An example of looming. The subject vehicle and a neighbouring vehicle are on a crossing course.}
\label{fig:collison_course}
\end{figure}

Note that two pairs of reference points are already theoretically sufficient but we use four pairs in Equation \eqref{eq:angular velocity} and Equation \eqref{eq:looming} for a more reliable result in corner cases. Additionally, \textit{Looming detection} is directly valid when the subject vehicle only has translational motion with constant acceleration and thereby follows a straight path. When the subject vehicle has a yaw rate ($\dot{\psi}$), and follows a curved path the theory still stands based on a conformal mapping. \textit{Looming detection} forms the foundation for potential collision judgement by human drivers and the quantification of avoidance difficulty in this study.

\subsubsection{Collision avoidance difficulty} \label{Chap: avoidance difficulty}
We define a safe velocity set $\boldsymbol{V}$, which consists of elements (i.g., all possible subject velocity vectors) that meet Equation \eqref{eq:looming_NonCollision} or \eqref{eq:distance changing rate_NonCollision} based on the position of the two vehicles $\boldsymbol{p}_s$,  $\boldsymbol{p}_s$ and the velocity of the neighbouring vehicle $\boldsymbol{v}_n$ at the current moment in Equation \eqref{eq:velocity set}. 

\begin{equation}
    \forall \boldsymbol{v}_s \in \boldsymbol{V} \Rightarrow  \min \dot{\boldsymbol{\theta}}_{sj_1,nj_2} \cdot \max \dot{\boldsymbol{\theta}}_{sj_1,nj_2} \geqslant 0 \; (j_1,j_2 \in \{l,r\}, )\; \text{or} \; \dot{d}_{s,n}\geqslant0, \; 
\label{eq:velocity set}
\end{equation}
the equal sign stands when $\boldsymbol{v}_s$ is at the boundary of velocity set $\boldsymbol{V}$. 

\begin{figure}[H]
    \centering
    \includegraphics[width=0.8\textwidth]{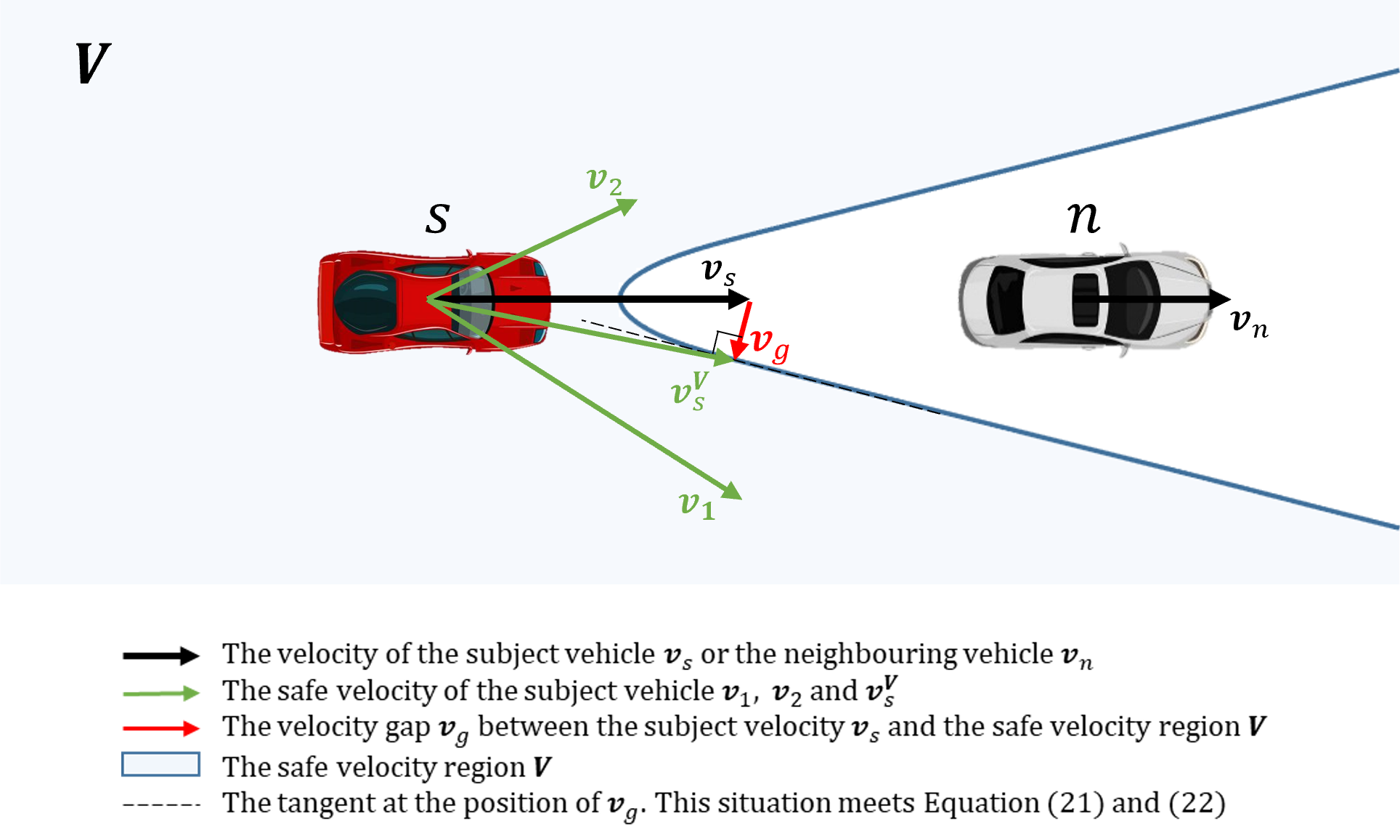}
    \caption{An example to show the collision avoidance difficulty. In this case, the subject vehicle (red) is following a leading vehicle (white, $\SI{50}{m}$ ahead, $\boldsymbol{v}_n=\SI{8.33}{m/s}$) with a higher velocity ($\boldsymbol{v}_s=\SI{16.67}{m/s}$). Equation \eqref{eq:looming_NonCollision} and \eqref{eq:distance changing rate_NonCollision} define the safe velocity set $\boldsymbol{V}$ as the blue area, e.g., if the current subject velocity is any one of the elements in $\boldsymbol{V}$ (e.g., $\boldsymbol{v}_1$ and $\boldsymbol{v}_2$), the neighbouring vehicle (white) is not looming. Namely, the collision avoidance difficulty is zero. In this example, since the subject vehicle is driving faster than the leading vehicle, the current subject velocity $\boldsymbol{v}_s \notin \boldsymbol{V}$, indicating that the neighbouring is looming regarding the subject vehicle. The distance from the subject velocity $\boldsymbol{v_s}$ to the safe velocity set $\boldsymbol{V}$ (the safety boundary) is $\boldsymbol{v}_g$ (the red arrow), the length of which is the defined collision avoidance difficulty. }
    \label{fig:multi options}
\end{figure}

We define the collision avoidance difficulty $||\boldsymbol{v}_g||$ as the 2D distance from the current subject velocity $\boldsymbol{v}_s$ to the boundary of the safe velocity set $\boldsymbol{V}$ (Equation \eqref{eq:velocity gap}) (See Figure \ref{fig:multi options} for illustration). If the current subject velocity $\boldsymbol{v}_s$ is within the set $\boldsymbol{V}$, the velocity gap is zero, which means the collision avoidance difficulty is zero. Hence, the collision avoidance function $\boldsymbol{\mathcal{A}}$ is defined as
\begin{equation}
        \boldsymbol{\mathcal{A}} = ||\boldsymbol{v}_g||=||\boldsymbol{v}_{s}^{\boldsymbol{V}} - \boldsymbol{v}_s||
\label{eq:velocity gap}
\end{equation}
where
\begin{equation}
    \boldsymbol{v}_{s}^{\boldsymbol{V}}=\arg\min_{\boldsymbol{v}_{s}^{\boldsymbol{V}} \in \boldsymbol{V}}||\boldsymbol{v}_{s}^{\boldsymbol{V}} - \boldsymbol{v}_s||
\label{eq:the arguement of the minimum}
\end{equation}
In the equations, $\boldsymbol{v}_g$ is a vector pointing from the current subject velocity $\boldsymbol{v_s}$ to the boundary of the safe velocity set $\boldsymbol{V}$; $\boldsymbol{v}_{s}^{\boldsymbol{V}}$ represents the subject velocity in set $\boldsymbol{V}$, which meets Equation \eqref{eq:the arguement of the minimum}. In this study, the technique of grid search is employed to identify $\boldsymbol{v}_g$ that satisfies both Equation \eqref{eq:velocity gap} and Equation \eqref{eq:the arguement of the minimum}.

\subsection{Perceived velocity function $\mathcal{V}$ of a neighbouring vehicle and the subject vehicle considering acceleration and uncertainties}\label{chap:probabilistic approaching velocity}
The collision avoidance difficulty calculated using the actual (deterministic) motion information (i.e., $\boldsymbol{p}_i$ and $\boldsymbol{v}_i$) (Equation \ref{eq:velocity gap}) presented in the previous section can already account for most of the perceived risk, which will be shown in Section \ref{chap:Results} (Figure \ref{fig:Rsq_Merging}(e) and Figure \ref{fig:Rsq_OV}(e)). However, human perception and behavioural uncertainty are not considered. Therefore, we still cannot consider the collision avoidance difficulty defined above as the perceived risk (\textbf{Assumption 1}). In this section, we define a perceived velocity function $\mathcal{V}$, which considers the acceleration and manoeuvre uncertainties of the subject vehicle and neighbouring vehicles, representing human drivers' perception process of the velocity for the collision avoidance difficulty computation. This perceived velocity function corresponds to perceived risk. The perceived velocity consists of three parts: the actual velocity, the velocity derived from acceleration and the velocity derived from uncertainties. 

\subsubsection{Perceived velocity derived from acceleration  \textemdash the acceleration-based velocity}
Previous studies have shown that human drivers take into account accelerations of the subject and other vehicles \citep{chandler1958traffic}. A model of a driver trying to avoid a collision would lose 20\% of its accuracy if the acceleration of other vehicles was not considered \citep{sultan2004drivers}. For instance, if the leading vehicle starts to brake during car-following, you might initially feel that the situation is dangerous, even if the vehicle is not approaching too quickly. However, after the subject vehicle also brakes, and despite the leading vehicle still getting closer, you may feel less danger.

To account for this, we introduce an accumulation time $t_{i,a}$ and include a velocity perceived by human drivers based on the current acceleration as part of the perceived velocity function $\mathcal{V}$ (Equation \eqref{eq:perceived velocity}). 
\begin{equation}
    \Delta\boldsymbol{v}_{i,a}=\boldsymbol{a}_i \cdot t_{i,a}, \;\; i \in \{s, n\}
\end{equation}
where $\Delta\boldsymbol{v}_{i,a}$ represents the part of perceived velocity caused by acceleration; $\boldsymbol{a}_i$ is the current acceleration, and $t_a$ is an accumulation time for computation that varies for the subject vehicle and the neighbouring vehicle. 

\subsubsection{Perceived velocity derived from uncertainties \textemdash the imaginary velocity}

\textbf{Assumption 1} assumes that uncertainties cause extra perceived risk. For instance, when we pass by a car in the adjacent lane, we unconsciously shift to the other side of the lane to keep away from the car for safety. Accordingly, we define an imaginary velocity perceived by human drivers based on the uncertainties as a part of the perceived velocity function $\mathcal{V}$ (Equation \eqref{eq:perceived velocity}. 

The imaginary velocity in human driver's mind caused by the uncertainties of each vehicle in interaction makes the situation being perceived more dangerous (\textbf{Assumption 2}, \textbf{Assumption 2a} and \textbf{Assumption 2b}). Figure \ref{fig:imaginary velocity set} shows an example of the imaginary velocity and its influence on the velocity set $\boldsymbol{V}$. 

\begin{figure}[H]
\centering
\begin{subfigure}[b]{0.75\textwidth}
  \centering
\includegraphics[width=\textwidth]{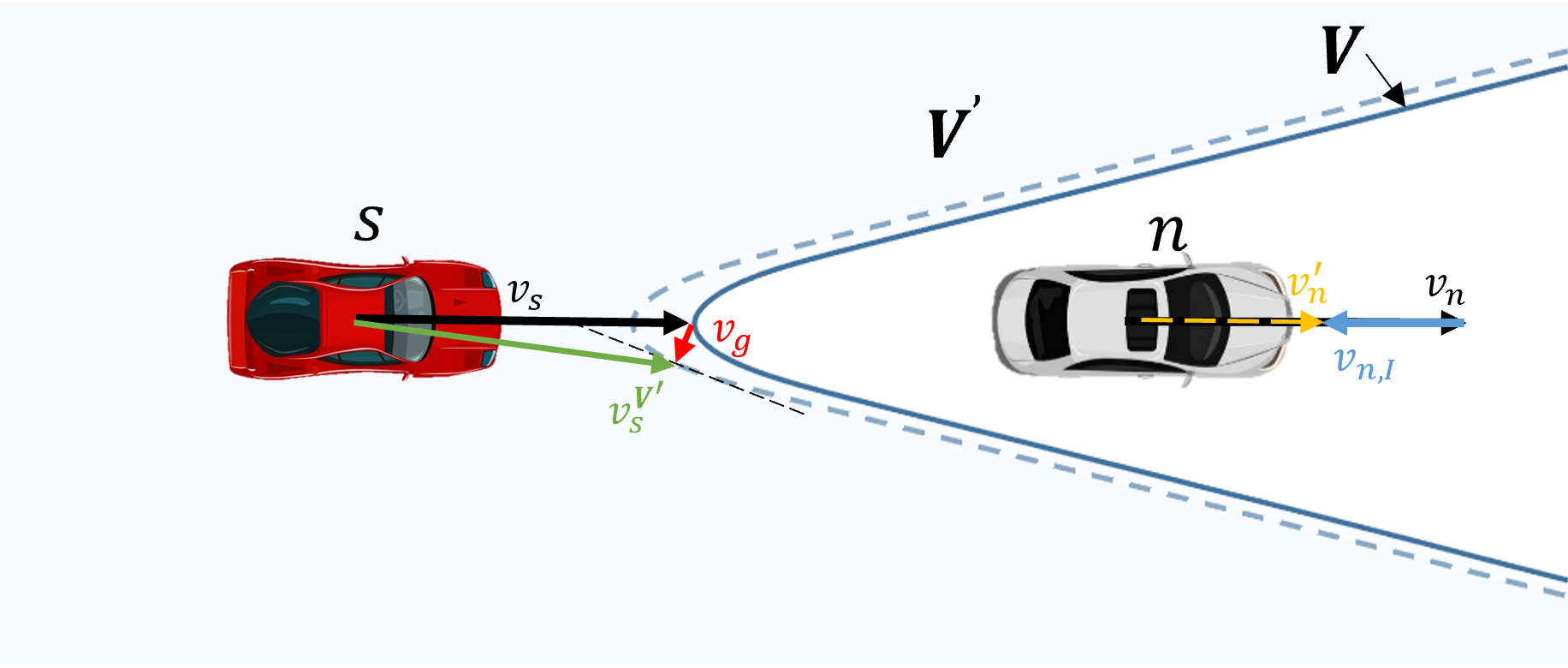}
\subcaption[]{The imaginary velocity, the final perceived velocity of the neighbouring vehicle and their influence on $\boldsymbol{V}$ . $\boldsymbol{v}_s \in \boldsymbol{V}$ indicates collision avoidance difficulty is originally zero. However, the human driver inside the subject vehicle perceives an imaginary velocity $\boldsymbol{v}_{n,I}$ (the blue arrow on the neighbouring vehicle) of the leading vehicle considering the uncertainties, changing the neighbouring vehicle's velocity from $\boldsymbol{v}_n$ to $\boldsymbol{v}_n'$ (the yellow dashed arrow on the neighbouring vehicle, which is the perceived velocity of the neighbouring vehicle). Correspondingly, the velocity set $\boldsymbol{V}$ becomes $\boldsymbol{V'}$ that is smaller than $\boldsymbol{V}$ based on the perceived velocity $\boldsymbol{v}_n'$, which leads to $\boldsymbol{v}_s \notin \boldsymbol{V'}$ causing extra perceived risk.}
\end{subfigure}%
\end{figure}

\begin{figure}[H]\ContinuedFloat
    \centering
\begin{subfigure}[b]{0.75\textwidth}
  \centering
\includegraphics[width=\textwidth]{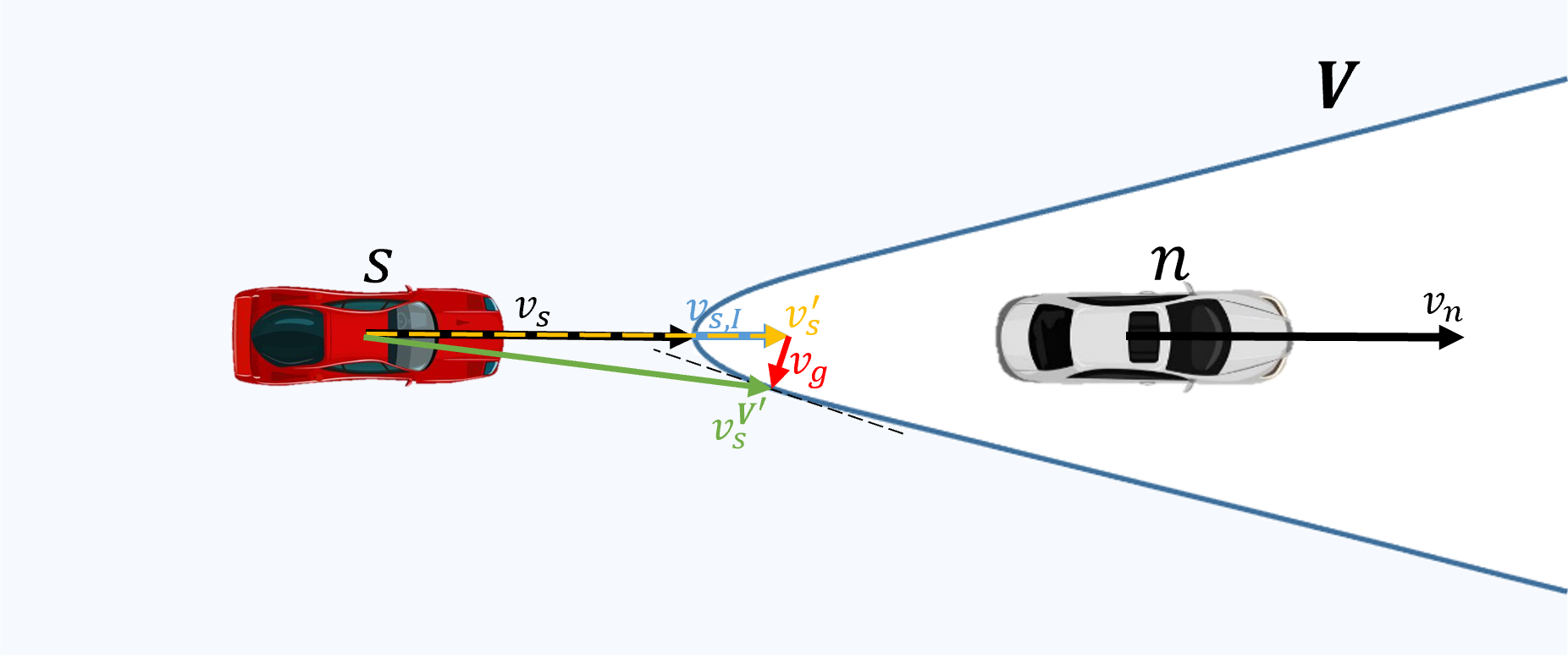}
\subcaption[]{The imaginary velocity and the final perceived velocity of the subject vehicle. $\boldsymbol{v}_s \in \boldsymbol{V}$ indicates collision avoidance difficulty is originally zero. However, the human driver inside the subject vehicle perceives an imaginary velocity  $\boldsymbol{v}_{s,I}$ (the blue arrow on the subject vehicle) due to human drivers' inaccurate control or imperfect control of driving automation, making the subject velocity change from $\boldsymbol{v}_s$ to $\boldsymbol{v}_s'$ (the yellow dashed arrow on the subject vehicle, which is the perceived velocity of the subject vehicle). The perceived velocity $\boldsymbol{v}_s'\notin \boldsymbol{V}$, causing extra perceived risk }
\end{subfigure}%
\end{figure}

\begin{figure}[H] \ContinuedFloat
\centering
\begin{subfigure}[b]{0.75\textwidth}
  \centering
\includegraphics[width=\textwidth]{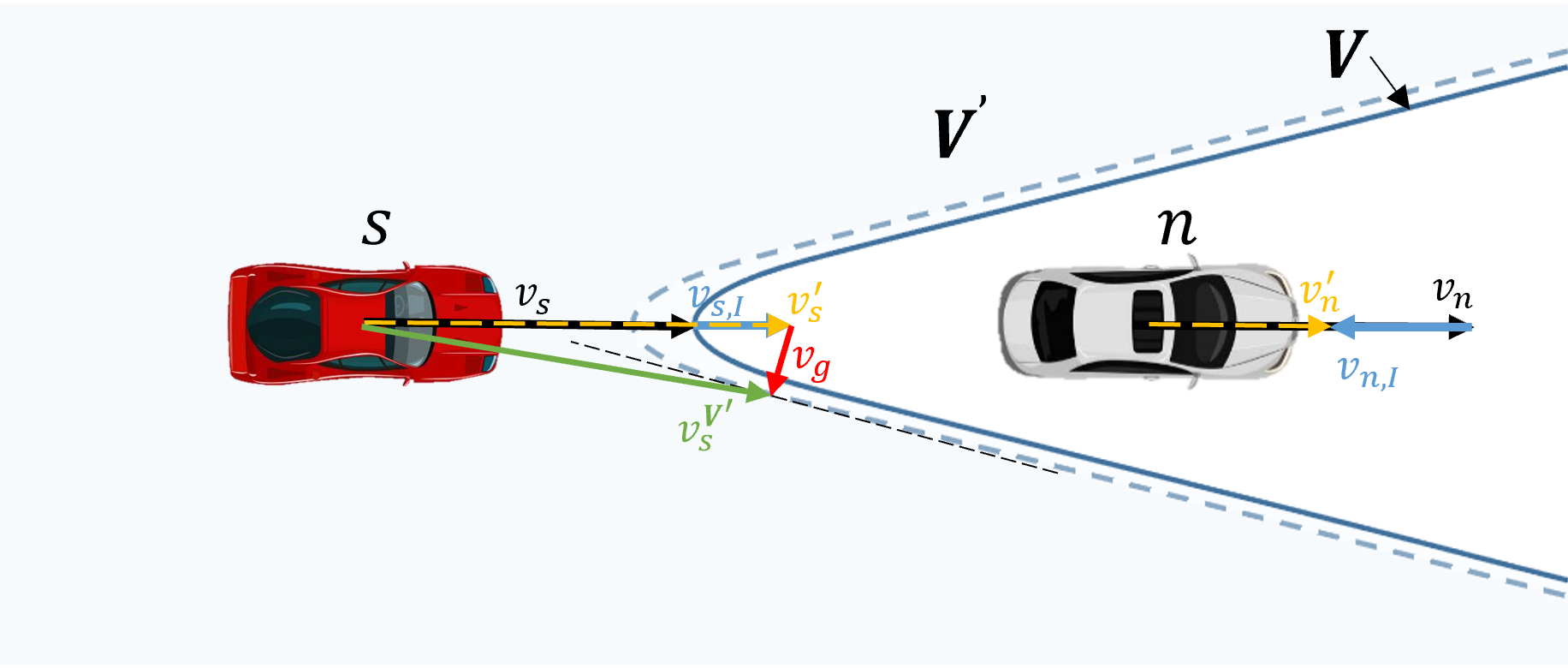}
\subcaption[]{The perceived velocity set $\boldsymbol{V'}$ based on the perceived velocity. With the larger perceived subject velocity $\boldsymbol{v}_s'$ and the smaller perceived velocity set $\boldsymbol{V'}$ simultaneously, $\boldsymbol{v}_s' \notin \boldsymbol{V'}$. $\boldsymbol{v}_g$ is the distance from the perceived subject velocity to the boundary of the perceived safe velocity set $\boldsymbol{V'}$,  which is the perceived risk in this study. }
\end{subfigure}%
\end{figure}

\begin{figure}[H] \ContinuedFloat
\centering
\begin{subfigure}[b]{0.68\textwidth}
    \centering
\includegraphics[width=\textwidth]{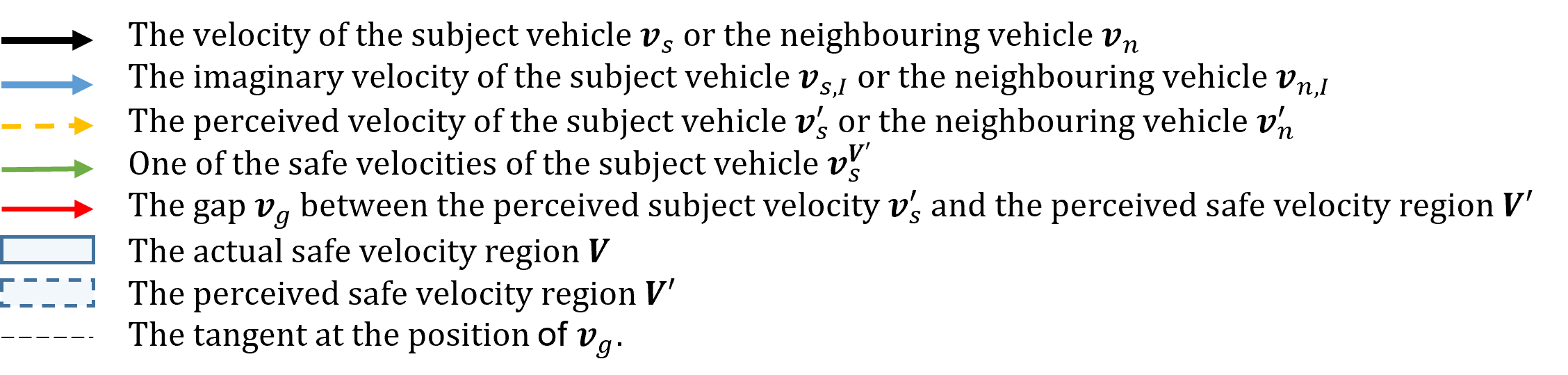}
\subcaption[]{The legend of subfigures (a), (b) and (c)}
\end{subfigure}
    \caption{An example of the imaginary velocity and its influence on the perceived velocity and the velocity set $\boldsymbol{V}$. The subject vehicle is following a leading vehicle (\SI{50}{m} ahead) with the same velocity $\boldsymbol{v}_s=\boldsymbol{v}_n = \SI{16.67}{m/s}$ and we have the velocity set $\boldsymbol{V}$ according to Equation \eqref{eq:velocity set}. In all cases, $\boldsymbol{v}_s \in \boldsymbol{V}$ indicating that the collision avoidance difficulty is originally zero. The velocity $\boldsymbol{v}_{i,a}$ is not considered due to zero acceleration. This scenario illustrates that uncertainties can cause drivers to perceive vehicle velocities differently from their actual values, which can increase the perceived difficulty of collision avoidance, and thus, perceived risk. Note that uncertainties also cause possible velocities in other directions, making the velocity set $\boldsymbol{V}$ larger, namely making the situation safer. For example, the leading vehicle changes to another lane during a car-following. In these cases, uncertainties do not increase perceived risk, which differs from human drivers' typical perception of uncertainties. This behaviour is not taken into account for imaginary velocity in this study. }
    \label{fig:imaginary velocity set}
\end{figure}

The imaginary velocity should make the situation more dangerous, so that it should follow the line connecting the subject vehicle and the neighbouring vehicle, which is $\frac{\boldsymbol{p}_s-\boldsymbol{p}_n}{||\boldsymbol{p}_s-\boldsymbol{p}_n||}$. See \ref{chap:Appendix C} for a detailed explanation.

Based on the discussion above, we have
\begin{equation}
    \boldsymbol{v}_{i,I} =l \cdot \frac{\boldsymbol{p}_s-\boldsymbol{p}_n}{||\boldsymbol{p}_s-\boldsymbol{p}_n||} \cdot r
\label{eq:relation between relative velocity}
\end{equation}
where $\boldsymbol{v}_{i,I} \left( \; i \in \{s, n\} \right)$ is the imaginary velocity; $l$ is the length of the imaginary velocity vector; $\frac{\boldsymbol{p}_s-\boldsymbol{p}_n}{||\boldsymbol{p}_s-\boldsymbol{p}_n||}$ is a unit vector pointing from the neighbouring vehicle to the subject vehicle; $r$ determines the direction of the imaginary velocity where $r=1$ represents the direction from the neighbouring vehicle to the subject vehicle, and $r=-1$ represents the opposite direction. 

According to \textbf{Assumption 2}, the uncertainties are presented as an uncertain acceleration, which is assumed to follow Gaussian distributions, and the acceleration will stay constant in a short accumulation time period \citep{Ko2010AnalysisVehicles}. Hence, given a specific accumulation time, the imaginary velocity $\boldsymbol{v}_{i,I} \left( \; i \in \{s, n\} \right)$ also follows Gaussian. With the consideration of physical restrictions of the vehicle velocity, the Gaussian is
\begin{equation}
\begin{gathered}
    v_{i,I,X} \sim ~ \mathcal{D}(v_{i,I,X}|0, \sigma_{i,X}, fb, bb)\\
    v_{i,I,Y} \sim ~ \mathcal{D}(v_{i,I,Y}|0, \sigma_{i,Y}, lb, rb)
\end{gathered}
\end{equation}
where $v_{i,I,X}$ and $v_{i,I,Y}$ are the imaginary velocity in $X$ and $Y$ directions; $\mathcal{D}$ is the probability density function of the imaginary velocity in each direction. $fb$, $bb$, $lb$, $rb$ are the forward, backward, left and right bound of the density function for the imaginary velocity, which are set to \SI{30}{m/s}, \SI{-10}{m/s}, \SI{6}{m/s} and \SI{-6}{m/s} respectively in this study. The truncated distribution $\mathcal{D}$ becomes
\begin{equation}
\begin{gathered}
   \mathcal{D}(v_{i,I,X}|0, \sigma_{i,X}, fb, bb) = \Bigg \{
\begin{array}{l}
   \frac{\frac{1}{\sigma _{i,X}}N(\frac{v_{i,I,X}}{\sigma _{i,X}})}{\boldsymbol{\mathcal{N}}(\frac{fb}{\sigma _{i,X}})-\boldsymbol{\mathcal{N}}(\frac{bb}{\sigma _{i,X}})} , \;\; bb \leqslant {v}_{i,I,X} \leqslant fb, i \in \{s,n\} \\
   0, \;\; \text{otherwise}
\end{array} \\
   \mathcal{D}(v_{i,I,Y}|0, \sigma_Y, lb, rb) = \Bigg \{
\begin{array}{l}
\frac{\frac{1}{\sigma _{i,Y}}N(\frac{v_{i,I,Y}}{\sigma _{i,Y}})}{\boldsymbol{\mathcal{N}}(\frac{lb}{\sigma _{i,Y}})-\boldsymbol{\mathcal{N}}(\frac{rb}{\sigma _{i,Y}})} , \;\;  rb \leqslant {v}_{i,I,Y} \leqslant lb, i \in \{s,n\} \\
    0, \;\; \text{otherwise}
\end{array}
\end{gathered}
\end{equation}
where $N$ is the probability density function of the Gaussian distribution and $\boldsymbol{\mathcal{N}} $ is its cumulative distribution function.

To obtain the final imaginary velocity, its length and the corresponding probability should be considered simultaneously. Hence, we use the mathematical expectation as the length of the imaginary velocity, which can be calculated as follows
\begin{equation}
\begin{aligned}
   E(||\boldsymbol{v}_{i,I}||)&=\int\limits_0^{+\infty} \mathcal{P} \left(\boldsymbol{v}_{i,I} \Bigg| \frac{\boldsymbol{p}_s-\boldsymbol{p}_n}{||\boldsymbol{p}_s-\boldsymbol{p}_n||}\right)\cdot l \; dl \\
   &=\int\limits_0^{+\infty} \mathcal{D}(v_{i,I,X}|0, \sigma_{i,X}, fb, bb) \cdot \mathcal{D}(v_{i,I,Y}|0, \sigma_{i,Y}, lb, rb) \cdot \frac{1}{\mathcal{P} \left(\frac{\boldsymbol{p}_s-\boldsymbol{p}_n}{||\boldsymbol{p}_s-\boldsymbol{p}_n||}\right)} \cdot l \;  dl
   \label{eq: imaginary velocity length}
\end{aligned}
\end{equation}

This conditional probability is denoted by $\mathcal{P} \left(\boldsymbol{v}_{i,I} \Bigg| \frac{\boldsymbol{p}_s-\boldsymbol{p}_n}{||\boldsymbol{p}_s-\boldsymbol{p}_n||}\right)$ . To ensure that this direction-specific probability is considered, we divide the product of the two probability density functions, $\mathcal{D}(v_{i,I,X}|0, \sigma_{i,X}, fb, bb) $ and $ \mathcal{D}(v_{i,I,Y}|0, \sigma_{i,Y}, lb, rb)$, by the aforementioned conditional probability $\mathcal{P} \left(\frac{\boldsymbol{p}_s-\boldsymbol{p}_n}{||\boldsymbol{p}_s-\boldsymbol{p}_n||}\right)$. This division effectively normalises the probability densities and allows for the proper calculation of the mathematical expectation of the length of the imaginary velocity  $E(||\boldsymbol{v}_{i,I}||)$.

Accordingly, the imaginary velocity is 
\begin{equation}
    \boldsymbol{v}_{i,I}= E(||\boldsymbol{v}_{i,I}||)\cdot \frac{\boldsymbol{p}_s-\boldsymbol{p}_n}{||\boldsymbol{p}_s-\boldsymbol{p}_n||} \cdot r
\label{eq: adjusted relative velocity}
\end{equation}
Note that this imaginary velocity is not the most probable one but it is the probabilistic average in the most dangerous direction which is assumed to dominate the risk perceived by human drivers. Although the integral is expressed in an analytical format, \textit{integral} function is used in MATLAB for numerical evaluation.

\subsubsection{The final perceived velocity}
\begin{figure}[H]
    \centering
    \includegraphics[width=0.7\textwidth]{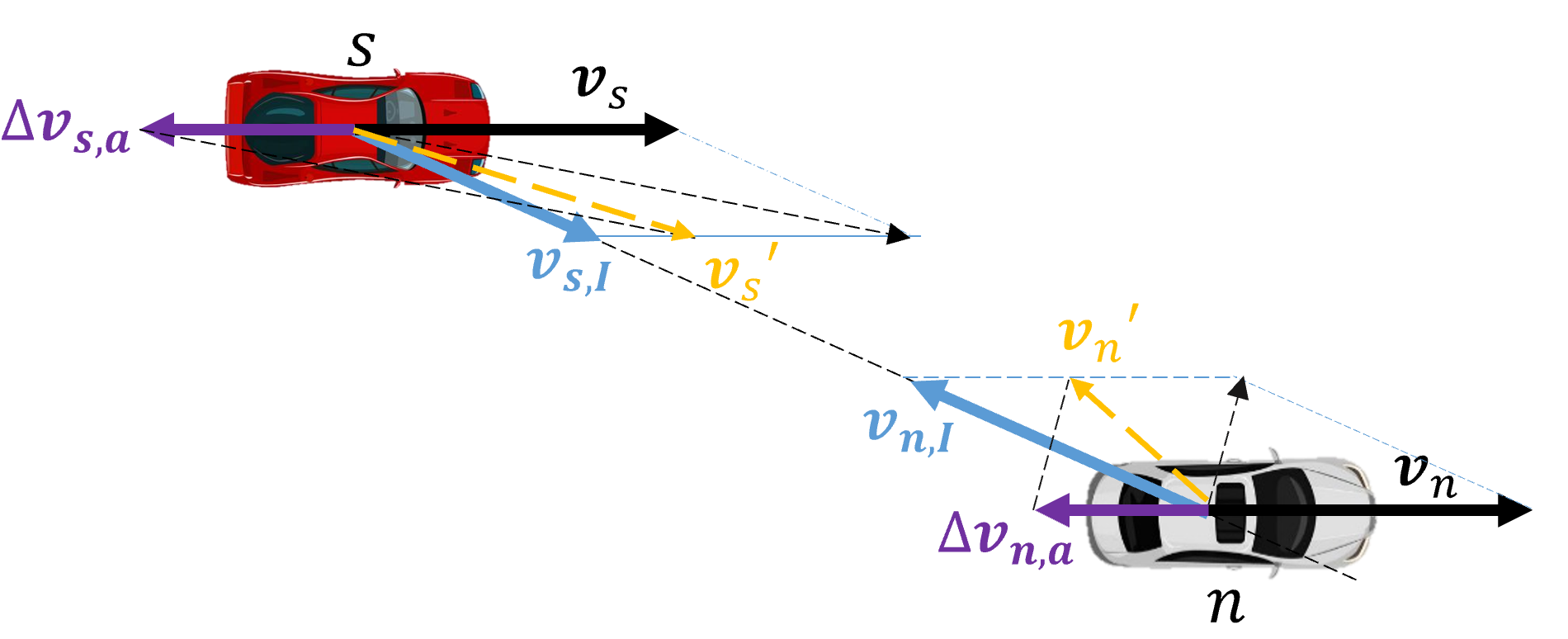}
    \caption{The relationship between the actual velocity $\boldsymbol{v}_i$ the imaginary velocity $\boldsymbol{v}_{i,I}$, the acceleration caused velocity change and the perceived velocity $\boldsymbol{v}_i'$ of the subject vehicle $s$ and the neighbouring vehicle $n$ ($i\in\{s,n\}$). In this case, the subject vehicle (red) is passing by a neighbouring vehicle (white). Both vehicles are decelerating causing acceleration-based velocities $\Delta \boldsymbol{v}_{s,a}$ and $\Delta \boldsymbol{v}_{n,a}$ (the yellow arrows). The imaginary velocities $\boldsymbol{v}_{s,I}$ and $\boldsymbol{v}_{n,I}$ are pointing to each other. The final perceived velocity $\boldsymbol{v}_s'$ and $\boldsymbol{v}_n'$ contains the contribution of the acceleration-based velocity and the imaginary velocity. }
\label{fig:imaginary velocity}
\end{figure}

We finally incorporate the actual velocity $\boldsymbol{v}_i$, the acceleration-based velocity $\Delta\boldsymbol{v}_{i,a}$ and the imaginary velocity $\boldsymbol{v}_{i,I}$ into the perceived velocity function $\mathcal{V}$ as shown in Equation \eqref{eq:perceived velocity}. This integrated perceived velocity function now considers the acceleration and uncertainties, thus contributing to extra perceived risk. Figure \ref{fig:imaginary velocity} illustrates the relationship between the actual velocity $\boldsymbol{v}_i$, the imaginary velocity $\boldsymbol{v}_{i,I}$, the acceleration-based velocity $\Delta\boldsymbol{v}_{i,a}$ and the final perceived velocity $\boldsymbol{v}_i'$. The final perceived velocity is utilised for computing perceived risk.

\subsection{Weighting function $\mathcal{W}$}\label{chap:weighting function}
The subject velocity significantly influences perceived risk, as it affects the accident rate and the consequence of a crash. 
The relationship between velocity and crash outcome is related to the kinetic energy ($E_k=\frac{1}{2}mv^2$) released during a collision but the relationship is not a simple linear mapping. A scaling function ranging on [0,1] is needed to show the relationship between the subject velocity and perceived risk. Previous studies tried to examine the relationship between the subject velocity and the crash outcome based on real-world crash data and found that a power function best fits the relationship \citep{aarts2006driving}. We employ an exponential function proposed by \cite{finch1994speed} to describe such a relationship: 

\begin{equation}
    \mathcal{W} = \left(\frac{||\boldsymbol{v}_s||}{v_{ref}}\right)^{\alpha}
\end{equation}
where $\boldsymbol{v}_s$ is the subject velocity;  $v_{ref}$ is a reference velocity and it can be set as the velocity limit in specific conditions; $\alpha$ is the coefficient. Given a speed limit in a specific scenario, the $\mathcal{W}\propto v^{\alpha}$ ranging on $[0, 1]$, which can be used as a weight for the final perceived risk based on functions $\mathcal{A}$ and $\mathcal{V}$ as in Equation \eqref{eq:general structure}. 
\subsection{PCAD Model parameters}
Table \ref{tab:PCAD parameters} summarises the parameters of the PCAD model to be calibrated. Details regarding the avoidance difficulty function $\mathcal{A}$, the perceived velocity function $\mathcal{V}$, and the weighting function $\mathcal{W}$ can be found in sections \ref{chap: avoidance difficulty}, \ref{chap:probabilistic approaching velocity}, and \ref{chap:weighting function}, respectively.

\begin{table}[H]
    \centering
    \caption{The key parameters of PCAD model}
\resizebox{\textwidth}{!}{%
    \begin{tabular}{p{0.25\textwidth}<{\centering}|p{0.75\textwidth}<{\centering}}
    \hline
Parameters     & Explanations \\ \hline
$\sigma_{i,X}$, $\sigma_{i,Y}$ & Standard deviations of Gaussian distributions for the imaginary velocity of other road users or the subject vehicle \\ \hline
$t_{i,a}$ & An accumulation time used for the computation of the acceleration-based velocity \\ \hline
$\alpha$ & The exponent of the Weighting function for the influence of subject velocity on perceived risk. \\ \hline
    \end{tabular}
\label{tab:PCAD parameters}
}
\end{table}
\section{Analytical model properties} \label{Qualitative properties of models}
This section offers an analysis of the PCAD model and the three baseline models. Table \ref{tab:Model properties} summarises model properties, covering aspects such as output dimension, the usage of distance, relative motion, acceleration, subject speed, motion uncertainties, crash consequences, and usability on curved lanes. In summary, PCAD is a comprehensive model based on Risk Allostasis Theory which considers all aspects listed in Table \ref{tab:Model properties}. It is a 2-D model capturing both longitudinal and lateral perceived risk, and can also be used on curved lanes.

For an intuitive understanding, we visualise the perceived risk variations of the four models in a 2-D coordinate system describing the relative position of a neighbouring vehicle. As demonstrated in Figure \ref{fig:risk field visualization}, perceived risk varies with different relative velocities (Figure \ref{fig:risk field visualization RelVel}), different decelerations (Figure \ref{fig:risk field visualization acceleration}), and different subject velocities (Figure \ref{fig:risk field visualization EgoVel}). Figure \ref{fig:risk field explanation} provides the legend for these diagrams.

The PCAD model indicates that perceived risk amplifies as an object or neighbouring vehicle nears the subject vehicle, demonstrating a sharp rise both longitudinally and laterally.  The non-linear relationship caused by non-linear \textit{looming detection} in PCAD prevails in the other three typical models but is described by different functions such as Gaussian (i.g., the lateral risk in PPDRF and DRF), Exponential (i.g., the potential risk in PPDRF), logarithmic (i.g., the risk in RPR) and Quadratic functions (i.g., the longitudinal risk in DRF). Note that RPR cannot capture perceived risk in the lateral direction since it is only defined in the same traffic lane. 

PCAD shows that human drivers perceive more risk when approaching an object faster. Compared to the other three models, PCAD and PPDRF can output different perceived risk values facing different relative velocities (Figure \ref{fig:risk field visualization RelVel}). RPR and DRF do not include velocity information of the neighbouring vehicles or objects, so their predicted risk is independent of relative speed, although they can be tuned to fit different relative speed. 

Reacting to neighbouring vehicles' velocity changes (e.g., braking) is a common task in the daily drive. PCAD can clearly describe this relationship (Figure \ref{fig:risk field visualization acceleration}) where $a_n = \SI{-8}{m/s^2}$ leads to the highest perceived risk and $a_n = 0$ causes the lowest perceived risk. RPR and PPDRF also support this feature but DRF cannot indicate the change of perceived risk caused by a neighbouring vehicle's deceleration due to a lack of acceleration information in the model.

The subject velocity significantly influences perceived risk. In Figure \ref{fig:risk field visualization EgoVel}, PCAD demonstrates that, given the same following gap, human drivers perceive more risk with a higher subject velocity, which is similar to PPDRF and DRF. However, RPR does not contain subject speed information in the model and cannot capture the perceived risk variance in this condition. 

\begin{table}[H]
    \resizebox{\textwidth}{!}{%
    \centering
    \begin{threeparttable} 
    \caption{Model features}
    \label{tab:Model properties}

    \begin{tabular}{p{5cm}<{\centering}|p{2cm}<{\centering}|p{2cm}<{\centering}|p{2cm}<{\centering}|p{2cm}<{\centering}}
    \hline
     \multicolumn{1}{c|}{}& \multicolumn{1}{c|}{PCAD} & \multicolumn{1}{c|}{RPR} &\multicolumn{1}{c|}{PPDRF}&\multicolumn{1}{c}{DRF} \\ \hline
    \centering Dimension &  2-D &  1-D & 2-D & 2-D\\ \hline
    \centering Distance & \CIRCLE  &\CIRCLE   &\CIRCLE  &\CIRCLE \\ \hline
    \centering Using relative velocity   &  \CIRCLE &  -  & \CIRCLE & - \\ \hline
    \centering Using acceleration &  \CIRCLE &  \CIRCLE & \CIRCLE & -\\ \hline
    \centering Using subject velocity   & \CIRCLE & - & \CIRCLE&\CIRCLE\\ \hline
    \centering Considering crash consequence  & \CIRCLE & - & \CIRCLE&\CIRCLE\\ \hline
    \centering Considering motion uncertainties & \CIRCLE & - & \CIRCLE&-\\ \hline
    \centering Usable on curved lanes & \CIRCLE & - & \CIRCLE&\CIRCLE\\ \hline
    \end{tabular}
\begin{tablenotes}
      \small
      \item \CIRCLE indicate "yes" and $-$ indicate "no". 
    \end{tablenotes}
    \end{threeparttable}
}
\end{table}
\begin{figure}[H]
\begin{subfigure}{\textwidth}
  \centering
    \includegraphics[width=0.45\textwidth]{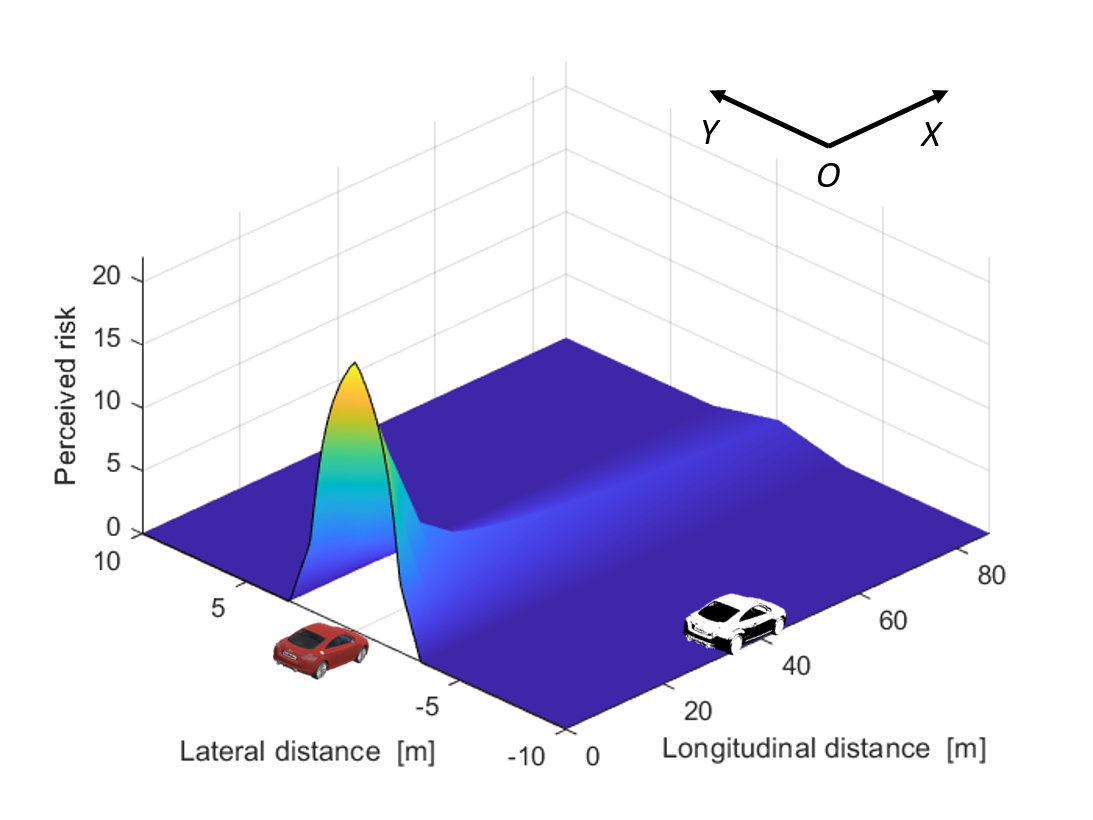}
    \includegraphics[width=0.05\textwidth]{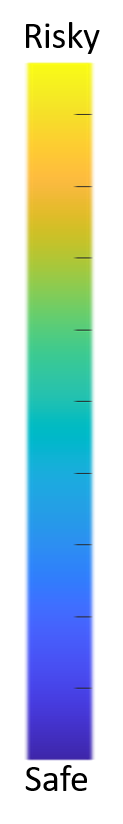}
    \includegraphics[width=0.45\textwidth]{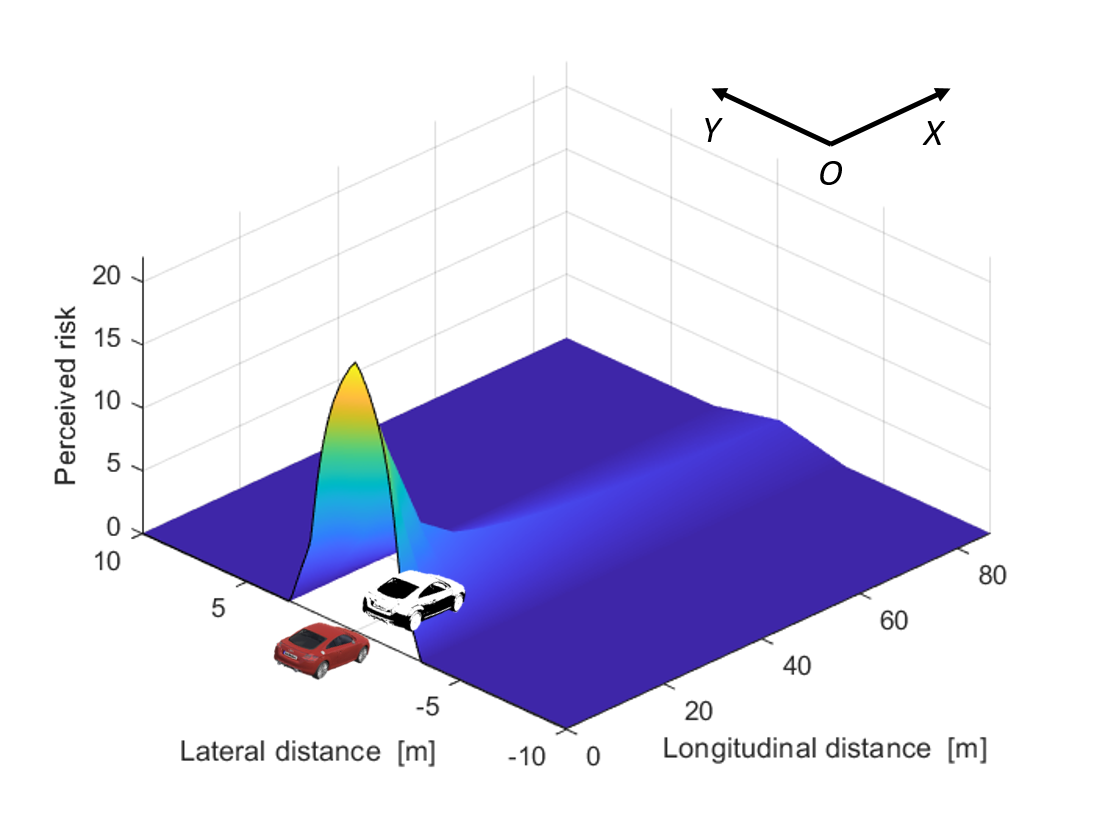}
    \caption{Legend and example explanation of figures (b), (c), (d). Surfaces represent the computed perceived risk value of different models as function of the relative 2D position of subject vehicle front and neighbouring vehicle rear end. In all figures the subject vehicle front (Red) is at the origin (0,0) with the subject velocity $\boldsymbol{v}_{s,X} = \SI{100}{km/h}$ along the \textit{X} direction. In this example the neighbouring vehicle (White) longitudinal velocity $\boldsymbol{v}_{n,X} = \SI{50}{km/h}$ along with \textit{X} direction on the 2-D coordinate. In the left subfigure, the neighbouring vehicle is at (40, -10) which means the neighbouring vehicle is in the front right of the subject vehicle with a lateral clearance, for example, in an adjacent lane. The perceived risk is almost zero in this condition; In the right subfigure, the neighbouring vehicle is at (10, 0), indicating that the subject vehicle is approaching the leading vehicle with the relative velocity $\Delta\boldsymbol{v} = \SI{50}{km/h}$ and the longitudinal gap is only $\SI{10}{m}$. The perceived risk is high in this case. }
    \label{fig:risk field explanation}
\end{subfigure} 
\end{figure}
\begin{figure}[H]\ContinuedFloat
    \centering
\begin{subfigure}{\textwidth}
  \centering
    \includegraphics[width=0.45\textwidth]{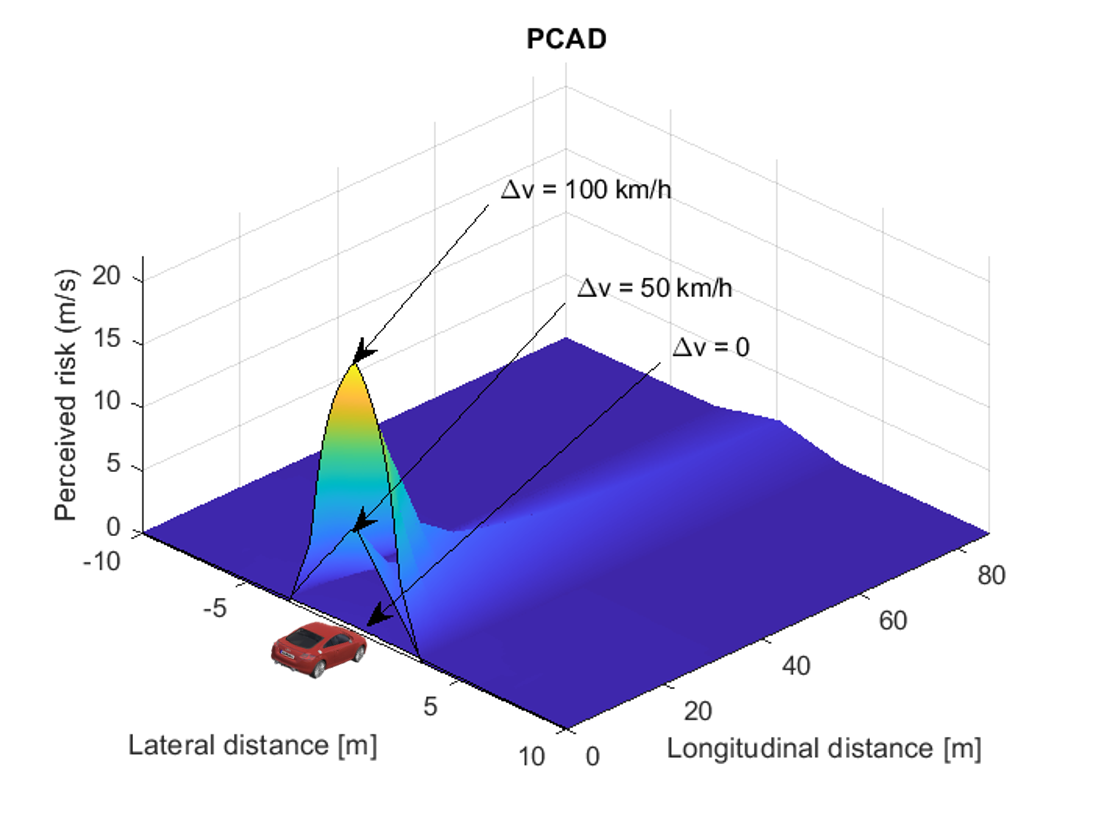}
    \hfill
    \includegraphics[width=0.45\textwidth]{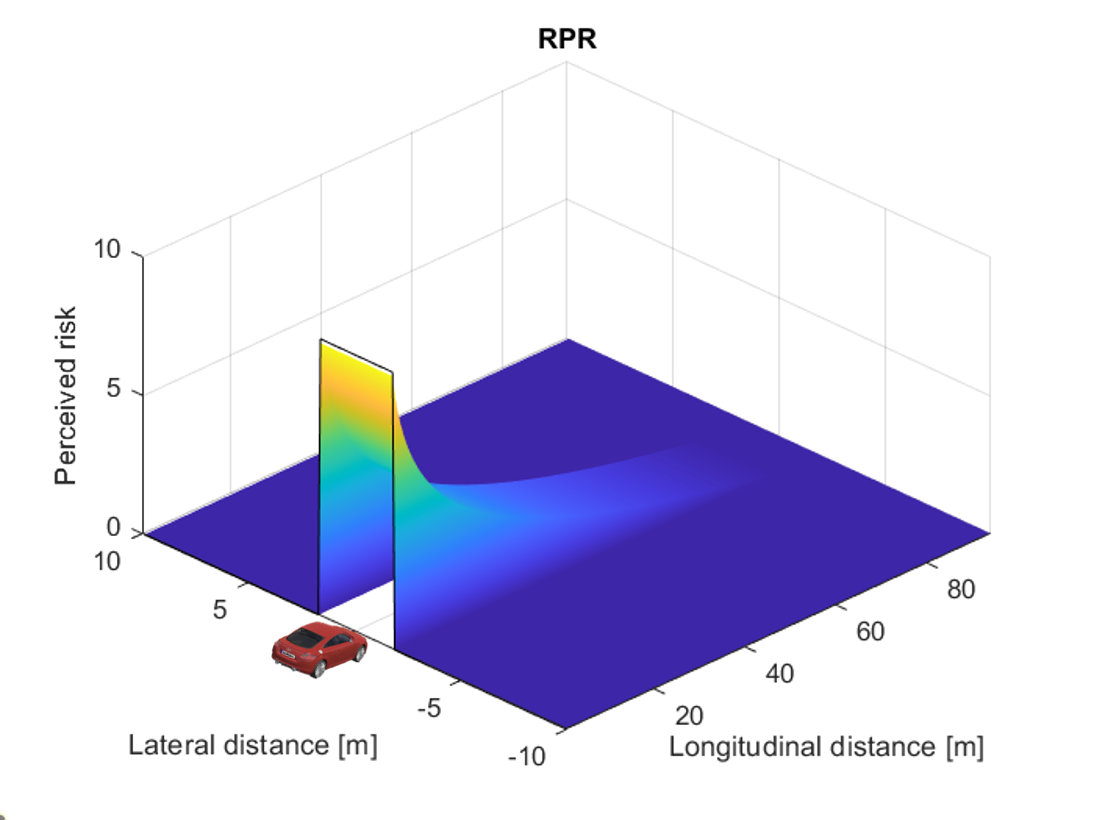}
\end{subfigure}
\end{figure}

\begin{figure}[H]\ContinuedFloat
     \centering  
\begin{subfigure}{\textwidth}
    \includegraphics[width=0.45\textwidth]{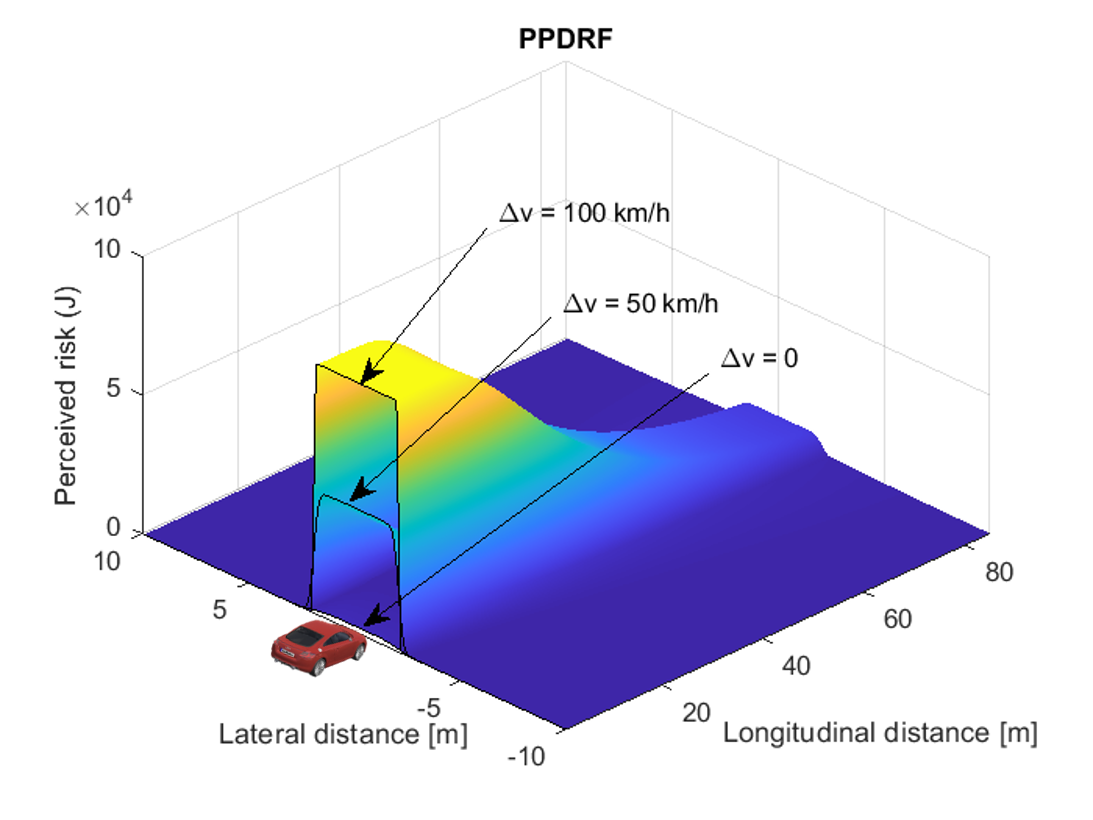}
    \hfill
    \includegraphics[width=0.45\textwidth]{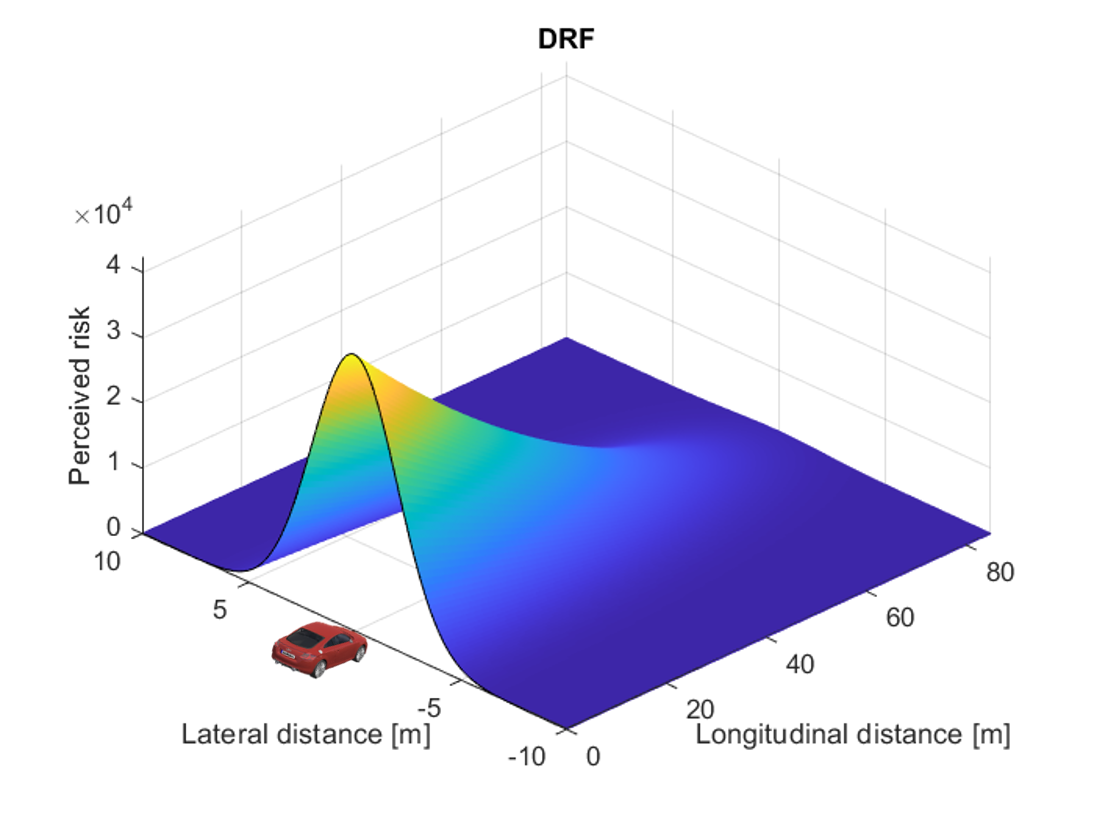}
    \caption{The effect of relative velocity on human driver's perceived risk to traffic vehicles on a 2-D plane. The subject vehicle is at the original point $(0,0)$. The constant subject vehicle velocity is $v_{s,X}$ = 100 km/h but the neighbouring vehicle has different velocities {$v_{n,X}$}; no acceleration of the subject vehicle and the neighbouring vehicle $a_{s,X} = a_{n,X} = 0$. Note that PPDRF's kinetic risk instead of potential risk is used here for a more understandable visualisation.}
    \label{fig:risk field visualization RelVel}
\end{subfigure} 
\end{figure}

\begin{figure}[H] \ContinuedFloat
    \centering
\begin{subfigure}{\textwidth}
    \centering
    \includegraphics[width=0.45\textwidth]{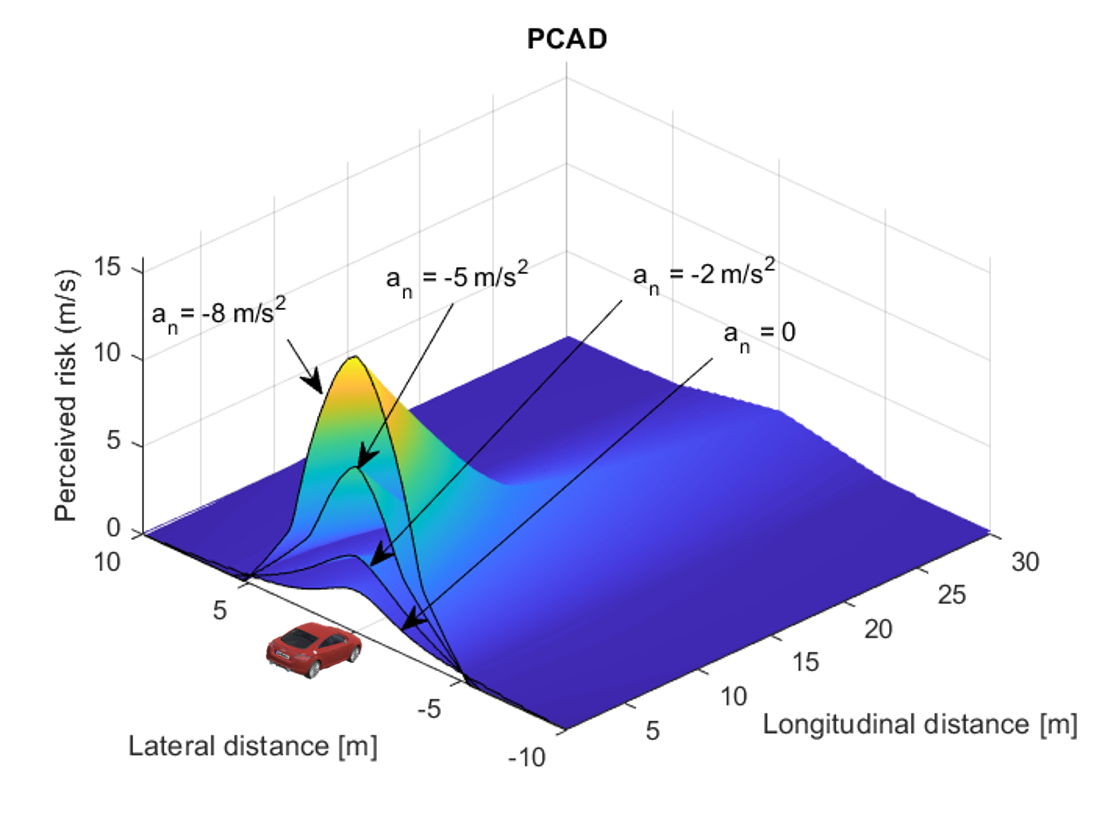}
    \hfill
    \includegraphics[width=0.45\textwidth]{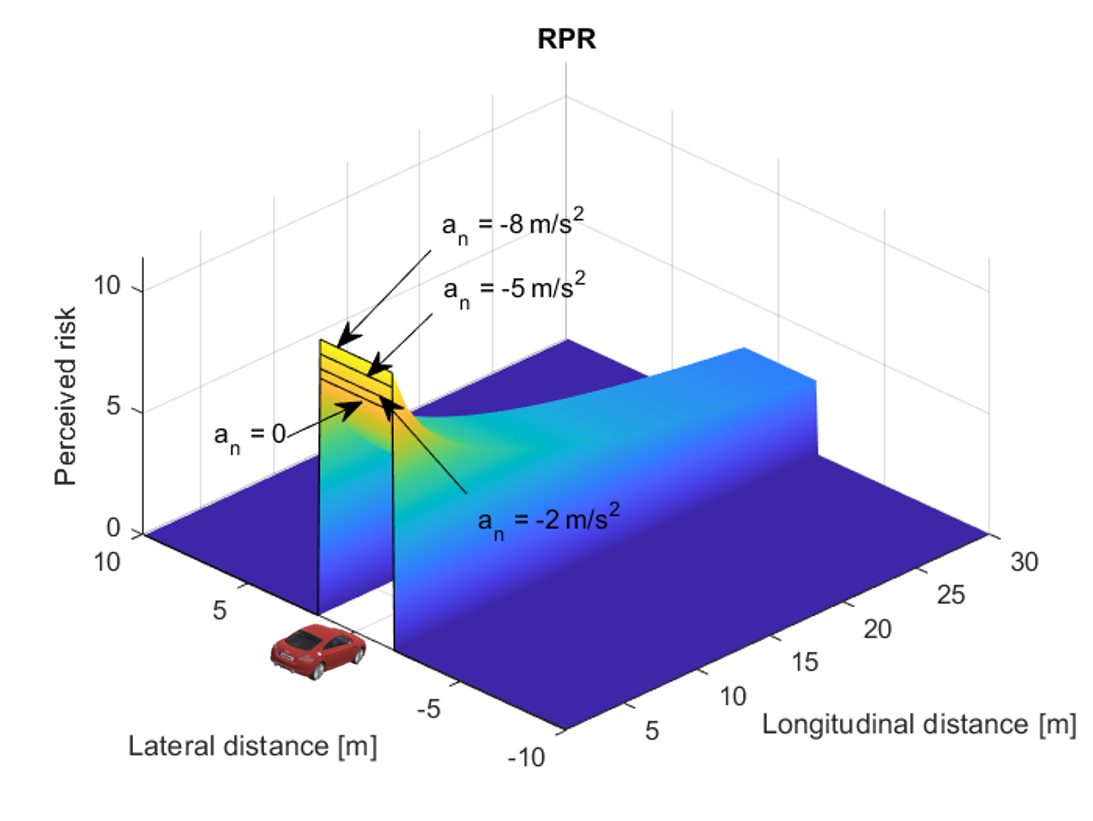}
\end{subfigure}
\end{figure}

\begin{figure}[H] \ContinuedFloat
    \centering
\begin{subfigure}{\textwidth}
    \includegraphics[width=0.45\textwidth]{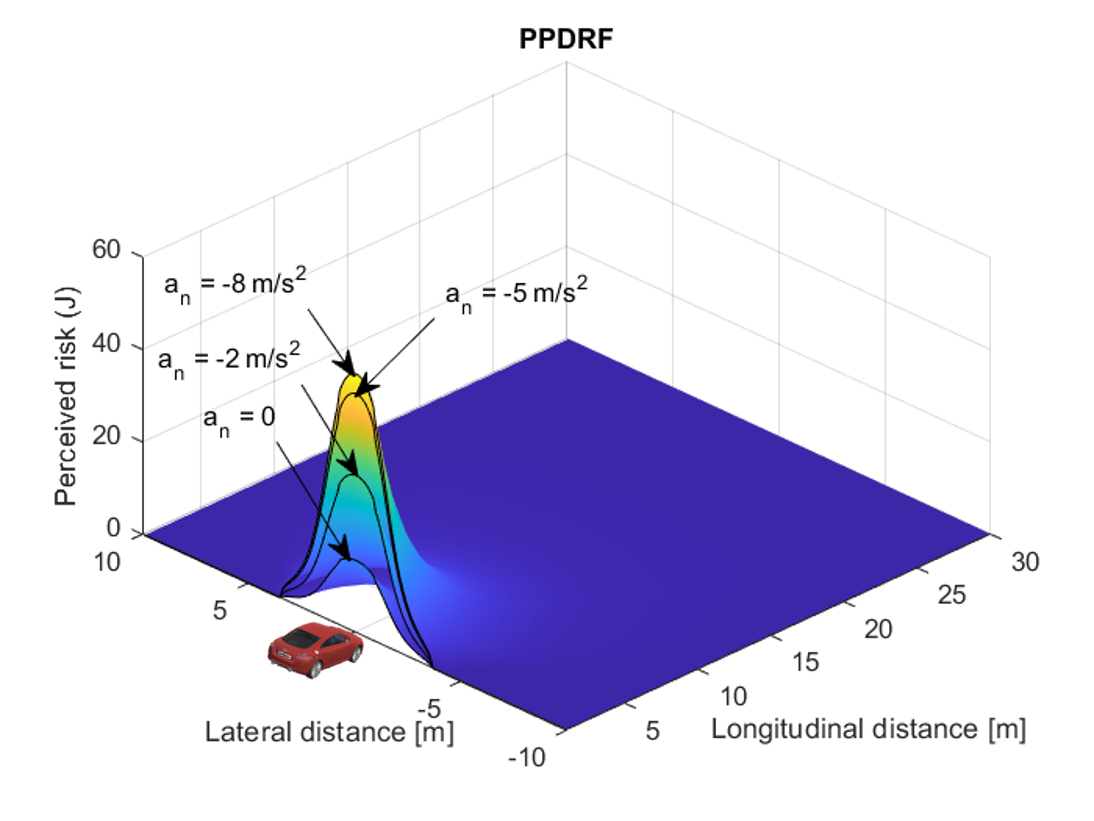}
    \hfill
    \includegraphics[width=0.45\textwidth]{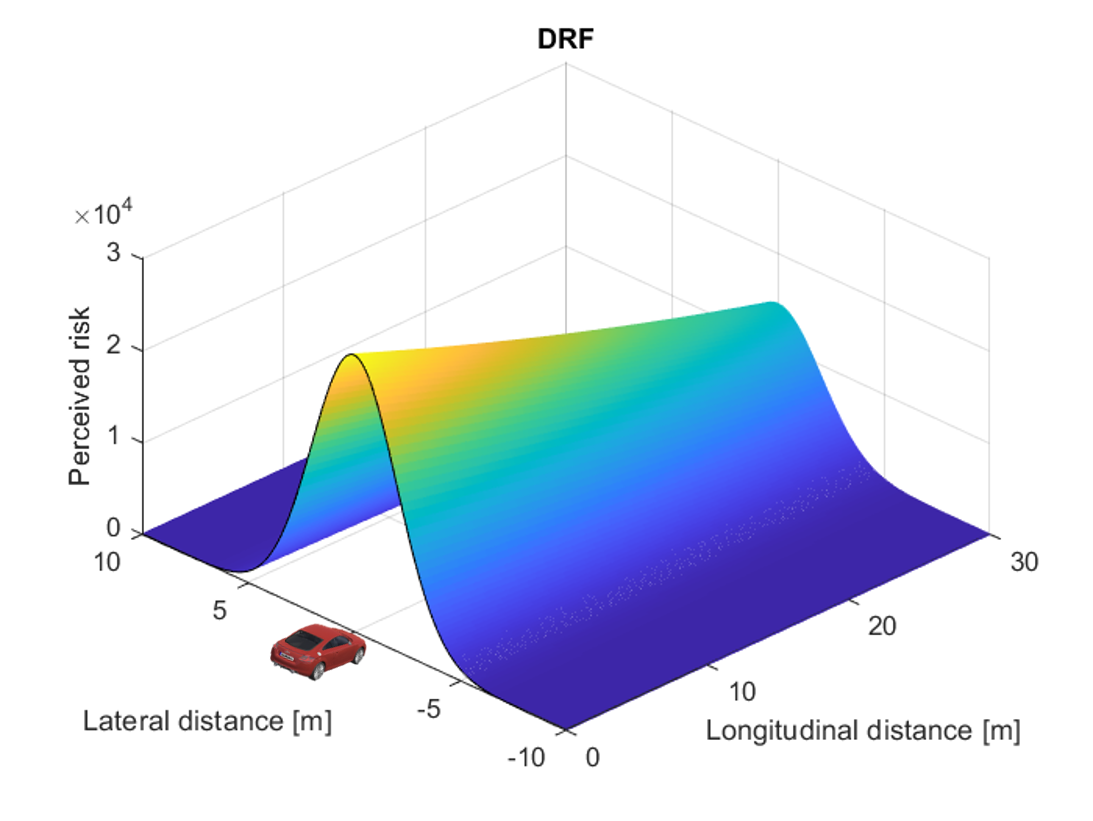}
    \caption{The effect of neighbouring vehicle's acceleration on human driver's perceived risk to traffic vehicles on a 2-D plane. The subject vehicle is at the original point $(0,0)$. The velocity of the subject vehicle and the neighbouring vehicle are equal $v_{s,X} = v_{n,X} =$ 100 km/h; the subject vehicle has no acceleration $a_{s,X} = 0$ but the neighbouring vehicle's acceleration ($a_{n,X}$) varies.}
    \label{fig:risk field visualization acceleration}
\end{subfigure}

\begin{subfigure}{\textwidth}
  \centering
    \includegraphics[width=0.45\textwidth]{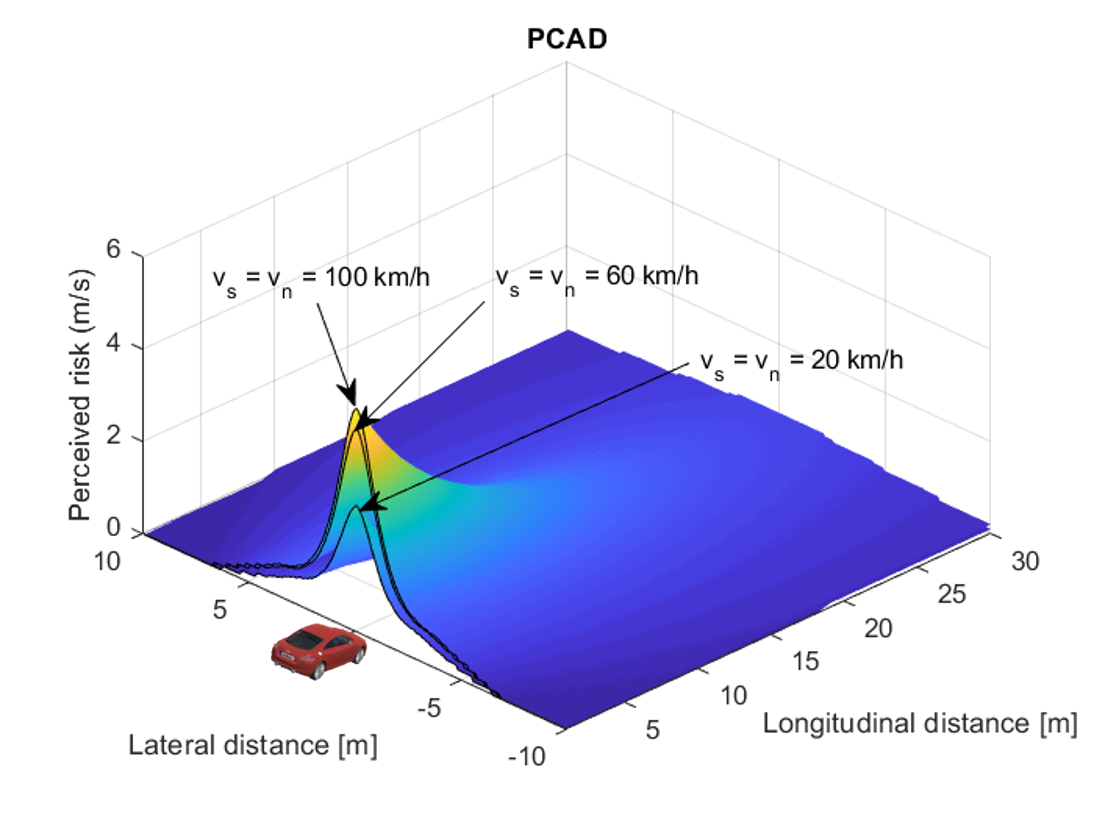}
    \hfill
    \includegraphics[width=0.45\textwidth]{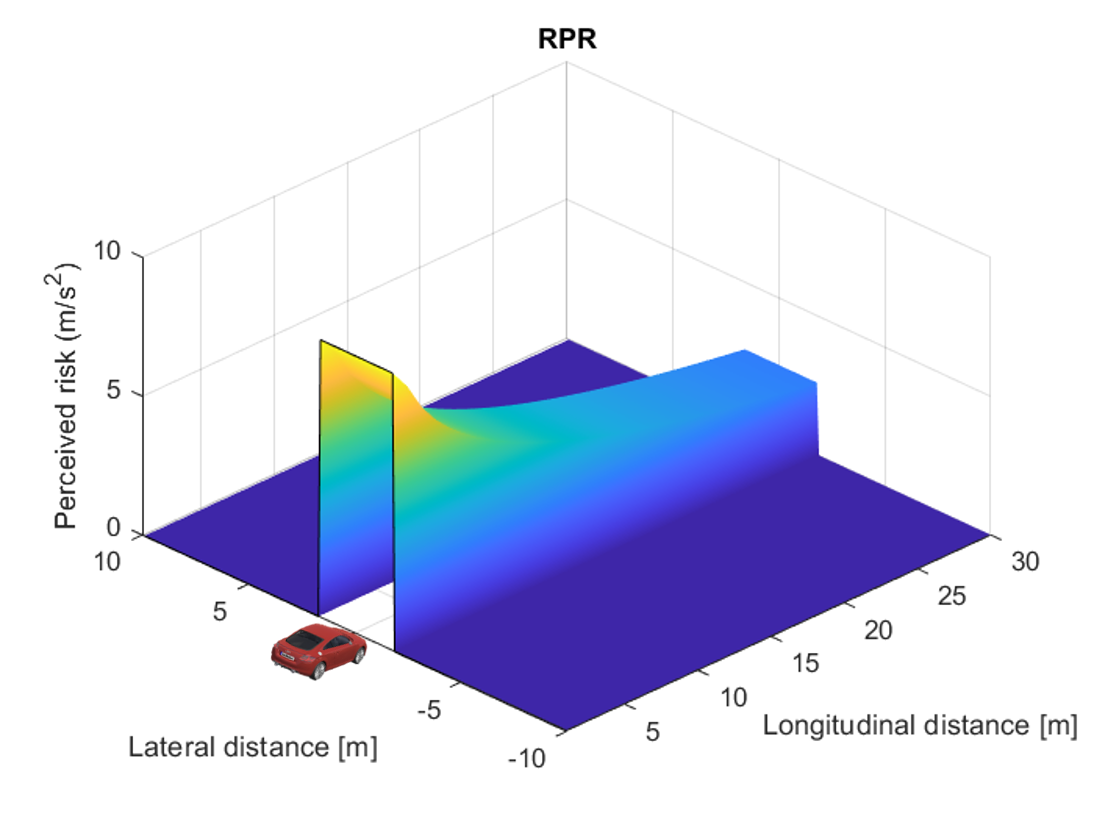}
\end{subfigure}
\end{figure}

\begin{figure}[H] \ContinuedFloat
    \centering
\begin{subfigure}{\textwidth}
    \includegraphics[width=0.45\textwidth]{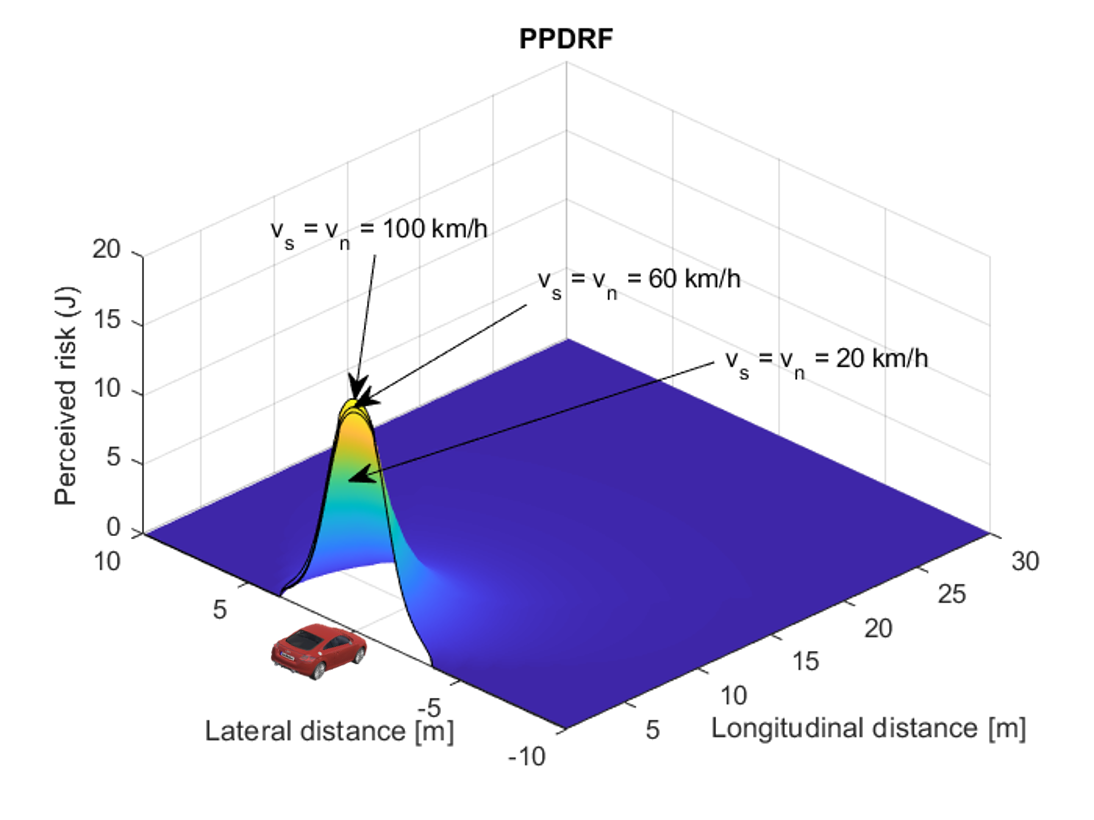}
    \hfill
    \includegraphics[width=0.45\textwidth]{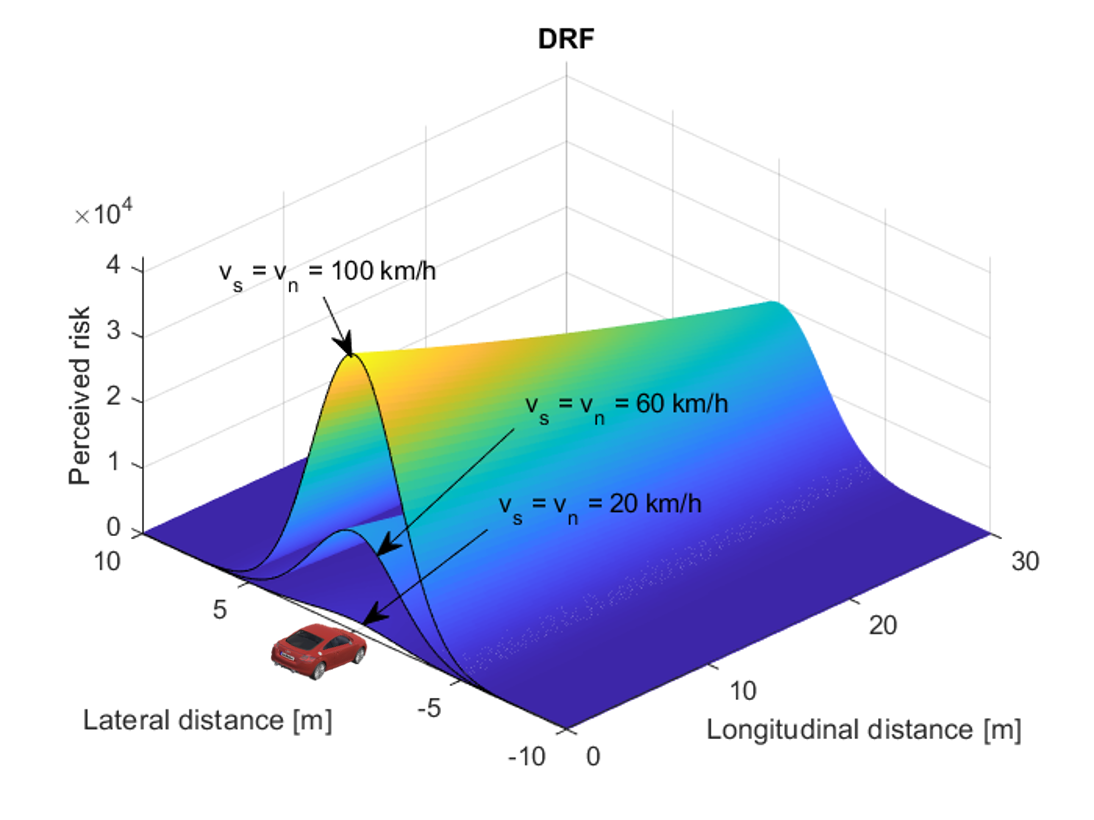}
    \caption{The effect of subject velocity. Human driver's perceived risk to traffic vehicles on a 2-D plane. The subject vehicle is at the original point $(0,0)$. The subject vehicle and the neighbouring vehicle have equal velocity ($v_{s,X} = v_{n,X}$) varying from \SI{20}{km/h}-\SI{100}{km/h}; no acceleration of the subject vehicle and the neighbouring vehicle $a_{s,X} = a_{n,X} = 0$.}
    \label{fig:risk field visualization EgoVel}
\end{subfigure} 
\caption{The effect of relative velocity (b), the acceleration of the neighbouring vehicle (c) and the subject velocity (d) with legend in (a)}
\label{fig:risk field visualization}
\end{figure}

\section{Model evaluation method} \label{chap:Methodology}
To conduct a comparative evaluation of the proposed model and the baseline models, model calibration with empirical data is indispensable. This section details the experimental datasets, calibration method, and performance indices for the models. 
\subsection{Dataset introduction} \label{Chap: Dataset introduction}
\begin{figure}[H]
    \centering
    \includegraphics[width=0.9\textwidth]{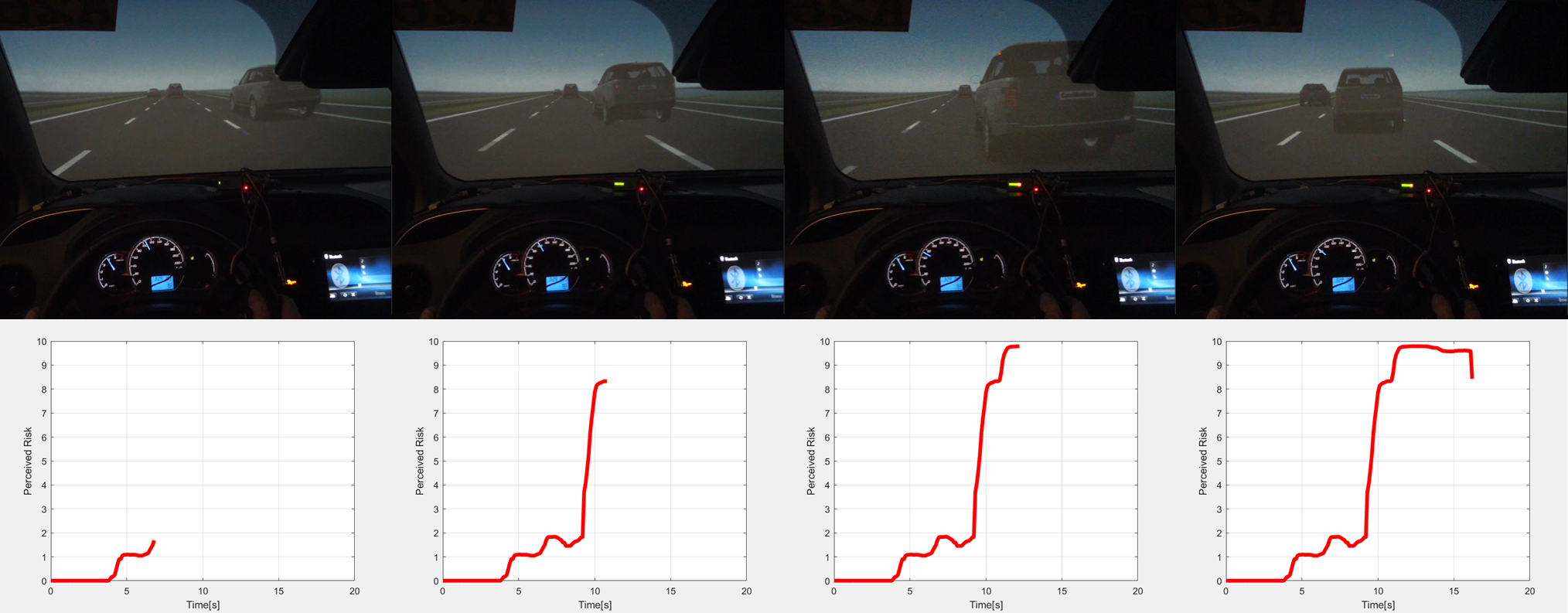}
    \caption{The experiment where Dataset Merging was collected. Upper row: Video stream of a merging with hard braking event simulated in the experiment. Lower row: Corresponding perceived risk values indicated by a participant with the pressure sensor. }
    \label{fig:video_perceivedrisk}
\end{figure}

We employ two datasets for model calibration and evaluation. 
The first dataset (\textbf{Dataset Merging})  was collected in our previous simulator experiment where the subject automated vehicle reacts to merging and hard-braking vehicles. The experiment simulated 18 merging events with different merging distances and braking intensities on a 2-lane highway \citep{He2022}. Figure \ref{fig:video_perceivedrisk} shows an example of the simulated events during the experiment. The participants were asked to monitor the scenario as fall-back ready drivers for an SAE Level 2 automated vehicle. They used a pressure sensor on the steering wheel to provide perceived risk ratings from 0-10 continuously in the time domain (see the lower row in Figure \ref{fig:video_perceivedrisk}), which are the continuous perceived risk data. After each event, the participants were also asked to give a verbal perceived risk rating from 0-10 regarding the previous event, which is the discrete event-based perceived risk data. The corresponding kinematic data (e.g. position, speed and acceleration of the subject vehicle and neighbouring vehicles) were collected simultaneously. 

\begin{figure}[H]
    \centering
    \includegraphics[width=0.5\textwidth]{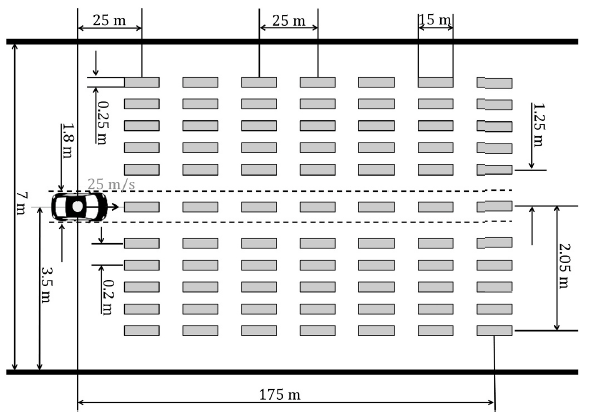}
    \caption{Dataset Obstacle Avoidance, with stationary obstacle positions from \citet{Kolekar2020WhichField}}
    \label{fig:ObstacleDistributionKolekar}
\end{figure}

The second dataset (\textbf{Dataset Obstacle Avoidance}) includes drivers’ verbal perceived risk ratings and steering angle signals when the participants face static obstacles suddenly appearing in front the subject vehicle driving at $\SI{25}{m/s}$ in manual driving mode \citep{Kolekar2020WhichField}. Figure \ref{fig:ObstacleDistributionKolekar} shows the distribution of the obstacles. The corresponding vehicle kinematic data and the positions of the obstacles were recorded at the same time. 

The following reference data is utilised for model calibration:
\begin{itemize}
    \item Dataset Merging: the \textbf{event-based perceived risk} and the \textbf{peak of the continuous perceived risk} in specific events 
    \item Dataset Obstacle Avoidance: the \textbf{event-based perceived risk}, and the \textbf{peak of steering wheel angle} in specific events
\end{itemize}

Figures \ref{fig:PCAD output in a merging event} and \ref{fig:PCAD output in an obstacle avoidance event} in \ref{chap:Appendix A} illustrate the kinetic data from the two datasets, along with the continuous risk predicted by PCAD.

\subsection{Model calibration} \label{Model calibration}
While our aim is to develop general models considering the average characteristics of all participants, we cannot ignore the influence of group features and scenarios. To optimise performance, we perform a dataset-level calibration of parameters for all models. We have $RMSE_i$ defined as
\begin{linenomath}
\begin{equation}
RMSE_q=\sqrt{\frac{\sum_{k=1}^{K}\left(\hat{y}_{k}-y_{k}\right)^2}{K}}  \label{eq:rmse}
\end{equation}
\end{linenomath}
Here, $RMSE_q$ denotes the root mean square error between the collected perceived risk data and the model output. For Dataset Merging, $q=event$ and $q=peak$ represent the $RMSE$ for event-based perceived risk and the peak of continuous perceived risk respectively; for Dataset Obstacle Avoidance, $q=event$ and $q=peak$ denote the $RMSE$ for event-based perceived risk and the maximum steering wheel angle separately. In Equation \eqref{eq:rmse}, $\hat{y}_k$ represents the model output, while $y_k$ refers to the perceived risk data. The variable $k$, which falls within the set of ${1,2,3,...,K}$, represents the event number in the specific dataset. $K$ signifies the number of available events in different datasets, with $K=414$ for Dataset Merging and $K=2496$ for Dataset Obstacle Avoidance. Note that the first sample point of the kinematic data when the obstacle suddenly appears in Dataset Obstacle Avoidance is used for the calibration since the participants were asked to give a verbal perceived risk rating as soon as the obstacle appeared. 

The calibration aims to minimise $\sum RMSE_q$ for all models by tuning the key model parameters based on perceived risk and corresponding kinematic data. Given the differences in ranges of perceived risk data across two datasets and the outputs of four models, we employ min-max feature scaling. This approach scales perceived risk data and model output to a range of [0,10] during model calibration and comparison, as shown in Equation \eqref{eq:linear scaling}. 
\begin{equation}
    \hat{z}_k = \frac{z_k- z_\text{min}}{z_{\text{max}} - z_\text{min}} \times 10
\label{eq:linear scaling}
\end{equation}
where $\hat{z}_k$ is the scaled model output or perceived risk data; $z_\text{max}$ and $z_\text{min}$ are the maximum and minimum value of the model outputs or perceived risk data of a specific participant regarding one dataset. 

\subsection{Performance indicators} \label{Performance indices}
We use five indicators to evaluate the model performance: Correlation, Model error, Detection rate, Computation cost, and Linear Time Scaling Factor.
\begin{itemize}
\item {\textbf{Correlation}}\label{chap:R-square}
The predicted perceived risk has to be correlated with the event-based perceived risk. We use R-Square to quantify how well model outputs fit the real perceived risk. Since the perceived risk output by the models is linearly rescaled to 0-10 (Equation \eqref{eq:linear scaling}), different ranges of perceived risk data or model outputs have no influence on R-Square in our case.
\item {\textbf{Model error}} \label{chap:computation accuracy}
We use Root Mean Squared Error (RMSE) to quantify a model's overall Model error, which is the same as the model calibration criterion (Equation \eqref{eq:rmse}). This indicator reflects the model's ability to compute the overall perceived risk in a certain dataset. A model with a smaller $RMSE$ can more accurately predict the overall perceived risk for a given scenario. 

\item {\textbf{Detection rate}}\label{chap:detection rate}
The Detection rate represents the model's ability to detect a risky event that is also perceived as dangerous by human drivers. We defined Detection rate as in Equation \eqref{eq:rate_detection}
\begin{linenomath}
\begin{equation}
    R_{det} = \frac{K_{detected}}{K_{event}} \times 100\%
    \label{eq:rate_detection}
\end{equation}
\end{linenomath}
where $K_{detected}$ represents the number of events where the model manages to detect the risk with non-zero output; $K_{event}$ is the total number of the events where human drivers gave perceived risk ratings in a certain dataset with $K_{event} = 414$ for Dataset Merging and $K_{event} = 2496$ for Dataset Obstacle Avoidance. A model with a higher detection rate can recognise more events that are also perceived as dangerous by human drivers. 
\item {\textbf{Computation cost}}
It is essential that all models possess real-time risk computation capability, so the computation cost is critical. More complex models may offer a better performance in other aspects such as Model errors but tend to take longer to compute. We define the computation cost as the model's computation time per computation step. If the time consumption per computation step exceeds the on-board computation capability, it means that the computation of perceived risk cannot be completed in real-time on-board. 
\end{itemize}

The above metrics validate the event-based perceived risk. We also compared the continuous perceived risk measured for Dataset Merging. However, we observed that participants pressed the button following a fixed pattern regardless of the actual real-time risk level. This pattern suggests that the timing of their responses was more likely influenced by the given instructions and their interpretation, rather than reflecting a valid measure of continuous perceived risk over time. Consequently, these responses, although appearing as a 'continuous perceived risk', do not offer reliable time-domain information. Due to this lack of time-domain validation, we have chosen not to report on the validation of continuous perceived risk.

\section{Model evaluation results}\label{chap:Results}
In this section, we illustrate the applicability of the four models and evaluate their performance with the performance indicators introduced previously regarding the two datasets.

\subsection{Model calibration results}
According to the model structure and dataset features, the parameters to be calibrated are listed in Table \ref{tab:Calibrated parameter value}. The calibration is conducted separately for the two datasets. To facilitate analysis, we linearly rescale the perceived risk to 0-10 in Dataset Obstacle Avoidance to match the range in Dataset Merging based on Equation \eqref{eq:linear scaling}.

\begin{table}[H]
\resizebox{\textwidth}{!}{%
    \centering
    \begin{threeparttable} 
    \caption{Calibrated parameters for all models}
\label{tab:Calibrated parameter value}
\begin{tabular}{M{1.5cm}|M{2cm}|M{7cm}|M{2cm}|M{2cm}}
\hline
Model             & Parameters & Explanation & Values for Dataset Merging & Values for Dataset Obstacle Avoidance  \\ \hline
\multirow{7}{*}{PCAD} &$\sigma_{n,X}$& The standard deviation in $X$ of the velocity Gaussian of a neighbouring vehicle   & 4.28 & / \\ \cline{2-5} 
                     & $\sigma_{n,Y}$& The standard deviation in $Y$ of the velocity Gaussian of a neighbouring vehicle  & 3.86 & /        \\ \cline{2-5} 
                    & $\sigma_{s,X}$& The standard deviation in $X$ of the velocity Gaussian of the subject vehicle  & 0.80 & 6.58       \\ \cline{2-5} 
                    & $\sigma_{s,Y}$& The standard deviation in $Y$ of the velocity Gaussian of the subject vehicle  & 1.70 & 1.20      \\ \cline{2-5} 
                     & $t_{s,a}$& The accumulation time for the acceleration-based velocity of the subject vehicle & 0.13 & /      \\ \cline{2-5} 
                    & $t_{n,a}$& The accumulation time for the acceleration-based velocity of a neighbouring vehicle & 0.01 & /      \\ \cline{2-5} 
                    & $\alpha$& The exponent of the power function in weighting function  & $0.52^{[1]}$ &  / \\ \hline
\multirow{3}{*}{RPR} &$C_0$& The intercept in the regression model   & 12.10 & 20.70 \\ \cline{2-5} 
                     & $C_1$&The coefficient of gap to the leading vehicle & -3.70 &-3.68        \\ \cline{2-5} 
                     & $C_2$& The coefficient of leading vehicle's braking intensity& -0.36 & 0      \\ \hline

\multirow{3}{*}{PPDRF} & $\tilde{\sigma}_x$ &The standard deviation of longitudinal acceleration distribution of neighbouring vehicle & 2.01& /  \\ \cline{2-5} 
                    & $\tilde{\sigma}_y$ &The standard deviation of lateral acceleration distribution of neighbouring vehicle  & 0.02& / \\ \cline{2-5} 
                      & $D$ & The steepness of descent of the potential field& / &  0.14     \\ \hline
\multirow{4}{*}{DRF} &$s$ & The steepness of the height parabola of the risk field &0.15&  0.005    \\ \cline{2-5} 
                    &$t_{la}$& Human driver's preview time & 1.20 &  8.12 \\ \cline{2-5}
                    &$m$& The rate of the risk field width expanding & \SI{3.98e-8}{} &  \SI{3.66e-4}{}  \\ \cline{2-5}
                    &$c$& The initial width of the DRF & 0.45 &  1.10     \\ \hline
\end{tabular}%
 \begin{tablenotes}
      \small
      \item $^1$ This is the calibrated value regarding the specific dataset. Due to the lack of subject velocity change, $\alpha$ has limited influence on Dataset Merging. $\alpha$ ranging on [0, 2.5] leads to an R-square ranging on [0.80, 0.90]. For Dataset Obstacle Avoidance, $\alpha$ was set to 0 since it almost has no influence. Additionally, the $v_{ref}$ in Weighting function $\mathcal{W}$ was set to \SI{27.78}{m/s} for Dataset Merging and \SI{25}{m/s} for Dataset Obstacle Avoidance.

    \end{tablenotes}
    \end{threeparttable}
x}
\end{table}
 
\subsection{Performance evaluation results}
We test the four models using both datasets including the perceived risk data and the corresponding kinematic data with the calibrated parameters shown in Table \ref{tab:Calibrated parameter value}. The following sections present different aspects of performance. 
\subsubsection{Correlation}
The correlation between predicted and measured event-based perceived risk data plays a crucial role in assessing the performance of risk assessment models. This is particularly important given the uncertainty in defining the unit of perceived risk. Figure \ref{fig:Rsq_Merging} and Figure \ref{fig:Rsq_OV} display the correlation between the predicted perceived risk and event-based perceived risk for Dataset Merging and Dataset Obstacle Avoidance respectively. The adjusted R-Square is calculated based on the averaged event-based perceived risk across the same event type (\CIRCLE in Figure \ref{fig:Rsq_Merging} and Figure \ref{fig:Rsq_OV} ). 

In both datasets, the PCAD model demonstrates a superior correlation with event-based perceived risk data compared to other models. Furthermore, the regression models RPR and DRF exhibit strong performance in Dataset Merging and Dataset Obstacle Avoidance, respectively, where they were originally developed.

\begin{figure}[H]
\centering
\begin{subfigure}{0.46\textwidth}
  \centering
\includegraphics[width=\textwidth]{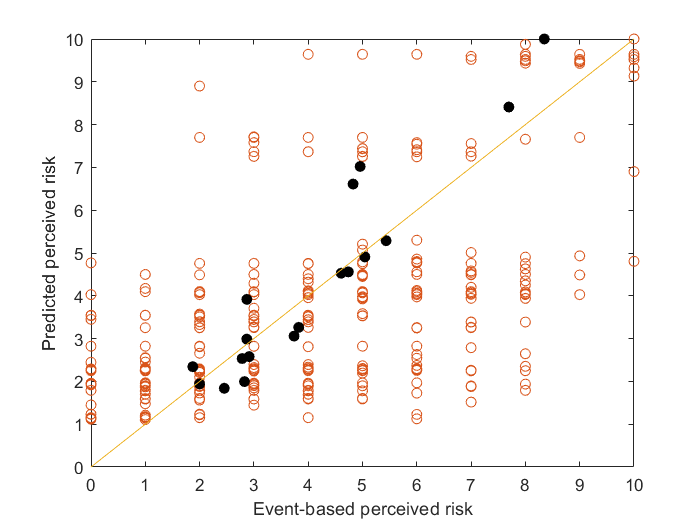}
\subcaption[]{PCAD (Adjusted R-Square = 0.90)}
\end{subfigure}
\begin{subfigure}{0.46\textwidth}
  \centering
\includegraphics[width=\textwidth]{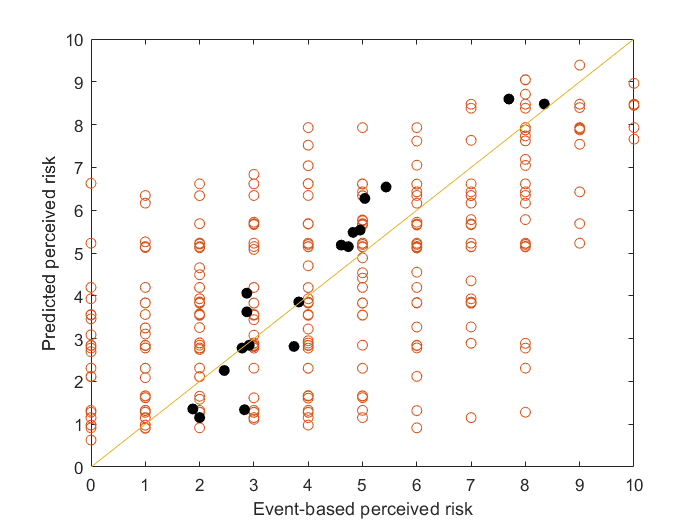}
\subcaption[]{RPR (Adjusted R-Square = 0.90)}
\end{subfigure} \\%
\begin{subfigure}{0.46\textwidth}
  \centering
\includegraphics[width=\textwidth]{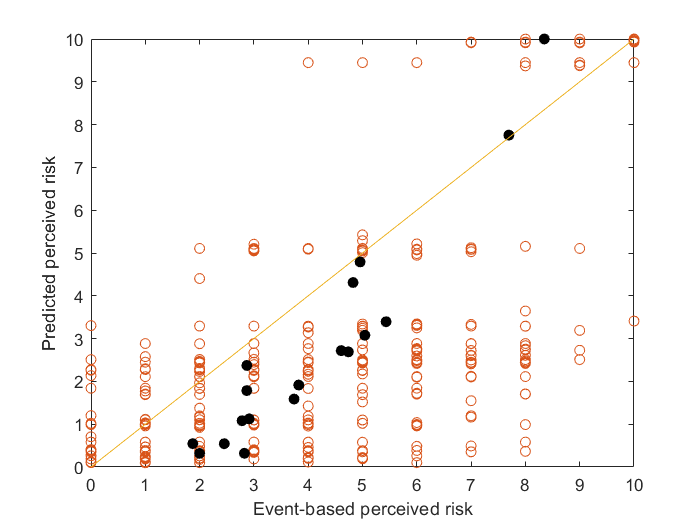}
\subcaption[]{PPDRF (Adjusted R-Square = 0.90)}
\end{subfigure}
\begin{subfigure}{0.46\textwidth}
  \centering
\includegraphics[width=\textwidth]{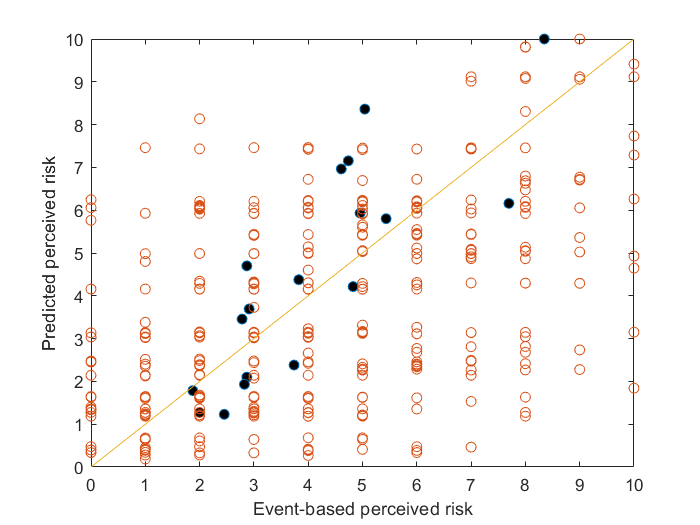}
\subcaption[]{DRF (Adjusted R-Square = 0.67)}
\end{subfigure} 
\end{figure}

\begin{figure}[H] \ContinuedFloat
\centering
\begin{subfigure}{0.46\textwidth}
     \centering
\includegraphics[width=\textwidth]{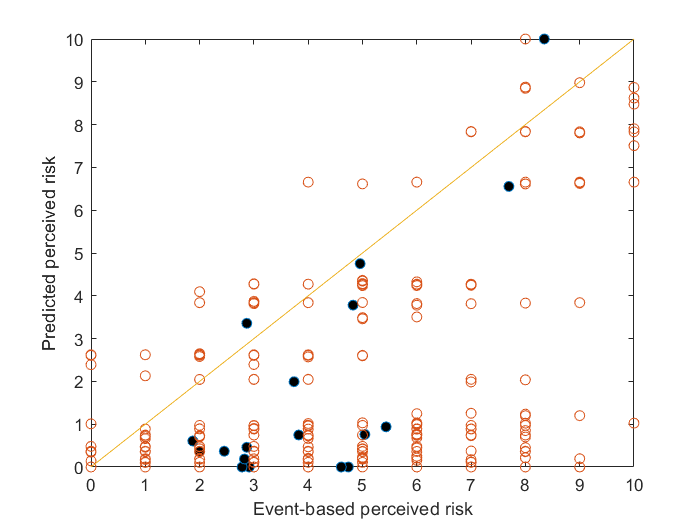}
\subcaption[]{PCAD without uncertainties (Adjusted R-Square = 0.58)}
\label{fig:Rsq_Merging_without_uncertainties}
\end{subfigure}
\caption{Predicted and measured event-based perceived risk in Dataset Merging. ``\Circle''  indicates individual event-based perceived risk and ``\CIRCLE''  indicates the averaged event-based perceived risk across the same event type. }
\label{fig:Rsq_Merging}
\end{figure}

\begin{figure}[H]
\centering
\begin{subfigure}{0.46\textwidth}
  \centering
\includegraphics[width=\textwidth]{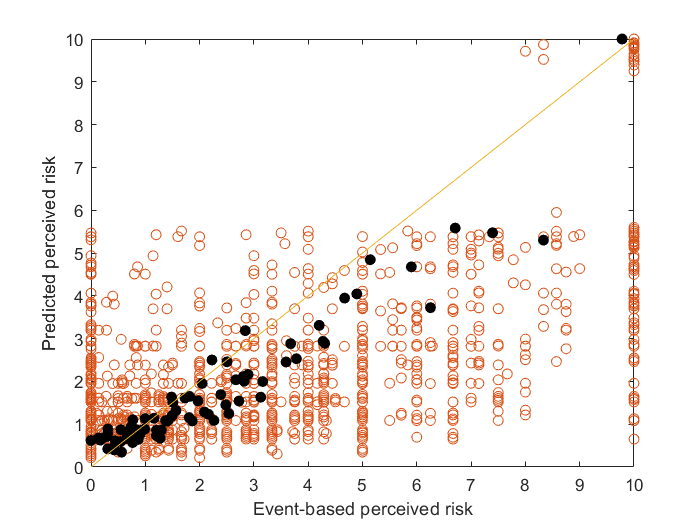}
\subcaption[]{PCAD (Adjusted R-Square = 0.90)}
\end{subfigure}%
\begin{subfigure}{0.46\textwidth}
  \centering
\includegraphics[width=\textwidth]{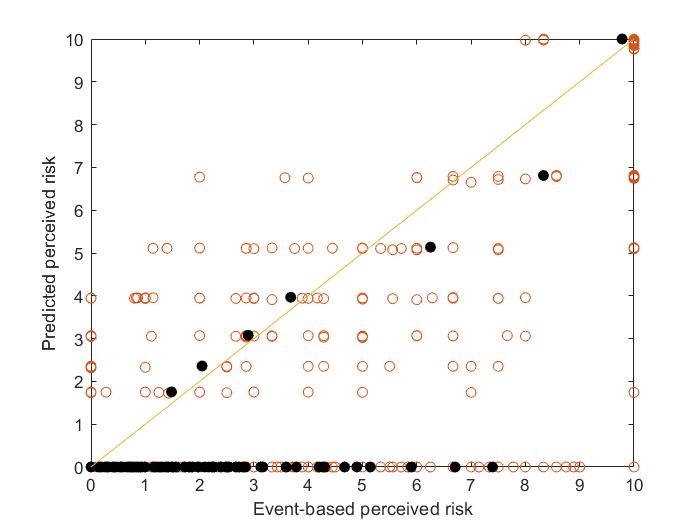}
\subcaption[]{RPR (Adjusted R-Square = 0.38)}
\end{subfigure}%

\begin{subfigure}{0.46\textwidth}
  \centering
\includegraphics[width=\textwidth]{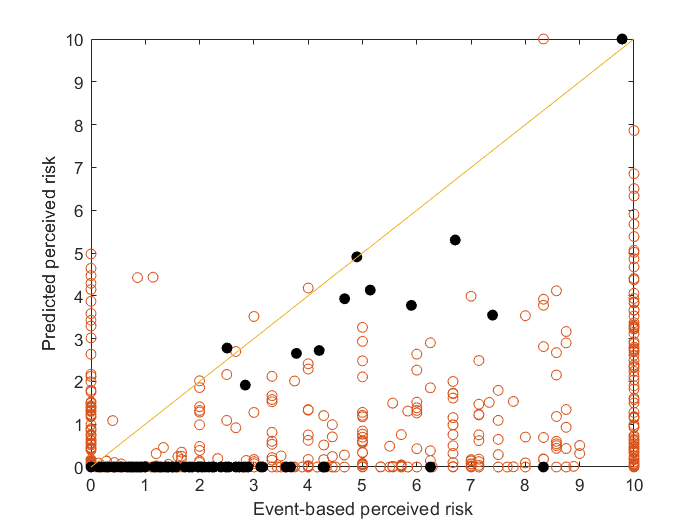}
\subcaption[]{PPDRF (Adjusted R-Square = 0.50)}
\end{subfigure}
\begin{subfigure}{0.46\textwidth}
  \centering
\includegraphics[width=\textwidth]{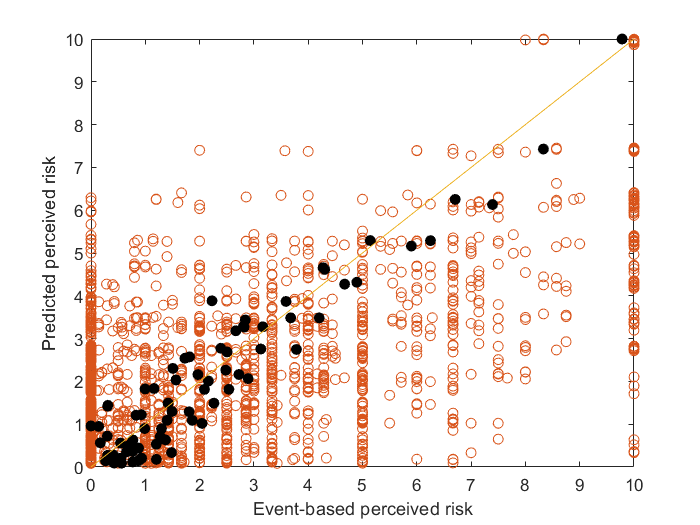}
\subcaption[]{DRF (Adjusted R-Square = 0.90)}
\end{subfigure}
\end{figure}

\begin{figure}[H] \ContinuedFloat
\centering
\begin{subfigure}{0.46\textwidth}
    \centering
\includegraphics[width=\textwidth]{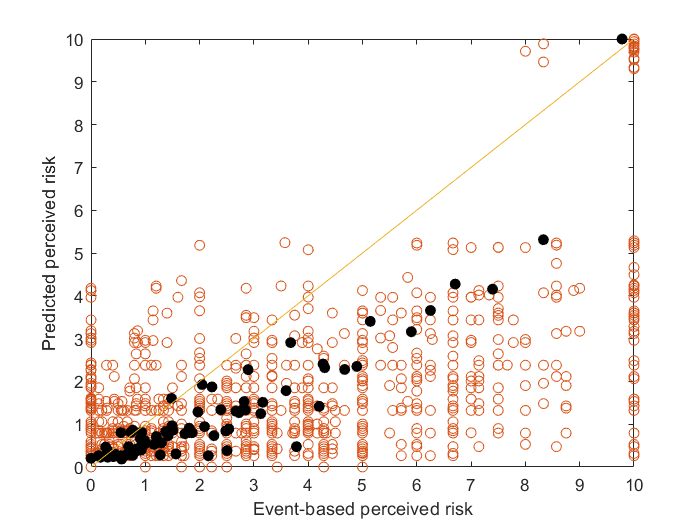}
\subcaption[]{PCAD without uncertainties (Adjusted R-Square = 0.81)}
\label{fig:Rsq_OV_without_uncertainties}
\end{subfigure}
\caption{Predicted and measured event-based perceived risk in Dataset Obstacle Avoidance. ``\Circle''  indicates raw event-based perceived risk and ``\CIRCLE''  indicates the averaged event-based perceived risk across the same event type. Note that in (c), there are many dots with marginal values but they are actually detected by PPDRF.}
\label{fig:Rsq_OV}
\end{figure}
\subsubsection{Model error}
As discussed in Section \ref{chap:computation accuracy}, the Root Mean Square Error ($RMSE$) is an indicator of the overall Model error. Table \ref{tab:model performance} presents the RMSE values for all four models across the two datasets.

The $RMSE$ values reveal that the PCAD model achieves a comparable performance level to the regression models (e.g., RPR in Dataset Merging and DRF in Dataset Obstacle Avoidance), albeit with a slightly larger model error. In Dataset Merging, the lower $RMSE$ values for PCAD and RPR suggest better performance, as these models directly incorporate the neighbouring vehicle's acceleration. In Dataset Obstacle Avoidance, the lower $RMSE$ values for PCAD and DRF indicate superior performance, as these models also consider lateral perceived risk, resulting in reduced model error when applied to a 2-D dataset (Table \ref{tab:model performance}). The PPDRF model, adapted from PDRF, was originally designed to evaluate actual collision risk in traffic, and demonstrates moderate performance in both datasets.

\subsubsection{Detection rate}
As per Equation \ref{eq:rate_detection}, the detection rates for the four models across both datasets are presented in Table \ref{tab:model performance}. In Dataset Merging, the merging vehicle primarily poses longitudinal risk in the same lane. Consequently, all models are capable of detecting dangerous events, regardless of whether they are 1-D or 2-D models. However, in Dataset Obstacle Avoidance, the obstacles are dispersed across a 2-D plane. As a result, only models that account for lateral risk can effectively identify dangerous vehicles outside the forward path. This leads to a lower detection rate for the RPR model, while the other three models are able to recognise all dangerous events that human drivers also perceive as risky. Note that Figure \ref{fig:Rsq_OV}(e) displays outputs with marginal values that appear to be zero but are, in fact, detected by PPDRF.

\subsubsection{Computation cost}
Table \ref{tab:model performance} presents the computation cost for different models, tested on a workstation with an Intel Core i7-8665U 1.9Ghz processor and 8GB RAM. Generally, models that take more factors into account require longer computation times. In both datasets, RPR is the fastest model, as it only involves logarithmic calculations.

In Dataset Merging, PCAD is the most time-consuming model since it relies on a grid search to identify the optimal velocity gap vector to the safe velocity region. PPDRF requires spatial overlap computations and multiple integrals over variations of acceleration probability density function in the overlap area, making it a time-intensive process. Although DRF involves discretising a 2D area of an object or a vehicle with a grid and summing the risk values over each grid cell to obtain the final perceived risk, its overlap computations are simpler than those of PPDRF, as the risk field and severity field are static and no motion prediction of neighbouring vehicles is needed.

In Dataset Obstacle Avoidance, PPDRF takes less time than DRF and PCAD, as it only computes potential risk, which is a simpler process compared to the kinetic risk computation in Dataset Merging.

\begin{table}[H]
\resizebox{\textwidth}{!}{%
    \centering
    \begin{threeparttable} 
    \caption{Model performance represented by the performance indicators}
\begin{tabular}{c|c|c|c|c|c}
\hline
                               Dataset     & Performance indicators&   PCAD         & RPR      & PPDRF   & DRF$^{[6]}$  \\ \hline
\multirow{6}{*}{Dataset Merging}    & $RMSE_{event}$   & 2.25    & 2.18    & 2.76 &2.58  \\ \cline{2-6} 
                                    & $RMSE_{peak}$ &   3.41     & 3.39   & 3.73 &3.35  \\ \cline{2-6} 
                                    & Adjusted R-Square   &   0.90 & 0.90   & 0.90 & 0.67 \\ \cline{2-6} 
                                    & Detection   rate  & 1.00   & 1.00    & 1.00& 1.00  \\  \cline{2-6}
                                    & Time consumption (ms)$^{[1]}$  & 2.79         & $1.77\times10^{-4}$    & $6.14^{[5]}$  &1.30 \\  
                                    \hline
\multirow{6}{*}{Dataset Obstacle Avoidance} & $RMSE_{event}$  & 2.27 & 3.20   & 3.34  & 2.17 \\ \cline{2-6} 
                                    & $RMSE_{peak}$  & 2.71 & 3.84   & 4.02 & 2.61 \\ \cline{2-6} 
                                    & Adjusted R-Square   &    0.90      & 0.38   & 0.50 & 0.90 \\ \cline{2-6} 
                                    & Detection   rate  & 1.00   & $0.09^{[4]}$  & 1.00 & 1.00 \\   \cline{2-6}
                                    & Time consumption (ms)  $^{[2]}$  &  $6.70^{[3]}$  & $2.01\times10^{-4}$   & $1.08\times10^{-2}$ & 1.22\\ 
                                    \hline
\end{tabular}%

 \begin{tablenotes}
      \small
      \item $^1$ The average value of computing 124614 steps.
      \item $^2$ The average value of computing 349440 steps.
      \item $^3$ PCAD consumed more time in Dataset Obstacle Avoidance because the searching algorithm worked in a larger searching area to find the velocity gap $\boldsymbol{v}_g$.
      \item $^4$ Only the vehicles directly in front of the vehicle can be detected by RPR, which leads to a low detection rate. See \citep{Kolekar2020WhichField} for more experiment details.
      \item $^5$ PPDRF consumed much more time in Dataset Merging because the model contains numerical integral when facing moving vehicles. 
      \item $^6$ The best performance was obtained only when the subject velocity was fixed as $\SI{27.78}{m/s}$ and $\SI{25}{m/s}$ for a constant preview distance. 
    \end{tablenotes}
    \label{tab:model performance}
    \end{threeparttable}
}
\end{table}

\subsubsection{Summary of model performance evaluation}
\begin{figure}[H]
\centering
\begin{minipage}{0.46\textwidth}
  \centering
\includegraphics[width=\textwidth]{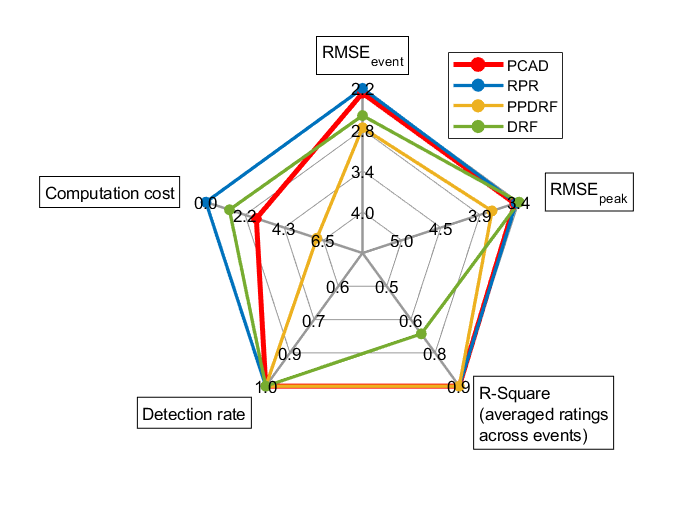}
\subcaption[]{Performances in Dataset Merging}
\end{minipage}%
\begin{minipage}{0.46\textwidth}
  \centering
\includegraphics[width=\textwidth]{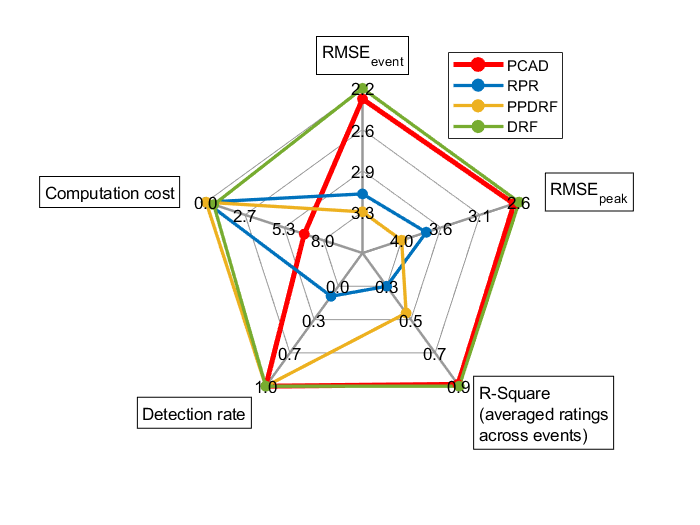}
\subcaption[]{Performances in Dataset Obstacle Avoidance}
\end{minipage}
\caption{Radar charts of model performance indicators in two datasets}
\label{fig:radar chart}
\end{figure}

Based on the results discussed above, we utilise radar charts to illustrate the performance of each model across various aspects, as shown in Figure \ref{fig:radar chart}. Generally, PCAD demonstrates strong performance in terms of overall model error, R-square, and detection rate. However, the primary drawback of PCAD is its high computation cost, which results from its complexity. The regression models (i.e., RPR in Dataset Merging and DRF in Obstacle Avoidance) exhibit the best performance in their respective datasets. We remark that the advantage of PPDRF in capturing the motion uncertainties of the surrounding vehicles vanishes in the second dataset due to the specific experimental setting. As a result, the PPDRF models used in the two datasets are two different models. This largely explains the poor performance of PPDRF, albeit it clearly showed advantages in the analytical model properties. 

It is worth mentioning that PCAD demonstrates excellent and consistent performance across both datasets. We also conducted cross-validation between the two datasets, in which the four models and their parameters were calibrated using one dataset and then used to predict perceived risk in the other dataset. PCAD performs quite well even without re-calibration. As seen in Figure \ref{fig:radar chart_validation}, Figure \ref{fig:Rsq_validation1}, and Figure \ref{fig:Rsq_validation2} in \ref{chap:Appendix B}, PCAD maintains its strong performance in cross-validation. Additionally, this suggests that the calibration process has a low risk of overfitting, further highlighting the robustness of the PCAD model. 

\section{Discussion} \label{chap:discussions}
In this paper, we present a novel computational perceived risk model based on the Risk Allostasis Theory \citep{Fuller1984,Fuller1999TheProcess,Fuller2011DriverAllostasis}. Our new model successfully quantifies event-based perceived risk and the peak of continuous perceived risk in both longitudinal and lateral directions. We validate the model on two datasets of human drivers' perceived risk, comparing its performance with three typical perceived risk models. Our work contributes to addressing the challenge of perceived risk computation for SAE Level 2 driving automation, while also illustrating the mechanisms underlying human drivers' risk perception.

Traditional perceived risk models consider collision probability and the collision consequence \citep{Naatanen1976Road-userAccidents}, such as DRF and PPDRF, while our PCAD model is developed based on the concept of potential collision judgement, which originates from aerospace and maritime experience \citep{Ward2015ExtendingScenarios}. PCAD demonstrates superior performance in estimating perceived risk in various driving conditions, aligning with the argument of \citet{mckenna1982human, Rundmo2017DoesExist} that drivers are incapable of monitoring infrequent event probabilities, thus supporting the underlying theory of PCAD. 

The demonstrated superior performance of our PCAD model unveils new insights into perceived risk. Firstly, PCAD considers all motion information in Table \ref{tab:Model properties}, highlighting the importance of position, velocity, and acceleration for risk perception. Secondly, the models that can capture lateral risk lead to a higher detection rate in Dataset Obstacle Avoidance, indicating that perceived risk is 2-D and human drivers perceive the risk from all directions in a 2-D plane when driving. Thirdly, motion uncertainties of the subject vehicle and other road users cause extra perceived risk, which is supported by \citet{Kolekar2020WhichField} and \citet{Ding2014}. Lastly, perceived risk in driving scenarios is a dynamic concept and varies with the changing traffic conditions as illustrated in Figure \ref{fig:risk field visualization}, which presents the perceived risk variances in three distinct driving conditions (i.g., different relative velocities, subject velocities and accelerations). This observation motivates the need for models, such as the proposed PCAD model, which can adjust to varying driving scenarios even without recalibration.

We note that our study still has limitations. There is only one traffic object considered in this study. If multi-road users or even infrastructure are added, PCAD still has the potential to estimate perceived risk. In that case, we need to compute the potential collision avoidance difficulty for multiple neighbouring vehicles, and derive the total perceived risk by addition, or by creating a combined safe velocity region. 

Dataset Merging covers human drivers perceived risk with SAE Level 2 driving automation where drivers need to monitor the system and environment and be ready to intervene. This makes this data suitable to assess perceived safety when using automation, but the lateral risk is lacking. 

Dataset Obstacle Avoidance contains human drivers perceived risk data in 2-D, including lateral perceived risk. However, this dataset is collected from human-automation transitions (i.g., human drivers' taking-over process in this case), which may cause bias in automated driving studies. Additionally, the objects in the experiment are fixed and suddenly displayed during driving. The additional perceived risk caused by surprise cannot be neglected. 

Dataset Merging measured perceived risk on a scale from 0 to 10 for no to very high risk, while Dataset Obstacle Avoidance captured perceived risk as a non-negative real number without predetermined upper limit. To facilitate comparison, all results from Dataset Obstacle Avoidance and all models were scaled to 0-10 to match the scale used in Dataset Merging. More specific scales of perceived risk can be developed for experimental studies, including factors such as accident risk and severity, and the driver's tendency or need to intervene and overrule the driving automation.

To further advance perceived risk modelling, we recommend collecting more perceived risk data in various scenarios by online surveys with videos, simulator experiments and on-road observation. In the meantime, the model potential for perceived risk estimation in curve driving and multi-object interaction at different driving automation levels will be studied. Moreover, internal HMIs have positive effects reducing human drivers' perceived risk and the perceived risk modelling will be further improved to fit different internal HMI conditions \citep{Kim2023,Jouhri2023}. Our PCAD model can also be used as a cost function, a constraint, or a reference of perceived risk in driving automation path planning, decision making, or controller design, enhancing trust \citep{Hu2021Trust-BasedPerformance} and acceptance.

\section{Conclusions}\label{chap:Conclusions and future work}
In this study, we have formulated, calibrated, and validated a novel computational perceived risk model, and compared its performance with three well-established models across two different datasets. Our findings reveal valuable insights into the understanding and quantification of perceived risk in various driving situations. The key conclusions drawn from our analysis are as follows:
(1) Driving task difficulty serves as an effective indicator of perceived risk; (2) Perceived risk is 2-dimensional, originating from both longitudinal and lateral directions, and exhibits a non-linear increase as the distance to surrounding vehicles decreases; (3) Incorporating uncertainties in the modelling process is crucial for a more accurate representation of perceived risk; (4) Perceived risk is dynamic and changes with specific driving conditions, highlighting the necessity for models that can capture these variations with the least calibration efforts. 

\section*{Authorship contribution statements}
\textbf{Xiaolin He:} Conceptualisation, Methodology, Software, Formal analysis, Writing - Original Draft, Writing - review \& editing, Visualisation. \textbf{Riender Happee:} Conceptualisation, Methodology, Writing - review \& editing, Supervision, Project administration, Funding acquisition. \textbf{Meng Wang:} Conceptualisation, Methodology, Writing - review \& editing, Supervision, Project administration, Funding acquisition. 
\section*{Declaration of interests}
The authors declare no conflict of interest.
\section*{Acknowledgements}
The authors would like to thank Toyota Motor Europe NV/SA for the support in conducting the simulator experiment. This research is also supported by the SHAPE-IT project funded by the European Union’s Horizon 2020 research and innovation programme under the Marie Skłodowska-Curie grant agreement 860410.

\bibliography{references}
\bibliographystyle{apalike}
\newpage

\appendix
\section{PCAD time history output}\label{chap:Appendix A}
\begin{figure}[H]
\centering
\begin{minipage}[t]{0.7\textwidth}
\centering
\includegraphics[width=\textwidth]{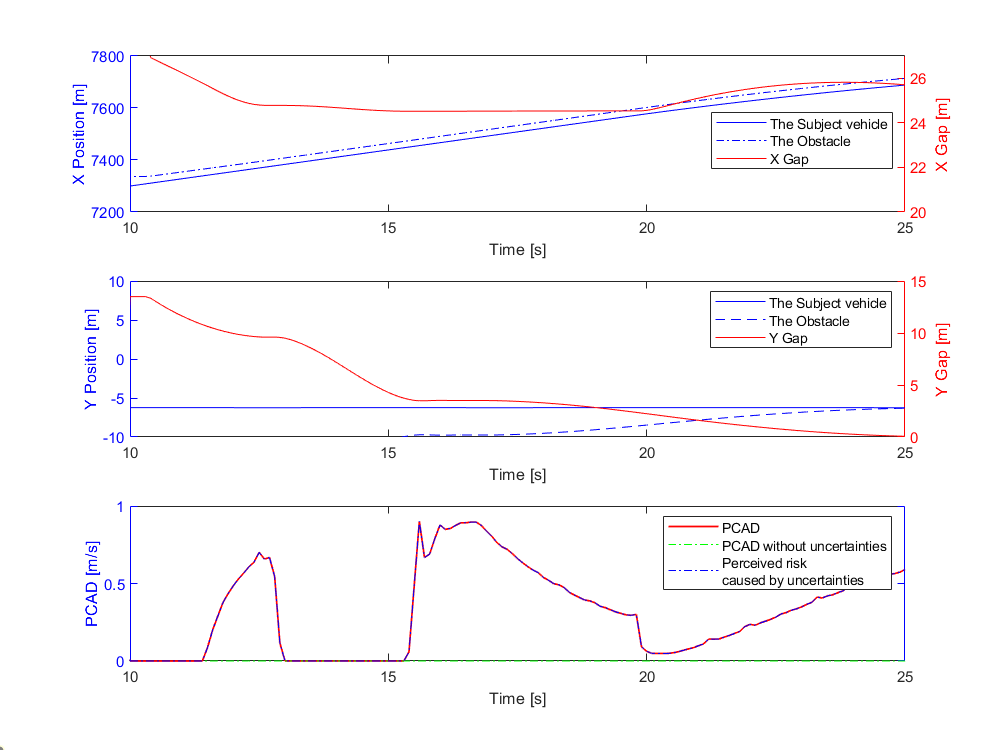}
\subcaption[]{Merging gap is $\SI{25}{m}$ and the deceleration of the merging vehicle is $\SI{-2}{m/s^2}$}
\end{minipage} \qquad
\begin{minipage}[t]{0.7\textwidth}
\centering
\includegraphics[width=\textwidth]{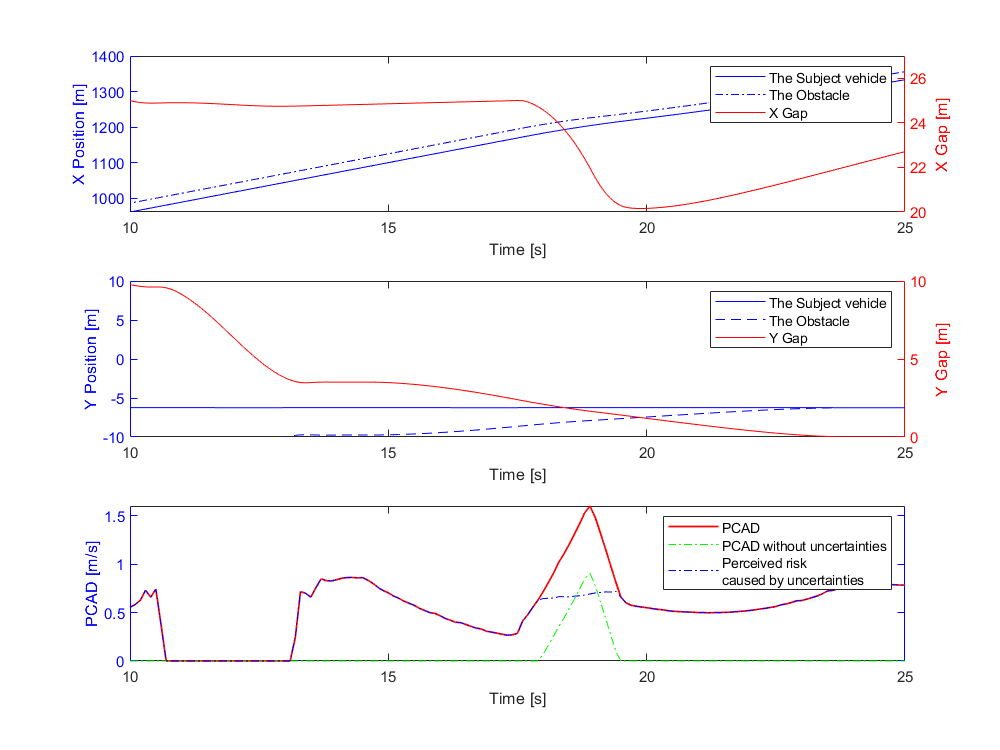}
\subcaption[]{Merging gap is $\SI{25}{m}$ and the deceleration of the merging vehicle is $\SI{-8}{m/s^2}$}
\end{minipage}%
\caption{PCAD historical output in merging and hard braking events}
\label{fig:PCAD output in a merging event}
\end{figure}

\begin{figure}[H]
\centering
\begin{minipage}[t]{0.7\textwidth}
\centering
\includegraphics[width=\textwidth]{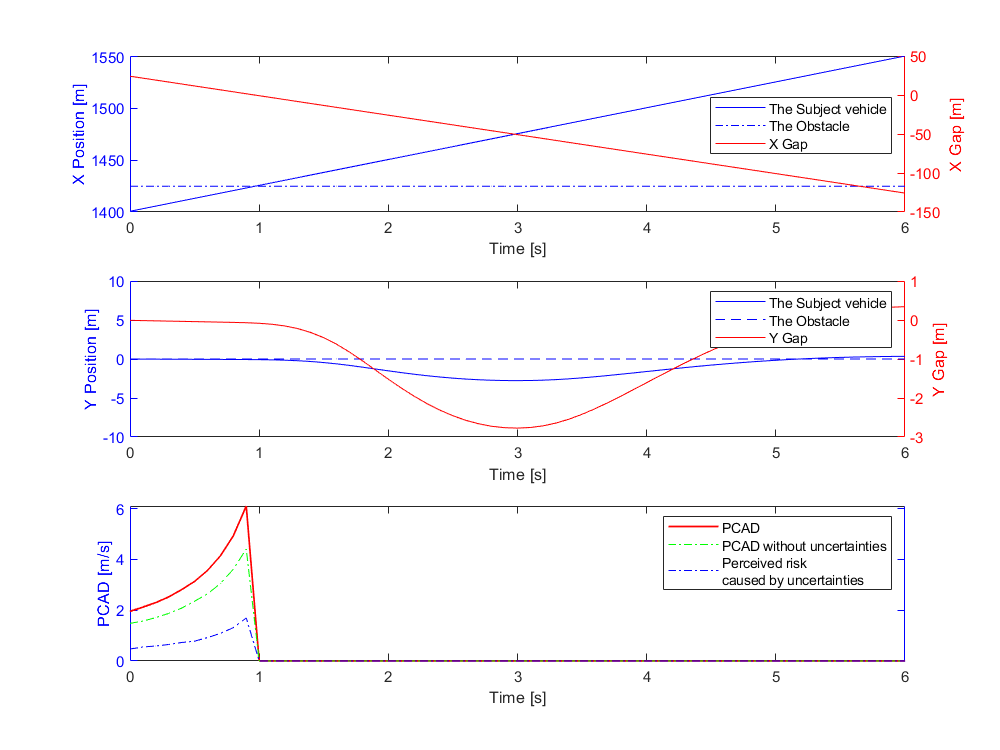}
\subcaption[]{The obstacle position is (25, 0), which means the obstacle is $\SI{25}{m}$ in front and the lateral offset is 0.}
\end{minipage} \qquad
\begin{minipage}[t]{0.7\textwidth}
\centering
\includegraphics[width=\textwidth]{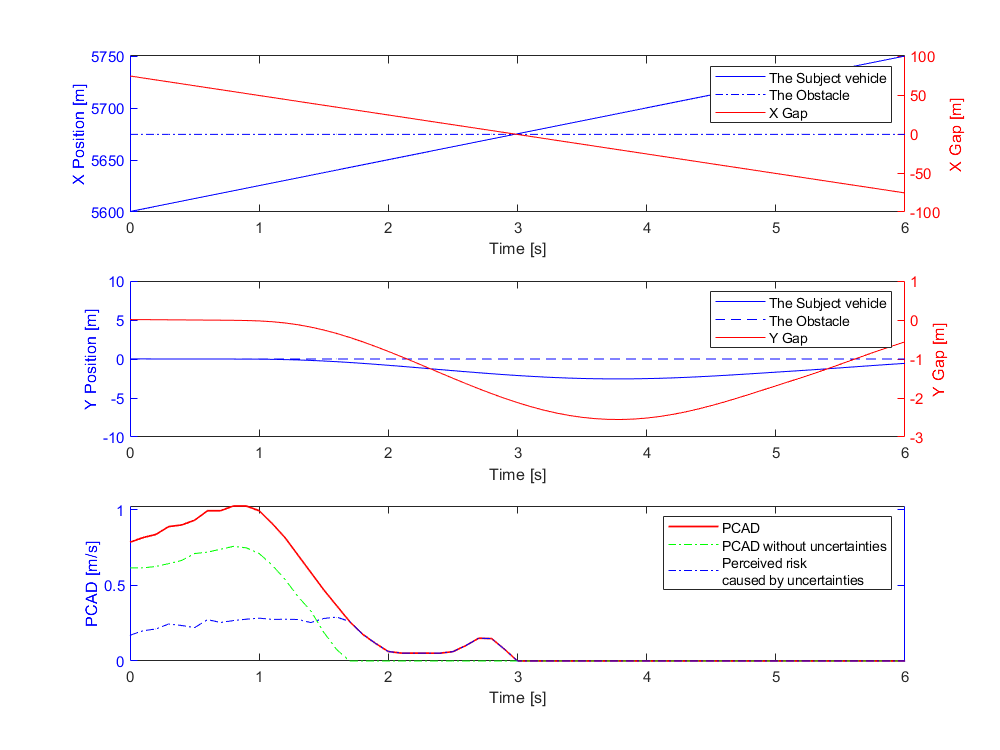}
\subcaption[]{The obstacle position is (75, 0), which means the obstacle is $\SI{75}{m}$ in front and the lateral offset is 0.}
\end{minipage}%
\caption{PCAD historical output in obstacle avoidance events}
\label{fig:PCAD output in an obstacle avoidance event}
\end{figure}

This Appendix presents the PCAD time history outputs in Dataset Merging (Figure \ref{fig:PCAD output in a merging event} and Dataset Obstacle Avoidance (Figure \ref{fig:PCAD output in an obstacle avoidance event}).

\clearpage

\section{Cross validation} \label{chap:Appendix B}
This Appendix presents the model performance in cross-validation between the two datasets.
\begin{table}[H]
\resizebox{\textwidth}{!}{%
    \centering
    \begin{threeparttable} 
    \caption{Model performance indicators for the cross-validation}
\begin{tabular}{c|c|c|c|c|c}
\hline
              Dataset              &Performance indicators&   PCAD         & RPR      & PPDRF   & DRF  \\ \hline
\multirow{6}{*}{\begin{tabular}[c]{@{}c@{}}
Dataset Merging\\ 
(with parameters calibrated\\
with Dataset Obstacle Avoidance)
\end{tabular}} & $RMSE_{event}$  & 2.37 & 7.72   & 4.86  & 5.14 \\ \cline{2-6} 
                                    & $RMSE_{peak}$  & 3.73 & 8.29   & 5.14 & 5.79  \\ \cline{2-6} 
                                    & Adjusted R-Square   &    0.79    & 0.88   & 0.20 & 0.00 \\ \cline{2-6} 
                                    & Detection   rate  & 1.00   & 1.00   & 1.00 & 1.00 \\   \cline{2-6}
                                    & Time consumption (ms)   &  3.25      & $2.09\times10^{-4}$   & $1.03\times10^{-2}$ & 1.01\\ 
                                    \hline

\multirow{6}{*}{\begin{tabular}[c]{@{}c@{}}
Dataset Obstacle Avoidance\\ 
(with parameters calibrated\\
with Dataset Merging)
\end{tabular}} & $RMSE_{event}$  & 2.28 & 3.20   & 3.48  & 3.19 \\ \cline{2-6} 
                                    & $RMSE_{peak}$  & 2.73 & 3.84   & 3.93 & 3.87  \\ \cline{2-6} 
                                    & Adjusted R-Square   &    0.90    & 0.38   & 0.11 & 0.42 \\ \cline{2-6} 
                                    & Detection   rate  & 1.00   & 0.09   & 1.00 & 1.00 \\   \cline{2-6}
                                    & Time consumption (ms)   &  5.58      & $1.98\times10^{-4}$   & $7.28$ & 1.82\\ 
                                    \hline
\end{tabular}%

    \label{tab:model performance for cross validation}
    \end{threeparttable}
}
\end{table}

\begin{figure}[H]
\centering
\begin{minipage}{0.46\textwidth}
  \centering
\includegraphics[width=\textwidth]{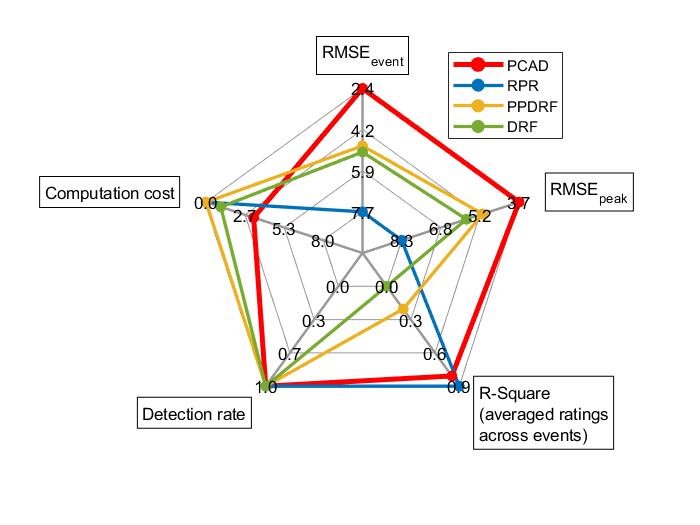}
\subcaption[]{Performances in Dataset Merging (with parameters calibrated with Dataset Obstacle Avoidance)}
\end{minipage} \qquad
\begin{minipage}{0.46\textwidth}
  \centering
\includegraphics[width=\textwidth]{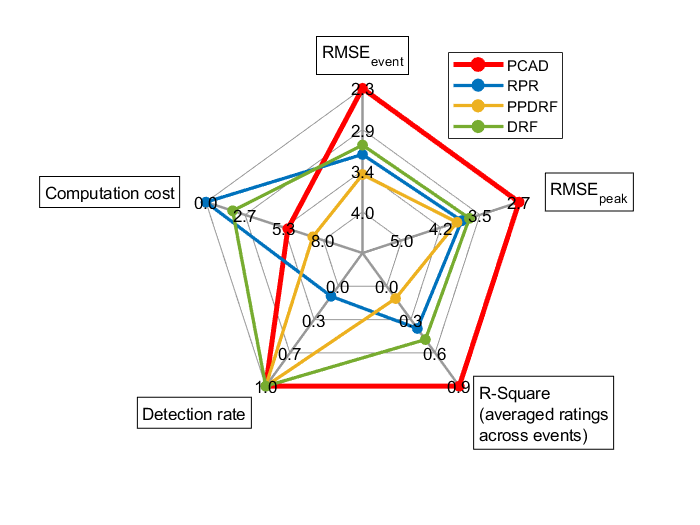}
\subcaption[]{Performances in Dataset Obstacle Avoidance (with parameters calibrated with Dataset Merging)}
\end{minipage}
\caption{Radar chart of model performance indicators for the cross-validation}
\label{fig:radar chart_validation}
\end{figure}

\begin{figure}[H]
\centering
\begin{subfigure}{0.46\textwidth}
  \centering
\includegraphics[width=\textwidth]{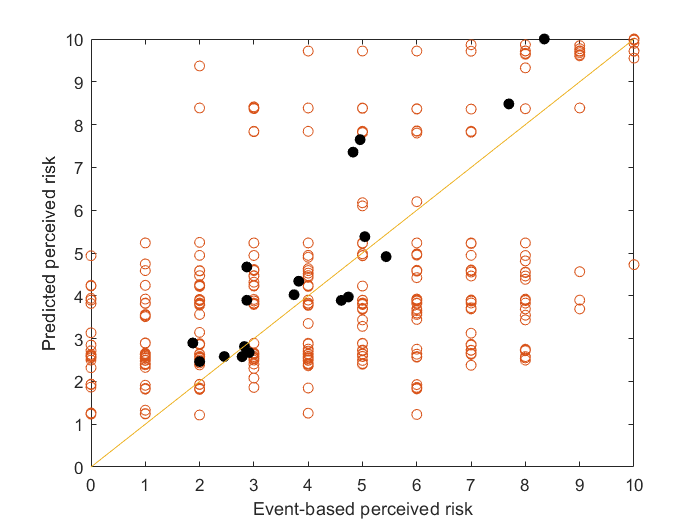}
\subcaption[]{PCAD (Adjusted R-Square = 0.79)}
\end{subfigure}%
\begin{subfigure}{0.46\textwidth}
  \centering
\includegraphics[width=\textwidth]{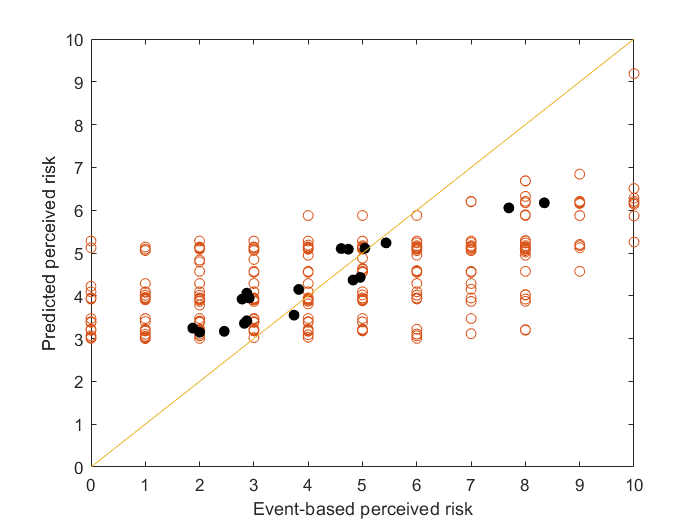}
\subcaption[]{RPR (Adjusted R-Square = 0.88)}
\end{subfigure}%

\begin{subfigure}{0.46\textwidth}
  \centering
\includegraphics[width=\textwidth]{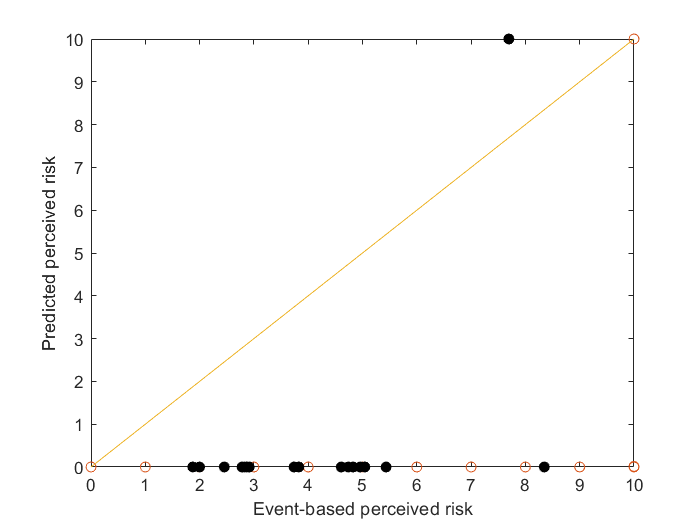}
\subcaption[]{PPDRF (Adjusted R-Square = 0.20)}
\end{subfigure}
\begin{subfigure}{0.46\textwidth}
  \centering
\includegraphics[width=\textwidth]{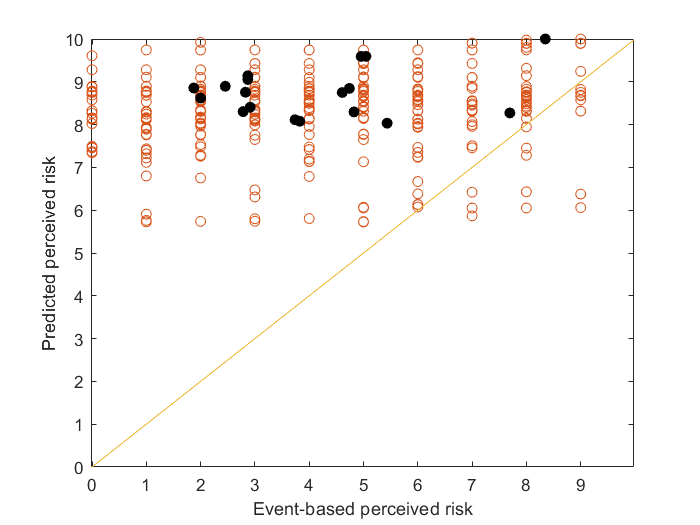}
\subcaption[]{DRF (Adjusted R-Square = 0.00)}
\end{subfigure}

\caption{Validation results in Dataset Merging with model parameters calibrated based on Dataset Obstacle Avoidance. '\Circle'  indicates raw event-based perceived risk and '\CIRCLE'  indicates the averaged event-based perceived risk across the same event type. }
\label{fig:Rsq_validation1}
\end{figure}

\begin{figure}[H]
\centering
\begin{subfigure}{0.46\textwidth}
  \centering
\includegraphics[width=\textwidth]{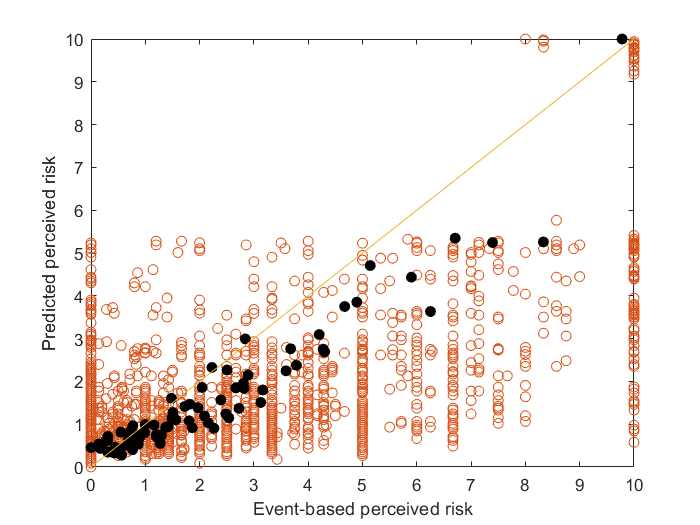}
\subcaption[]{PCAD (Adjusted R-Square = 0.90)}
\end{subfigure}%
\begin{subfigure}{0.46\textwidth}
  \centering
\includegraphics[width=\textwidth]{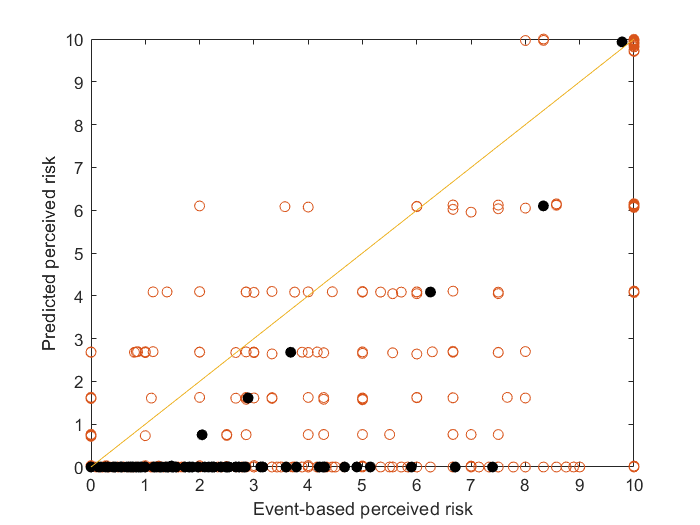}
\subcaption[]{RPR (Adjusted R-Square = 0.38)}
\end{subfigure}%

\begin{subfigure}{0.46\textwidth}
  \centering
\includegraphics[width=\textwidth]{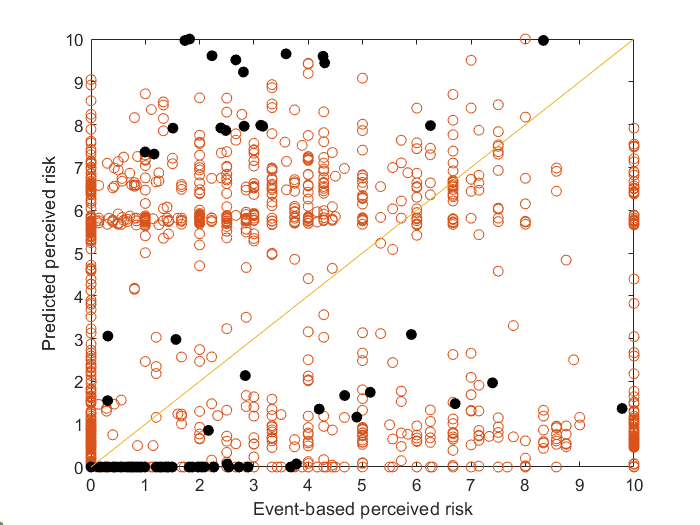}
\subcaption[]{PPDRF (Adjusted R-Square = 0.11)}
\end{subfigure}
\begin{subfigure}{0.46\textwidth}
  \centering
\includegraphics[width=\textwidth]{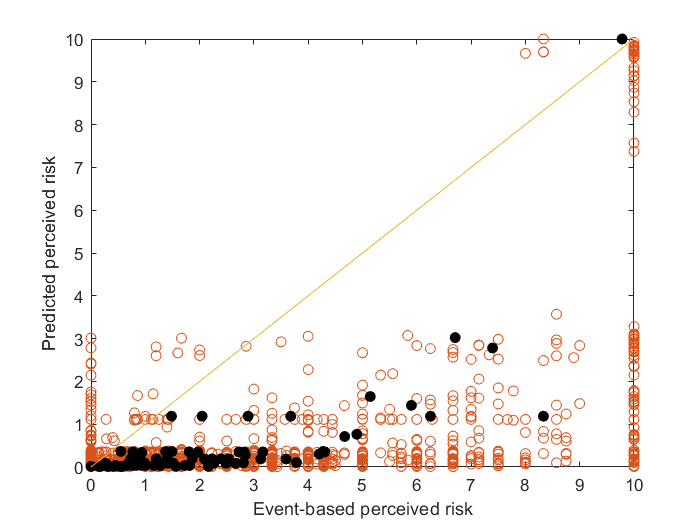}
\subcaption[]{DRF (Adjusted R-Square = 0.48)}
\end{subfigure}

\caption{Validation results in Dataset Obstacle Avoidance with model parameters calibrated based on Dataset Merging. '\Circle'  indicates raw event-based perceived risk and '\CIRCLE'  indicates the averaged event-based perceived risk across the same event type. }
\label{fig:Rsq_validation2}
\end{figure}

\clearpage

\section{The explanation of the imaginary velocity direction} \label{chap:Appendix C}
In this Appendix, we explain why the line connecting the subject vehicle and the neighbouring vehicle is selected as the direction of the imaginary velocity. 

With the imaginary velocity and the perceived velocity derived from acceleration, Equation \eqref{eq:angular velocity} and \eqref{eq:distance changing rate} should be changed to
\begin{equation}
\begin{aligned}
    \dot{\boldsymbol{\theta}}_{sj_1,nj_2}'&=\frac{\left(\boldsymbol{p}_{sj_1}-\boldsymbol{p}_{nj_2}\right) \times (\boldsymbol{v}_{sj_1}'-\boldsymbol{v}_{nj_2}')}{\left\|\boldsymbol{p}_{sj_1}-\boldsymbol{p}_{nj_2}\right\|^2}\\
    &=\frac{\left(\boldsymbol{p}_{sj_1}-\boldsymbol{p}_{nj_2}\right) \times \left[(\boldsymbol{v}_{sj_1}+\Delta\boldsymbol{v}_{s,a}+\boldsymbol{v}_{s,I})-(\boldsymbol{v}_{nj_2}+\Delta\boldsymbol{v}_{n,a}+\boldsymbol{v}_{n,I})\right]}{\left\|\boldsymbol{p}_{sj_1}-\boldsymbol{p}_{nj_2}\right\|^2} \\
    &= \dot{\boldsymbol{\theta}}_{sj_1,nj_2} + \Delta\dot{\boldsymbol{\theta}}_{sj_1,nj_2,a}+\dot{\boldsymbol{\theta}}_{sj_1,nj_2,I}, \  \ j_1,j_2 \in \{l,r\}
\end{aligned}
\label{eq: adjusted angular velocity}
\end{equation}
and
\begin{equation}
\begin{aligned}
        \dot{d}_{s,n}'&=\frac{1}{ d_{s,n}}(\boldsymbol{p}_{s}-\boldsymbol{p}_{n})^T\left(\boldsymbol{v}_s'-\boldsymbol{v}_n'\right)\\
        &=\frac{1}{ d_{s,n}}(\boldsymbol{p}_{s}-\boldsymbol{p}_{n})^T\left[(\boldsymbol{v}_s+\Delta\boldsymbol{v}_{s,a}+\boldsymbol{v}_{s,I})-(\boldsymbol{v}_n+\Delta\boldsymbol{v}_{n,a}+\boldsymbol{v}_{n,I})\right]\\
        &= \dot{d}_{s,n}+\dot{d}_{n,a}+\dot{d}_{s,n,I}
\end{aligned}
\label{eq:adjusted distance changing rate}
\end{equation}
where $\boldsymbol{v}_s'=\boldsymbol{v}_s+\Delta \boldsymbol{v}_{s,a}+\boldsymbol{v}_{s,I}$ and $\boldsymbol{v}_n'=\boldsymbol{v}_n+\Delta\boldsymbol{v}_{n,a}+\boldsymbol{v}_{n,I}$ are the perceived velocity with the imaginary velocity $\boldsymbol{v}_{s,I}$ and $\boldsymbol{v}_{n,I}$ for the subject vehicle and the neighbouring vehicle respectively. 

In order to make the situation more dangerous, the imaginary velocity has to create a situation that is opposite to Equation \eqref{eq:velocity set} namely
\begin{equation}
    \min \dot{\boldsymbol{\theta}}_{si,nj,I} \cdot \max \dot{\boldsymbol{\theta}}_{si,nj,I} \leqslant 0 \; (i,j \in \{l,r\}, )\; \text{and} \; \dot{d}_{s,n,I}<0, \; 
\label{eq: not in adjusted velocity set}
\end{equation}
where all relative bearing rate and the distance changing rate are only caused by the imaginary velocity. 

Comparing with Equation \eqref{eq:velocity set}, the optimal direction for the imaginary velocity based on Equation \eqref{eq:adjusted distance changing rate} to create a negative distance changing rate should be 
\begin{equation}
\begin{gathered}
    \left[\begin{array}{l}
         \frac{\partial \dot{d}_{s,n,I}}{\partial  v_{s,X,I}} \\
         \frac{\partial \dot{d}_{s,n,I}}{\partial  v_{s,Y,I}} 
    \end{array}\right]=
    \left[\begin{array}{l}
         \frac{\partial \dot{d}_{s,n}'}{\partial  v_{s,X,I}} \\
         \frac{\partial \dot{d}_{s,n}'}{\partial  v_{s,Y,I}} 
    \end{array}\right]=
    \frac{1}{d_{s,n}}\left[\begin{array}{l} 
\Delta \boldsymbol{p}_X \\
\Delta \boldsymbol{p}_Y
\end{array}\right] = \frac{\boldsymbol{p}_s-\boldsymbol{p}_n}{||\boldsymbol{p}_s-\boldsymbol{p}_n||} \\
    \left[\begin{array}{l}
         \frac{\partial \dot{d}_{s,n,I}}{\partial  v_{n,X,I}}  \\
         \frac{\partial \dot{d}_{s,n,I}}{\partial  v_{n,Y,I}} 
    \end{array}\right]=
    \left[\begin{array}{l}
         \frac{\partial \dot{d}_{s,n}'}{\partial  v_{n,X,I}}  \\
         \frac{\partial \dot{d}_{s,n}'}{\partial  v_{n,Y,I}} 
    \end{array}\right]
    =\frac{1}{d_{s,n}}\left[\begin{array}{l}
-\Delta \boldsymbol{p}_X \\
-\Delta \boldsymbol{p}_Y
\end{array}\right] = -\frac{\boldsymbol{p}_s-\boldsymbol{p}_n}{||\boldsymbol{p}_s-\boldsymbol{p}_n||}
\end{gathered}
\label{eq: imaginary velocity gradient}
\end{equation}
where $\Delta \boldsymbol{p}_X=\boldsymbol{p}_{s,X}-\boldsymbol{p}_{n,X}$, $\Delta \boldsymbol{p}_Y=\boldsymbol{p}_{s,Y}-\boldsymbol{p}_{n,Y}$. 

Equation \eqref{eq: imaginary velocity gradient} means that the distance direction is the optimal for the imaginary velocity of the subject vehicle and the neighbouring vehicle to create a negative distance changing rate in Equation \eqref{eq: not in adjusted velocity set}.

Simultaneously, the imaginary velocity should not cause extra relative bearing rate which makes the situation less dangerous. In other words, the imaginary velocity should follow the normal direction of the relative bearing rate gradient. Hence, the direction should be
\begin{equation}
\begin{gathered}
    \left[\begin{array}{l}
         \frac{\partial \dot{\theta}_{s,n,I}}{\partial  v_{s,X,I}} \\
         \frac{\partial \dot{\theta}_{s,n,I}}{\partial  v_{s,Y,I}} 
    \end{array}\right]\Bigg|_{\perp}=
    \left[\begin{array}{l}
         \frac{\partial \dot{\theta}_{s,n}'}{\partial  v_{s,X,I}} \\
         \frac{\partial \dot{\theta}_{s,n}'}{\partial  v_{s,Y,I}} 
    \end{array}\right]\Bigg|_{\perp}
    =\frac{1}{d_{s,n}^2}\left[\begin{array}{l} 
\Delta \boldsymbol{p}_Y \\
-\Delta \boldsymbol{p}_X
\end{array}\right] \Bigg|_{\perp} =  \frac{\boldsymbol{p}_s-\boldsymbol{p}_n}{||\boldsymbol{p}_s-\boldsymbol{p}_n||^2}\\
    \left[\begin{array}{l}
         \frac{\partial \dot{\theta}_{s,n}'}{\partial  v_{n,X,I}}  \\
         \frac{\partial \dot{\theta}_{s,n}'}{\partial  v_{n,Y,I}} 
    \end{array}\right] \Bigg|_{\perp}
    =\frac{1}{d_{s,n}^2}\left[\begin{array}{l}
-\Delta \boldsymbol{p}_Y \\
\Delta \boldsymbol{p}_X
\end{array}\right] \Bigg|_{\perp} = -\frac{\boldsymbol{p}_s-\boldsymbol{p}_n}{||\boldsymbol{p}_s-\boldsymbol{p}_n||^2}
\end{gathered}
\label{eq: angular velocity normal}
\end{equation}
which are exactly the directions shown in Equation \eqref{eq: imaginary velocity gradient}. That means the distance direction is the optimal direction for the imaginary velocity to create a negative distance changing rate and in the meantime not to cause less perceived risk.

\end{document}